\documentclass[journal=jacsat,manuscript=article]{achemso}
\setkeys{acs}{articletitle=true}
\usepackage{chemformula} 
\usepackage[T1]{fontenc} 
\usepackage{multirow}
\usepackage{array}
\usepackage{makecell}
\usepackage[version=3]{mhchem} 

\usepackage{graphicx}
\usepackage{xcolor}
\usepackage{lineno}
\usepackage{csquotes}
\usepackage{nicefrac}
\usepackage{caption}
\usepackage{subcaption}
\usepackage[]{graphicx}
\usepackage{url}
\PassOptionsToPackage{hyphens}{url}
\usepackage{hyperref}
\usepackage{soul} 

\usepackage{color}
\hypersetup{
        colorlinks=true,
        citecolor=blue,
        urlcolor  = black,
        linkcolor = blue
}
\usepackage{pifont} 
\usepackage{latexsym}
\usepackage{chemfig}
\usepackage{multicol}
\usepackage{setspace}
\usepackage{caption} 
\usepackage{float}
\usepackage{caption}
\usepackage{enumitem}
\usepackage{acro}
\usepackage{chngcntr}
\DeclareCaptionLabelSeparator{pipe}{|\,}
\usepackage{amsmath}

\author{Hendrik~Kra\ss}
\affiliation{Computer Science, University of Tübingen, Germany}
\alsoaffiliation{Chemical Engineering \& Applied Chemistry, University of Toronto, Toronto, Ontario M5S 3E5, Canada}

\author{Ju~Huang}
\affiliation{Chemical Engineering \& Applied Chemistry, University of Toronto, Toronto, Ontario M5S 3E5, Canada}

\author{Seyed~Mohamad~Moosavi}
\affiliation{Chemical Engineering \& Applied Chemistry, University of Toronto, Toronto, Ontario M5S 3E5, Canada}
\email{mohamad.moosavi@utoronto.ca}

\title{MOFSimBench: Evaluating Universal Machine Learning Interatomic Potentials In Metal--Organic Framework Molecular Modeling}

\DeclareAcronym{mof}{
    short = MOF,
    long  = metal-organic framework,
}

\DeclareAcronym{cof}{
    short = COF,
    long  = covalent organic framework,
}

\DeclareAcronym{dft}{
    short = DFT,
    long  = Density Functional Theory,
}

\DeclareAcronym{mlip}{
    short = MLIP,
    long  = machine learning interatomic potential,
}

\DeclareAcronym{umlip}{
    short = uMLIP,
    long  = universal machine learning interatomic potential,
}

\DeclareAcronym{dac}{
    short = DAC,
    long  = direct air capture,
}

\DeclareAcronym{mbe}{
    short=MBE,
    long=many-body expansion,
}

\DeclareAcronym{ace}{
    short=ACE,
    long=atomic cluster expansion,
}

\DeclareAcronym{md}{
    short=MD,
    long=molecular dynamics,
}

\DeclareAcronym{pbc}{
    short=PBC,
    long=periodic boundary conditions,
}

\DeclareAcronym{fs}{
    short=fs,
    long=femtosecond,
}

\DeclareAcronym{ps}{
    short=ps,
    long=picosecond,
}

\DeclareAcronym{ml}{
    short=ML,
    long=machine learning,
}

\DeclareAcronym{mpnn}{
    short=MPNN,
    long=message-passing neural network,
}

\DeclareAcronym{pes}{
    short=PES,
    long=potential energy surface,
}

\DeclareAcronym{eos}{
    short=EOS,
    long=equation of state,
}

\DeclareAcronym{uff}{
    short=UFF,
    long=Universal Force Field,
}

\begin{document}

\begin{abstract}

Universal machine learning interatomic potentials (uMLIPs) have emerged as powerful tools for accelerating atomistic simulations, offering scalable and efficient modeling with accuracy close to quantum calculations. However, their reliability and effectiveness in practical, real-world applications remain an open question. Metal-organic frameworks (MOFs) and related nanoporous materials are highly porous crystals with critical relevance in carbon capture, energy storage, and catalysis applications. Modeling nanoporous materials presents distinct challenges for uMLIPs due to their diverse chemistry, structural complexity, including porosity and coordination bonds, and the absence from existing training datasets. Here, we introduce MOFSimBench, a benchmark to evaluate uMLIPs on key materials modeling tasks for nanoporous materials, including structural optimization, molecular dynamics (MD) stability, the prediction of bulk properties, such as bulk modulus and heat capacity, and guest-host interactions. Evaluating over 20 models from various architectures on a chemically and structurally diverse materials set, we find that top-performing uMLIPs consistently outperform classical force fields and fine-tuned machine learning potentials across all tasks, demonstrating their readiness for deployment in nanoporous materials modeling.
Our analysis highlights that data quality, particularly the diversity of training sets and inclusion of out-of-equilibrium conformations, plays a more critical role than model architecture in determining performance across all evaluated uMLIPs.
We release our modular and extendable benchmarking framework at \url{https://github.com/AI4ChemS/mofsim-bench}, providing an open resource to guide the adoption for nanoporous materials modeling and further development of uMLIPs.

\end{abstract}

\section{Introduction} \label{sec:introduction}

Molecular modeling is a powerful tool for understanding the structure and interactions of molecules and materials, and it plays a crucial role in predicting material properties to accelerate discovery \cite{frenkel2023understanding, Tuckerman2023-ik,Noe2020}. However, computing interatomic interactions in a molecular simulation is bound to an accuracy-efficiency trade-off: \textit{Ab initio} quantum methods offer high accuracy, but are computationally expensive and do not scale well to large systems; in contrast, empirical potentials are computationally efficient, but often lack sufficient accuracy for practical applicability \cite{musil2021physics,huang2023central,xie_ultra-fast_2023}.
\Acp{mlip} have emerged as a promising tool to bridge this gap between efficiency and accuracy \cite{Behler2007, Deringer2017_gap, batatia_macehigherorderequivariant_2023, lysogorskiy_performant_2021_pace, xie_ultra-fast_2023, schutt_schnet_2017, Batzner2022_nequip, liao2024equiformerv2improvedequivarianttransformer, fu2025learningsmoothexpressiveinteratomic}. Their flexible functional forms enable the learning of complex atomic interactions and have enabled simulations of challenging phenomena that require scaled molecular simulations, including those that require simulating large systems \cite{musaelian_scaling_2023} and long time-scales, slow dynamics \cite{harper2025tracking}.
\Acp{mlip} have successfully modeled phenomena such as the liquid-liquid phase transition of hydrogen under pressure \cite{bischoff_hydrogen_2024} and the growth dynamics of carbon nanotubes \cite{Hedman2024nanotubes}, which were previously inaccessible using traditional methods. 

Despite these successes, MLIPs are often trained for specific materials or narrow classes of systems, requiring custom datasets, hyperparameter tuning, and extensive validation, significantly limiting their generality and increasing development costs.
To overcome these limitations, universal \acp{mlip} (\acsp{umlip}) have been introduced to benefit from the scale of data and models \cite{Chen2022_m3gnet, deng_chgnet_2023_mptraj, batatia_foundation_2024_mace}. These models are trained on large, diverse datasets, covering a wide range of materials and chemical elements, enabling more general deployment of \acp{mlip} in many domains \cite{Chen2022_m3gnet, deng_chgnet_2023_mptraj, batatia_foundation_2024_mace, barroso-luque_open_2024_omat24, yang_mattersim_2024,neumann2024orbfastscalableneural, rhodes_orb-v3_2025, Bochkarev_graph_2024_grace}. \acsp{umlip} have shown strong performance on predicting crystal stability \cite{riebesell_matbench_2023}, catalysis \cite{batatia_foundation_2024_mace}, and thermal conductivity \cite{póta_thermalconductivitypredictionsfoundation_2024_ksrme}, and have demonstrated good stability in long \ac{md} simulations \cite{neumann2024orbfastscalableneural}. 

The promise of \enquote{universal} interatomic potentials is compelling. However, their real-world deployment often encounters additional challenges, such as out-of-distribution (OOD) materials and tasks, which require robust testing of the models. The emergence of open benchmarks and evaluation frameworks, such as Matbench Discovery \cite{riebesell_matbench_2023}, has accelerated progress in the field by enabling rapid and consistent comparison of models. However, these general-purpose benchmarks primarily emphasize metrics closely aligned with the training objectives (e.g., energy, force, and stress fitting), which may not adequately reflect model performance on domain-relevant tasks or OOD scenarios.\cite{focassio_performance_2025} To bridge this gap, it is increasingly recognized that general benchmarks must be complemented with domain-specific evaluations tailored to the materials classes of interest. Recent efforts have begun incorporating downstream assessments, such as molecular dynamics (MD) stability and prediction of materials properties \cite{póta_thermalconductivitypredictionsfoundation_2024_ksrme,loew2024universalmachinelearninginteratomic_phonons,fu2023forces}, which may offer more reliable indicators of practical modeling performance.

This study focuses on applications of \acp{umlip} for \acp{mof}, a class of highly porous materials with applications in carbon capture \cite{lin2021scalable, ye2025architecting, boyd2019data, chen2025flexibility} energy storage \cite{chen2024tailoring, shin2023microscopic, gittins2024understanding}, and catalysis \cite{mourino2025exploring, yu2025tailored, fumanal2020charge}, and related nanoporous materials. \Acp{mof} are formed by self-assembly of metal nodes and organic linkers, resulting in modular porous frameworks with infinite design space \cite{moosavi2020understanding}. Their low symmetry and large unit cells make \textit{ab initio} quantum calculations prohibitively expensive for large-scale or long-timescale dynamic simulations. As a result, large-scale molecular modeling of \ac{mof}s has traditionally relied on classical force fields \cite{evans2016computational,islamov2023high,moosavi2018improving}. While specialized classical force fields have been developed for a few prototypical frameworks, such as HKUST-1~\cite{tafipolsky2010first} and the IRMOF-1 series \cite{dubbeldam2007exceptional}, they collectively cover only a small fraction of known MOFs. Consequently, the \ac{uff} \cite{rappe1992uff}, originally introduced in 1992, remains the most widely used interatomic potential for MOFs. As \ac{uff} was not parametrized for MOFs, several later efforts have adapted universal force fields to MOF systems, including UFF4MOF \cite{Coupry2016_uff4mof} and MOF-FF \cite{Bureekaew2013_mof-ff}, with the aim of including additional coordination environments through reparameterization. However, these force fields still fail to cover the diverse chemistries of MOFs as they retain a rigid functional form and rely on fixed parameters for specific coordination geometries \cite{boyd_force-field_2017}. For example, a copper center may require different angle parameters depending on its local coordination geometry, variations that these force fields cannot accommodate unless explicitly parameterized and manually specified. Despite these limitations, \ac{uff} and its derivatives remain widely used due to the absence of general-purpose alternatives.

\acp{umlip} provide an opportunity to address these shortcomings by offering quantum-level accuracy at a computational cost closer to classical force fields. Particularly, recent \acp{umlip} development papers report proof-of-principle demonstration on MOFs, showing that uMLIPs can reproduce \textit{ab initio} energies and forces with encouraging accuracy \cite{batatia_foundation_2024_mace} or show stable \ac{md} trajectories on selected structures \cite{neumann2024orbfastscalableneural,brabson2025comparing}.
However, whether this agreement extends to a larger variety of MOFs or derived properties, for example, elastic moduli, heat capacities, or guest-host interaction energies, remains unclear. Furthermore, efforts to fine-tune uMLIPs specifically on MOF data \cite{Lim_2025_DAC-SIM,elena2024machinelearnedpotentialhighthroughput_mace_mof_0}, while improving performance, restrict the chemical domain of validity and thus compromise the \enquote{universal} scope of the model. Consequently, a systematic, task-oriented benchmark is required to quantify the capabilities and limitations of current uMLIPs for MOF modeling to inform their broader adoption in nanoporous materials research.

In this work, we develop a comprehensive benchmark to assess the performance of \mbox{\acp{umlip}} in modeling MOFs and related nanoporous materials. We evaluate models on three key categories of tasks: 1) static modeling and geometry optimization, 2) dynamic modeling stability, including volume drift and changes in local coordination environments in \textit{NpT} \ac{md} trajectories, and 3) property prediction, including bulk modulus, constant-volume heat capacity, and host-guest interaction energy relevant to adsorption.
Reference data for these tasks is either obtained from our \ac{dft} calculations or curated from literature sources \cite{rosen_machine_2021_qmof, sours_predicting_2023, moosavi_data-science_2022, boyd_force-field_2017, Lim_2025_DAC-SIM}.
We evaluate over twenty \acp{umlip}, spanning a diverse range of architectures, including equivariant graph neural networks, graph network-based simulators, and transformer models, trained on different large-scale datasets. These are compared against established MOF potentials: the \ac{uff} force field \cite{rappe1992uff, Coupry2016_uff4mof} and a fine-tuned MOF-specific \acp{umlip} baseline \cite{elena2024machinelearnedpotentialhighthroughput_mace_mof_0, Lim_2025_DAC-SIM}. Our results provide critical insights into the current capabilities, limitations, and promising architectural directions for \acp{umlip} in \ac{mof} and nanoporous materials modeling. This benchmark is intended to guide the practical deployment of \acp{umlip} and inform their future development for accelerating the discovery and design of functional porous materials.


\section{Benchmark Design}

Our benchmark includes over 20 state-of-the-art \acp{umlip}, spanning a range of architectural choices and training datasets. The selected architectures include equivariant graph neural networks (GNNs) (MACE \cite{batatia_foundation_2024_mace, batatia_macehigherorderequivariant_2023}, MatterSim \cite{yang_mattersim_2024}, SevenNet \cite{park_scalable_2024_sevennet}), graph transformers (GT) (EquiformerV2 (eqV2) \cite{liao2024equiformerv2improvedequivarianttransformer, barroso-luque_open_2024_omat24}, eSEN \cite{fu2025learningsmoothexpressiveinteratomic}), graph network-based simulators (GNS) (orb-v2 \cite{neumann2024orbfastscalableneural}, orb-v3 \cite{rhodes_orb-v3_2025}), and the graph-basis function-based GRACE \cite{Bochkarev_graph_2024_grace} (see details in the method section). Including other models to the benchmark is straightforward. 
While most architectures rely on forces computed from energy gradients, orb-v2 \cite{neumann2024orbfastscalableneural} and eqV2 \cite{liao2024equiformerv2improvedequivarianttransformer, barroso-luque_open_2024_omat24} output predictions directly as a model output, resulting in non-conservative forces. This promises a computational speedup because no backward pass through the neural network is performed during inference, but forces are not guaranteed to be exact negative derivatives of the energy. 
All models, with the exception of MatterSim \cite{yang_mattersim_2024}, which was trained on a proprietary dataset, were trained on large-scale open datasets, including MPtraj~\cite{deng_chgnet_2023_mptraj}, Alexandria~\cite{schmidt_machinelearningassisted_2023_alexandria}, and/or OMat24 \cite{barroso-luque_open_2024_omat24}. While MPtraj and Alexandria primarily contain equilibrium conformations, OMat24 includes extensive out-of-equilibrium data, enabling broader coverage of the \ac{pes}.

Given the importance of long-range, non-covalent interactions, particularly van der Waals (vdW) forces, in nanoporous materials modeling, consistent treatment of dispersion interactions is critical \cite{formalik_benchmarking_2018}. Therefore, all \acp{umlip} evaluated here either apply D3 dispersion corrections \cite{Grimme2010_dftd3, takamoto2021pfp_dftd} at inference or were trained on D3-corrected reference energies \cite{formalik_benchmarking_2018}, ensuring comparability in their treatment of vdW forces.

Each uMLIP in our benchmark is evaluated on a consistent set of tasks critical for characterizing nanoporous materials: structural optimization, molecular dynamics stability, bulk property prediction, namely bulk modulus and the specific heat capacity, and interaction energy of guest molecules with the framework. Performance is assessed by comparing uMLIP predictions against DFT references.
For the structural optimization, molecular dynamics stability, and bulk modulus, we performed new DFT calculations on 100 structures, comprising 83 \acp{mof}, 7 COFs, and 10 zeolites, selected to represent diverse chemistries and topologies (see method section for details). This set primarily includes experimentally reported materials from QMOF \cite{rosen_machine_2021_qmof}, MOASEC-DB \cite{gibaldi_mosaec-db_2024}, IZA \cite{iza_database_baerlocher}, and CURATED-COF \cite{ongari_building_2019}, supplemented by a few hypothetical MOFs, sourced from QMOF \cite{rosen_machine_2021_qmof}.
For heat capacity predictions, we use the dataset of 231 structures with DFT reference values reported by Moosavi et al.~\cite{moosavi_data-science_2022}, with crystal structures from the CoRE-MOF database \cite{chung_advances_2019_coremof,zhao2025core}. Finally, host-guest interactions are evaluated using DFT interaction energies and forces of \ce{CO2} and \ce{H2O} with 26 MOFs from Lim et al. in the GoldDAC project~\cite{Lim_2025_DAC-SIM}, covering a range of challenging chemistries, including open metal sites. 

\begin{figure}[t]
    \centering
    \includegraphics[clip, width=\linewidth]{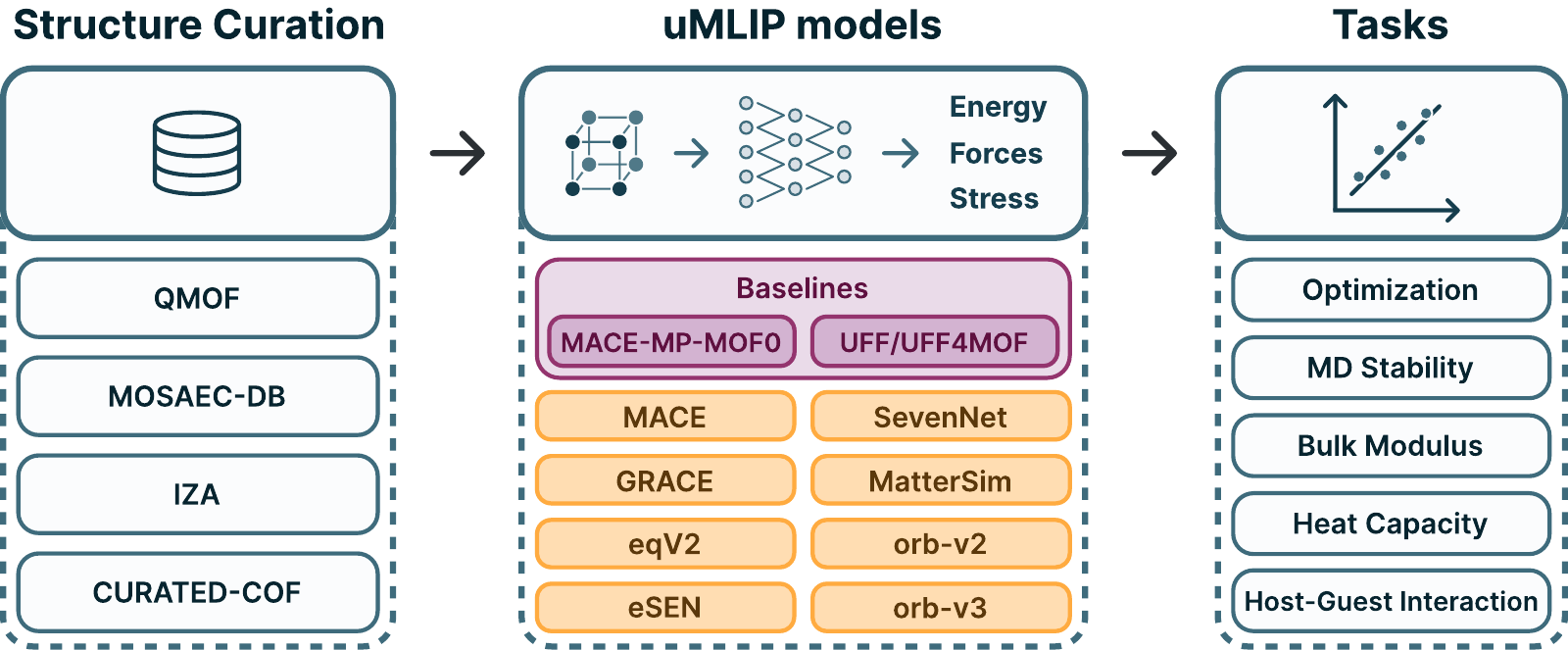}
    \caption{\textbf{Outline of benchmarking procedure.}
    Structures are curated from several databases, accounting for different pore sizes, metal variety, and framework types. Machine-learning potentials from several architectures are evaluated on a suite of tasks, benchmarked against DFT references.}
    \label{fig:pipeline}
\end{figure}

We compare \acp{umlip} performance against two baseline approaches. The first is the use of classical force fields, which remain the standard in MOF modeling due to their broad applicability and computational efficiency \cite{Park2024_generative, islamov2023high, Anderson2019_increasing}. For this baseline, we include results from the Universal Force Field (UFF) \cite{rappe1992uff} and its MOF-specific extension, UFF4MOF \cite{Coupry2016_uff4mof}. The second baseline is fine-tuned machine learning potentials. Elena et al.~\cite{elena2024machinelearnedpotentialhighthroughput_mace_mof_0} introduced \mbox{MACE-MP-MOF0} by fine-tuning the MACE-MP-0 potential on a curated dataset of 127 MOFs with DFT calculations to improve performance on MOF-specific tasks. This dataset includes out-of-equilibrium configurations generated through molecular dynamics and structural distortions. Similarly, Lim et al.~\cite{Lim_2025_DAC-SIM} introduced MACE-DAC-0 by fine-tuning the MACE-MP-0 model using host-guest interactions for gas adsorption simulations. Comparing \acp{umlip} against these baselines provides insight into their applicability and performance with respect to the state-of-the-art modeling methods for MOFs. A summary of structures, models, baselines, and tasks is provided in Figure~\ref{fig:pipeline}.

Ensuring strict comparability of architectures is a challenge, as not all architectures are available with checkpoints trained on all datasets. Therefore, in the main text of this article, we include results for models using available OMat24 \cite{barroso-luque_open_2024_omat24} checkpoints, which may include additional training or fine-tuning on MPtraj \cite{deng_chgnet_2023_mptraj} and Alexandria \cite{schmidt_machinelearningassisted_2023_alexandria}. 
We also include results of the MACE-MP-0a\cite{batatia_foundation_2024_mace} model to facilitate a comparison with the first of the kind \ac{umlip} from our selection of architectures. 
Additionally, we include the MatterSim model \cite{yang_mattersim_2024}, which was trained on a proprietary dataset, and two non-conservative models, orb-d3-v2\cite{neumann2024orbfastscalableneural} and eqV2-OMsA \cite{barroso-luque_open_2024_omat24}, to capture a broader architectural spectrum. This brings the main comparison set to nine universal models and MACE-MP-MOF0. Results for all 21 models evaluated in this benchmark are provided in the Supporting Information.

\subsection{Modeling Prototypical MOFs}\label{sec:prot_mofs}

\begin{figure}[ht]
    \centering
    \includegraphics[width=\linewidth]{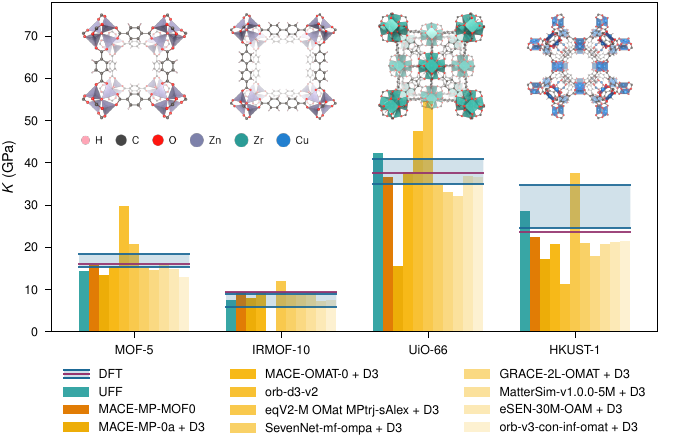}
    \caption{\textbf{Bulk modulus ($K$) predictions of four prototypical MOFs.}
    Purple lines show our DFT results, while blue lines and shaded areas represent DFT values from the literature. Turquoise represents results from the Universal Force Field (UFF) from Ref. \cite{boyd_force-field_2017}\,. \mbox{MOF-5} DFT: Refs. \cite{Bahr2007mechanical, Lukose2011structural, Samanta2006theoretical, Mattesini2006ab-initio, Zhou2006Lattice, elena2024machinelearnedpotentialhighthroughput_mace_mof_0}\,, \mbox{IRMOF-10} DFT: Refs. \cite{Kuc2007Metal-Organic, Yang2012ab-initio, elena2024machinelearnedpotentialhighthroughput_mace_mof_0}\,, \mbox{UiO-66} DFT: Refs. \cite{Wu2013Exceptional,elena2024machinelearnedpotentialhighthroughput_mace_mof_0}\,, \mbox{HKUST-1} DFT: Refs. \cite{Wu2013Exceptional, Lukose2011structural}\,.}
    \label{fig:bulk_modulus_esmof}
\end{figure}

Our analysis begins with four prototypical MOFs commonly found in the literature: MOF-5, IRMOF-10, UiO-66, and HKUST-1. These materials have been extensively studied through experiments \cite{lock_scrutinizing_2013, lock_elucidating_2010, zhou_origin_2008, wu_negative_2008} and DFT calculations \cite{Bahr2007mechanical, Lukose2011structural, Samanta2006theoretical, Mattesini2006ab-initio, Zhou2006Lattice, elena2024machinelearnedpotentialhighthroughput_mace_mof_0, Kuc2007Metal-Organic, Yang2012ab-initio, Wu2013Exceptional}, and consequently, they have been used in prior \acp{umlip} analyses \cite{neumann2024orbfastscalableneural, elena2024machinelearnedpotentialhighthroughput_mace_mof_0} to demonstrate proof-of-principle generalization to MOFs. Here, we extend these by comparing the bulk modulus predictions of various uMLIPs with DFT references. Figure~\ref{fig:bulk_modulus_esmof} presents these predictions alongside our DFT references, literature DFT values, and results from the empirical \ac{uff}\cite{rappe1992uff, boyd_force-field_2017} and the fine-tuned, MOF-specific MACE-MP-MOF0.

The uMLIPs show high variability across models and structures (Figure~\ref{fig:bulk_modulus_esmof} and see SI Table \ref{tab:bulk_modulus_esmof} for the full table of all 21 uMLIPs). For MOF-5 and IRMOF-10, almost all uMLIPs show good agreement with DFT, except for orb-d3-v2 and eqV2-OMsA, which exhibit significant deviations due to instability in their energy-volume profiles.
In particular, while most architectures show no signs of instability, the energy-volume profiles for these two models suffer from incorrectly identified minimum energy configurations in the initial optimization and unreliable energy increase at higher volumes. From an architecture point of view, these two models are distinct, where instead of obtaining forces by computing gradients from energy, forces are the direct output of the model. This approach, while introduced to improve computational efficiency, makes the models non-conservative in energy, leading to unstable \ac{eos} solutions. On the other hand, all other models correctly identified the minimum energy conformation in the initial optimization, and the energy consistently increased at scaled volumes (see SI Figure \ref{fig:bulk_modulus_eos_all_0}-\ref{fig:bulk_modulus_eos_all_2} for details).

For UiO-66 and HKUST-1, most uMLIPs underestimate bulk modulus, while the two non-conservative uMLIPs show unstable results. The systematic underestimation is consistent with the known softening of the \ac{pes} that arises when models are trained predominantly on equilibrium geometries \cite{Deng2025systematic}. For instance, MACE-MP-0a, which was trained only on the Materials Project dataset, severely underestimates the reference values because the predicted energy rise upon compression is too shallow. 
Other models, such as MACE-OMAT-0, eSEN-OAM, and SevenNet-mf-ompa, which were trained on the OMat-2024 dataset containing out-of-equilibrium configurations, exhibit smaller errors, indicating that the inclusion of non-equilibrium data partially mitigates the softening effect.

Remarkably, the fine-tuned model (MACE-MP-MOF0) provides the most accurate results across all four structures. In particular, in fine-tuning MACE-MP-MOF0, Elena et al. \cite{elena2024machinelearnedpotentialhighthroughput_mace_mof_0} used various sampling techniques, such as molecular dynamics simulations and strained unit cells, to include out-of-equilibrium configurations, which has resolved the smoothening effect.

The notable observation from this analysis of the small set is that while uMLIPs exhibit a large performance dispersion, in which some match DFT, others deviate substantially or even become unstable, the fine-tuned model delivers the most reliable results, followed by the \ac{uff} force field. This underscores two points. First, agreement on energies and forces for a handful of prototypical MOFs does not guarantee transferability to the derived properties. Second, there is as of yet no evidence that off-the-shelf uMLIPs systematically outperform a decades-old empirical force field, such as UFF. These observations motivate a broader benchmark that (i) spans a chemically and topologically diverse set of materials, and (ii) interrogates multiple task types, so that the strengths and limitations of each architecture can be rigorously analyzed.

\subsection{Static modeling and structure optimization}

\begin{figure}[ht!]
    \centering
    \includegraphics[width=\textwidth]{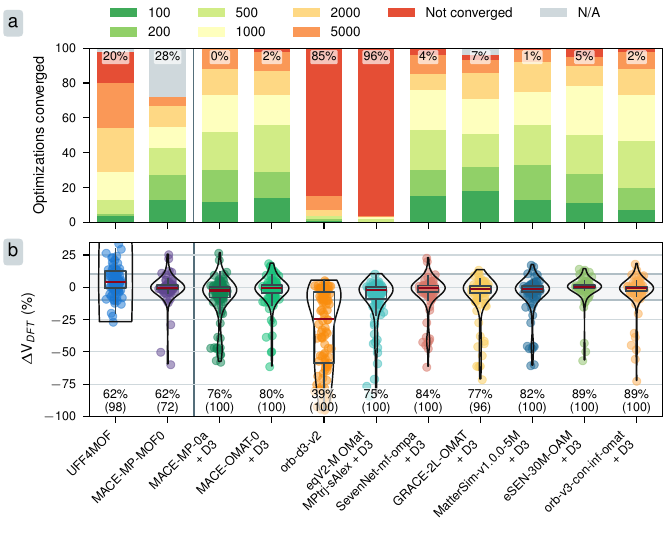}
    \caption{\textbf{Structural minimization convergence and volume comparison.}
    a) Structural minimization convergence grouped by number of optimization steps required to reach a force convergence criterion of $10^{-3}$\,eV/\AA. Numbers at the top indicate the percentage of N/A results and unconverged structural optimizations. b) Relative difference between uMLIP-optimized and DFT-optimized cell volumes. Numbers below violin plots denote the percentage of successfully computed structures with volume deviations to the DFT reference of less than 10\% and the number of successfully computed structures. The grey shaded region highlights a $\pm10$\,\% deviation threshold from the DFT-optimized volume.
    }
    \label{fig:opt_convergence_volume}
\end{figure}

We extend our analysis to a larger set of 100 \acp{mof}, COFs, and zeolites to systematically examine the behavior of \acp{umlip} across diverse framework chemistries. As the first task, we focus on structure optimization, a foundational task in atomistic modeling used to identify minimum-energy configurations, assess local stability, and reduce artifacts introduced by experimentally determined or hypothetical initial structures. To quantify model performance, we evaluated two key metrics: 1) convergence rate, which measures the fraction of structures for which the optimization was completed successfully, and 2) accuracy of convergence, which measures the fraction of structures for which the optimized geometry yields an accurate minimum-energy configuration compared to reference DFT optimized structures. 

Figure~\hyperref[fig:opt_convergence_volume]{\ref*{fig:opt_convergence_volume}a} illustrates the convergence behavior of structural minimizations for the main set of 100 structures, showing the distribution of optimization steps needed to reach a force threshold of $10^{-3}\,$eV/\AA, within at most 5,000 optimization steps. Similar to the previous case, the non-conservative models, orb-d3-v2 and eqV2-OMsA, exhibit the largest portions of non-converged optimizations after 5,000 steps at 85\,\% and 96\,\%, respectively. Small oscillations in the forces due to the direct predictions prevent a steady decrease to reach the convergence criterion, resulting in long-running optimizations. The other uMLIPs demonstrate high success rates, with approximately half of the structures converging within 500 optimization steps on average. GRACE-2L-OMAT encountered CUDA out-of-memory errors for a few structures, contributing to its slightly lower rate of successful computations. 

Compared to baseline models, UFF4MOF requires a larger number of optimization steps to reach the convergence criterion, whereas uMLIPs with conservative force formulations converge more quickly. This indicates that uMLIPs offer a smoother \ac{pes} compared to classical force fields.
Moreover, despite the fine-tuned model MACE-MP-MOF0 showing a high convergence rate for the supported materials, it exhibits the lowest number of successful optimizations due to limited element coverage. This model was fine-tuned on a reduced set of elements relative to the baseline universal model, resulting in 28 unsupported structures in the main set. Due to this restricted chemical scope, we do not treat MACE-MP-MOF0 as a \enquote{universal} potential in our evaluation; rather, we include it as a MOF-specific baseline model for comparative purposes.


Figure~\hyperref[fig:opt_convergence_volume]{\ref*{fig:opt_convergence_volume}b} compares the cell volumes of the uMLIP-minimized structures with the DFT-optimized volume, revealing variations in predictive accuracy across models. The non-conservative orb-d3-v2 has the lowest number of successfully computed and accurate (within 10\%) structures at 39\,\%. This shows that for this model, not only do the optimizations not converge, but the resulting configurations also significantly deviate from the DFT-optimized structures. The other non-conservative model, eqV2-OMsA, shows a lower share of diverging volumes, benefiting from the larger OMat24 training dataset. The orb-v3-omat model, relying on gradient-based forces as opposed to orb-d3-v2, shows a much higher percentage of successful and accurate optimizations (89\,\%). Other uMLIPs show relatively good performance. The best-performing model is the eSEN-OAM model, which exhibits the narrowest interquartile range and, alongside orb-v3-omat, the highest number of successful and accurate optimizations (89\,\%), suggesting its optimized geometries are generally closest to DFT. 

uMLIPs leveraging conservative forces show excellent performance compared to the classical force field baseline (UFF4MOF), which has a low number of completed and accurate structures at 62\,\%. Moreover, while the fine-tuned MACE-MP-MOF0 model shows very few outliers (structures deviating by >10\,\% in volume from DFT), its overall applicability is limited by the 28 unsupported structures noted earlier. Notably, even if we restrict our evaluations to the materials supported by the fine-tuned model, the eSEN model outperforms the fine-tuned model, showing stronger accuracy and applicability (see SI Figure \ref{fig:opt_all_intersection} for details). Therefore, we observe that on our benchmark, uMLIPs outperform available models used in the MOF molecular modeling community.

These results prompt the question of whether there is a pattern in the failed optimizations of uMLIP. Although multiple models show large optimized volume deviations for a few structures, no structure is an outlier for every model, which indicates that the instabilities of the \ac{pes} are different for every model (see SI Figure \ref{fig:failed_opt_structure_hist} for details). Furthermore, the distribution of elements in the outlier structures showed that no element was predominantly present in failed structures (see SI Figure \ref{fig:failed_symbol_distribution_hist}). The outliers present in all models could suggest possible \enquote{holes} in the potentials, where models predict incorrect forces for certain configurations due to high uncertainty, leading to unphysical configurations during optimization. We conclude that the deviations can likely be attributed to \ac{pes} irregularities instead of inherent issues in the crystal structures or training set.

\subsection{Dynamic modeling}

\begin{figure}[ht]
    \centering
    \includegraphics[width=\linewidth]{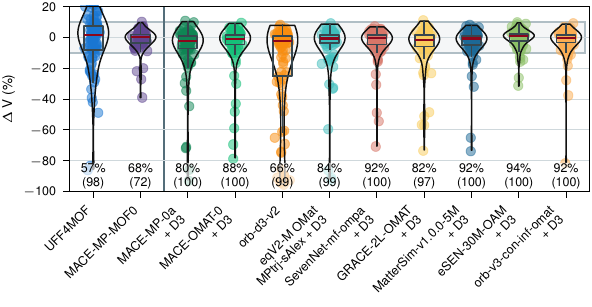}
    \caption{\textbf{MD volume change.}
    Relative volume difference between the beginning and end of 50\,ps long NpT simulations. Atom positions were first optimized, and the structures were equilibrated in the NVT ensemble. The grey shaded area indicates a threshold of $\pm10$\,\% for outliers. Numbers below the violin plots indicate the number of successfully computed structures with volume deviations from the initial structure of less than 10\,\% and the total number of successfully computed structures.
    }
    \label{fig:stability_volume}
\end{figure}

Evaluating machine learning potentials solely on static samples at 0~K for energy and force prediction is insufficient to establish their usability in molecular dynamics simulations (MD) \cite{fu2023forces}.
These simulations offer insight into the thermodynamics-dependent behavior of materials, and it is crucial that \acp{umlip} can perform these simulations robustly. In particular, irregularities or discontinuities in the \ac{pes} can lead to instabilities in MD trajectories, compromising physical fidelity. To assess the dynamic stability of \acp{umlip}, we performed NpT simulations on our main set of 100 structures, focusing on volumetric stability and changes in local coordination environments over time.

Figure~\ref{fig:stability_volume} illustrates the relative volume change between the initial and final simulation snapshots of 50\,ps of NpT MD at 300\,K and 1\,bar. Across models, several structures exhibit substantial volume changes exceeding 10\%. As observed in earlier tasks, the non-conservative models yield the highest number of unstable trajectories with large volume change. In contrast, eSEN-OAM demonstrates the best overall performance, followed by orb-v3-omat, MatterSim, and other architectures trained on out-of-equilibrium conformations. While architectural differences have a modest impact on MD stability, the training data appears to be the dominant factor. Models trained on out-of-equilibrium conformations, either through the OMat24 dataset \cite{barroso-luque_open_2024_omat24} or via active learning procedures (as in the case of MatterSim \cite{yang_mattersim_2024}), consistently deliver more stable simulations. This suggests that including non-equilibrium data during training is essential for robust MD performance.

Consistent with the optimization results, we observe no systematic pattern among the failed MD simulations. Notably, all outlier structures remain stable under at least one \ac{umlip}, suggesting that the large deviations arise from model-specific deficiencies in the learned \ac{pes} rather than from flaws in the initial crystal structures. Moreover, the number of outliers in the MD simulations is lower than in the optimization task. We attribute this to two factors. First, prior to MD simulations, all structures undergo atom-only optimization, keeping cell parameters fixed. This step avoids errors due to non-triangular cell shapes in ASE \cite{HjorthLarsen2017_ase} and ensures that simulations begin from configurations close to the experimental geometry. Second, unlike structure optimization, MD trajectories do not strictly follow the \ac{pes} gradient. As a result, even if transient unphysical configurations arise, the system can often relax back to a physically meaningful state over the course of the trajectory.

The most significant result in Figure~\ref{fig:stability_volume} is that the best performing \acp{umlip} consistently outperform the classical force fields (UFF4MOF and UFF, see SI Figure \ref{fig:stability_volume_copper_coordination_all} for full results). Under UFF4MOF, many frameworks collapse because key coordination geometries are missing from the parameter set, leading to unstable cell volumes. Furthermore, the fine-tuned model MACE-MP-MOF0 exhibits a low outlier rate but cannot be applied to several structures in the main set owing to its limited elemental coverage, which diminishes its overall utility (68\,\% of successful and accurate simulations). By contrast, the top universal models, namely eSEN-OAM, orb-v3-omat, and MatterSim, not only remain stable across a diverse range of chemistries but also surpass MACE-MP-MOF0 on the subset of structures for which that model is defined (see SI Figure \ref{fig:stability_volume_copper_coordination_all_intersection} for details). These findings demonstrate that broadly trained uMLIPs outperform current models, such as UFF4MOF, and can generalize at least as well as, and in some cases better than, MOF-specific fine-tuned potentials.


In addition to global volume stability, a potential must maintain the integrity of local atomic environments. This is particularly relevant in \Acp{mof}, which feature diverse metal coordination environments, where a metal can appear in various coordination geometries and oxidation states. 
Capturing such diversity is challenging for uMLIPs, as these models do not rely on explicit bonding topologies or predefined coordination assumptions, unlike classical force fields. 

\begin{figure}[t]
    \centering
    \includegraphics[width=\linewidth]{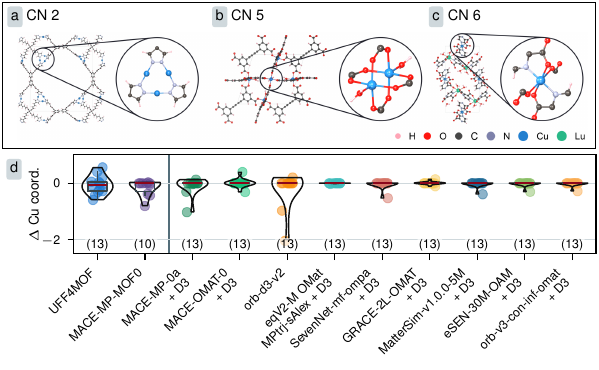}
    \caption{\textbf{Copper coordination change.}
    Visualization of several copper coordination environments with coordination numbers (CN) of a) 2, b) 5, and c) 6.
    d) Change in the average copper coordination number between the first and last snapshots of 30\,ps NpT simulations. Each point represents a single material. Numbers below violin plots indicate the number of successfully simulated Cu-MOFs for each model.}
    \label{fig:stability_copper_coordination}
\end{figure}

To probe the ability of uMLIPs to represent complex local environments, we designed a targeted evaluation in which the metal type (copper) is held constant but appears in different coordination environments across a set of 13 MOFs. These structures span copper coordination numbers ranging from 2 to 6, providing a rigorous test of each model's capacity to generalize across local chemical environments without explicit structural priors.

Figure~\ref{fig:stability_copper_coordination} shows the predicted change in the average Cu coordination numbers after 30\,ps of NpT \ac{md} (10\,ps at 300\,K, 10\,ps at 400\,K, and 10\,ps at 300\,K). Several universal models, including SevenNet-ompa, GRACE-OMAT, MatterSim, eSEN-OAM, and orb-v3-omat, demonstrate excellent stability across all tested structures. Interestingly, the fine-tuned MACE-MP-MOF0 model does not outperform the top-performing universal models in this task, suggesting that domain-specific fine-tuning does not necessarily improve the stability of local coordination environments. 

This result is particularly noteworthy because, unlike classical force fields, which rely on explicitly defined bonding networks and coordination geometries, uMLIPs operate without any predefined bonding assumptions. For instance, the original UFF force field lacks parameters for the square-planar Cu, which is a common coordination geometry in Cu-based MOFs featuring paddle-wheel motifs. While this limitation was addressed in UFF4MOF through explicit parametrization, such extensions do not generalize easily to new chemistries or coordination environments. More fundamentally, empirical force fields cannot truly function as ab initio models, as they depend on a predefined parameter set tailored to the system of interest. By contrast, uMLIPs learn interactions directly from data and require minimal system-specific input. Their ability to preserve local coordination environments without manual bonding assignments or extensive parameter tuning makes them not only more flexible but also significantly easier to deploy in automated and large-scale materials simulations.

\subsection{Modeling Bulk Properties}

\begin{figure}[ht]
    \centering
    \includegraphics[width=\linewidth]{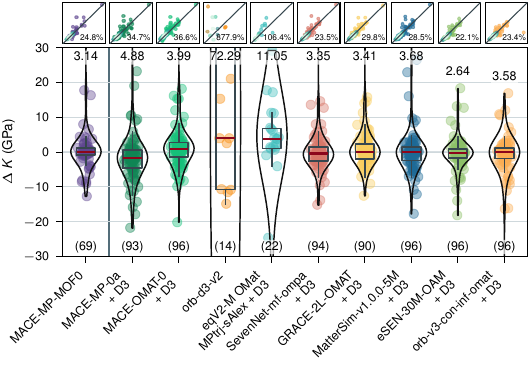}
    \caption{\textbf{Bulk modulus ($K$) prediction difference to DFT references.} 
    Parity plots at the top display the error distributions with DFT values on the x-axis and predictions on the y-axis. Numbers indicate the mean absolute percentage error. Values above violin plots are mean absolute errors. Numbers in parentheses below plots indicate the number of successfully computed predictions for each model.
    }
    \label{fig:bm_combined}
    
\end{figure}

Accurately predicting material properties is the ultimate goal for molecular modeling, and uMLIPs promise to combine high accuracy with computational efficiency. We evaluate this claim by predicting the bulk modulus for our main set of 100 structures and the heat capacity for a larger set of 231 MOF, COF, and zeolite structures (with DFT references from Ref. \cite{moosavi_data-science_2022}). Additional analysis of bulk modulus predictions for an extended set of zeolite structures is provided in the Supporting Information \cite{sours_predicting_2023} (Figure \ref{fig:bm_combined_zeolite_all}). These two properties probe different aspects of the \ac{pes}: the bulk modulus reflects the global curvature of the \ac{pes} with respect to volume changes, while the heat capacity is sensitive to the local curvature around equilibrium configurations. Together, they offer a complementary assessment of model fidelity across global and local scales. Since geometry optimization is a prerequisite for calculating both bulk modulus and heat capacity,\cite{moosavi2018improving,moosavi_data-science_2022} and given the poor performance of classical force fields in structure optimization tasks, we excluded them from this section of the evaluation.

Figure~\ref{fig:bm_combined} presents the distribution of bulk modulus ($K$) prediction errors, revealing a varying error range across the models. The fine-tuned MACE-MP-MOF0 (MAE 3.14\,GPa), SevenNet-ompa (MAE 3.35\,GPa), and eSEN-OAM (MAE 2.64 GPa) achieve the lowest mean absolute errors, with eSEN surpassing the fine-tuned baseline. In terms of distribution tightness, MACE-MP-MOF0, eSEN-OAM, and orb-v3-omat show the narrowest interquartile ranges for their error distributions. The non-conservative models show poor performance, where orb-d3-v2 exhibits the largest MAE by an order of magnitude (72.29\,GPa) compared to the conservative models, with eqV2-OMsA showing the second-largest MAE (11.05\,GPa). Both models have a large number of unsuccessful calculations. MACE-MP-0a shows a trend of underestimating the bulk modulus, which is not present in the MACE-OMAT-0 model. Its training on OMat24 features a more diverse set of non-equilibrium structures, enabling more accurate energy predictions at scaled volumes. For all models, several structures exhibited an unstable \ac{eos} fit, determined by a lowest-energy volume deviation of over 1\,\% from the volume minimum identified during optimization, and were consequently excluded. This filtering leads to increased accuracy by discarding predictions with high uncertainty.

While the eSEN-OAM and orb-v3-omat models achieve the lowest MAE, they still have a mean absolute percentage error (MAPE) of 22.1\,\% and 23.4\,\%, respectively, which indicates a wide error distribution relative to the range of values. A common reason for incorrect bulk modulus predictions across models is a low predicted energy at reduced volumes. Correct predictions show increasing energies when volumes are scaled away from the optimum. However, this trend is often broken in incorrect predictions after a given applied strain, and energies incorrectly decrease, sometimes with a non-continuous jump. This could further indicate possible instabilities in the learned \ac{pes}, arising from training data close to minimum-energy configurations, which are mitigated by training on OMat24, but not yet resolved.

\begin{figure}[t]
    \centering
    \includegraphics[width=\linewidth]{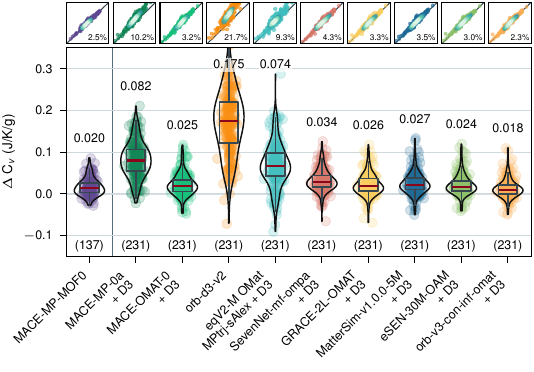}
    \caption{\textbf{Heat capacity ($C_V$) prediction difference to DFT references.} 
    Parity plots at the top display the error distributions with DFT values on the x-axis and predictions on the y-axis. Numbers indicate the mean absolute percentage error. Values above violin plots are mean absolute errors. Numbers in parentheses below plots indicate the number of successfully computed predictions for each model.
    }
    \label{fig:hc_combined}
    
\end{figure}

Heat capacity is another fundamental property of materials, particularly relevant for applications like process modeling in carbon capture \cite{moosavi_data-science_2022}. The heat capacity of a material is defined by atomic vibrations, and hence, it is a good measure of the local curvature of the \ac{pes} with respect to atomic displacements. Figure~\ref{fig:hc_combined} displays the errors in \ac{umlip} heat capacity predictions compared to DFT references from Ref. \cite{moosavi_data-science_2022}, which comprise a set of 231 MOFs, COFs, and zeolites. Similar to previous cases, orb-d3-v2 shows the largest error (MAE 0.175\,J/K/g) and strongest systematic shift. MACE-MP-0a (MAE 0.082\,J/K/g) follows and is surpassed by eqV2-OMsA (MAE 0.074\,J/K/g). MACE-MP-MOF0 (MAE 0.020\,J/K/g), eSEN-OAM (MAE 0.024\,J/K/g), and orb-v3-omat (MAE 0.018\,J/K/g) exhibit both the narrowest error distributions and the lowest MAEs, with orb-v3-omat surpassing the fine-tuned model in terms of MAE. orb-v3-omat furthermore achieves the lowest MAPE of 2.3\,\%, underscoring its strong performance and narrow error distribution for this property. 

A key observation across all uMLIPs is a systematic overestimation of heat capacity relative to the DFT values. The systematic shift further relates to \ac{pes} softening, where the learned \ac{pes} has a reduced curvature, arising from training data points biased towards minimum-energy configurations. Consequently, the model predicts a softer surface during extrapolation due to insufficient training examples. This underpredicted curvature of the \ac{pes} results in an overprediction of the heat capacity, and is related to the underprediction of phonon vibration frequencies that has also been observed in other studies of \acp{umlip} \cite{Deng2025systematic, Merchant2023gnome}. The models trained on the OMat24 dataset show a reduced systematic shift, highlighting the importance of diverse training data with non-equilibrium structures and more research to address these challenges to increase stability in \acp{umlip} further.

\subsection{Predicting Host-Guest Interactions}

Our evaluations thus far have focused on modeling pristine MOF structures, demonstrating the strong performance of \acp{umlip} in capturing the behavior of empty frameworks. A critical next step is to assess the models' performance in host–guest interactions, which are central to designing MOFs for applications in carbon capture, gas storage, and catalysis. Traditionally, these interactions are modeled using classical force fields, supplemented by electrostatic interactions, often computed from DFT-derived point charges or charge equilibration methods.\cite{wilmer2012extended,ongari2018evaluating,luo2024mepo} Despite known limitations, particularly in systems with open metal sites, where classical force fields often fail to capture complex interactions,\cite{dzubak2012ab} these methods remain in widespread use due to their computational efficiency. While resources, such as the OpenDAC dataset,\cite{sriram2024open} have begun to make large-scale DFT reference data available, such efforts remain limited,\cite{jin2025correspondence} and empirical force fields continue to dominate practical host-guest modeling.

Recent efforts have begun to explore the use of machine learning potentials for this task, either through system-specific training\cite{vandenhaute2023machine,liu2024machine,zheng2023quantum} or fine-tuning from general-purpose models. For instance, in the GoldDAC project, Lim et al.~\cite{Lim_2025_DAC-SIM} fine-tuned MACE-MP-0 using additional DFT-calculated interaction energies for \ch{CO2} and \ch{H2O} to introduce MACE-DAC-1, and evaluated its performance across 26 MOFs, spanning a range of challenging chemistries; notably, over 40\% of these structures featured open metal sites, which are particularly challenging for classical force field accuracy. Their results highlighted the potential of fine-tuned MLIPs for screening materials in direct air capture (DAC) applications. 

\begin{figure}[t]
    \centering
    \includegraphics[width=\textwidth]{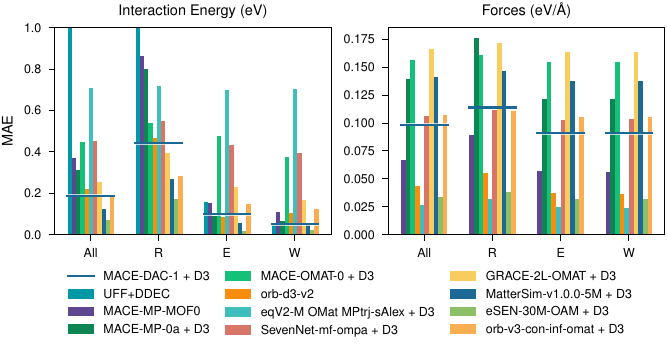}
    \caption{\textbf{Comparison of a) interaction energy errors and b) force errors.} Results were computed on the test set of GoldDAC \cite{Lim_2025_DAC-SIM}. Errors are separated into three interaction categories: (R) Repulsion, (E) Equilibrium, and (W) Weak-attraction.}
    \label{fig:dac}
\end{figure}

To assess how uMLIPs compare to both this fine-tuned model and classical approaches, we used the GoldDAC test dataset to evaluate interaction energy and force prediction accuracy. As shown in Figure~\ref{fig:dac}, the top-performing uMLIPs, namely MatterSim and eSEN-OAM, consistently outperform the fine-tuned MACE-DAC-1 model. MatterSim achieves lower errors in repulsion (R) and equilibrium (E) regimes, while eSEN-OAM exhibits uniformly lower errors across all interaction types. These uMLIPs also support a broader range of chemical elements compared to the fine-tuned model, enhancing their potential as general-purpose tools for host–guest interaction modeling. 

While dataset diversity emerged as the dominant factor in previous tasks, our results indicate that model architecture also plays a role in host-guest interaction modeling. For example, although the MACE-OMAT-0 model consistently outperformed MACE-MP-0 in earlier evaluations, this trend does not hold for host-guest interactions. Moreover, different architectures trained on the same OMat24 dataset exhibit a wide range of performance, highlighting the increased sensitivity of this task to architectural design. Notably, the non-conservative models, namely orb-d3-v2 and eqV2-OMsA, despite their poor performance in structure optimization and MD stability, show relatively strong force prediction accuracy in select host-guest configurations. This suggests that, although their overall reliability is limited, such models may offer advantages in specific applications and warrant further exploration, particularly where force accuracy for localized interactions is the primary concern.

A notable outcome is that both MatterSim and eSEN-OAM also outperform classical force fields with DFT-derived partial charges (UFF+DDEC), the most common method used for modeling host-guest interaction in high-throughput computational screening studies. These results provide early, yet strong evidence that uMLIPs are reaching a level of maturity suitable for practical application in adsorption modeling. However, their widespread adoption will depend on integration into established molecular simulation packages, such as RASPA,\cite{dubbeldam2016raspa} to enable scalable and efficient deployment in MOF screening workflows. Such integration will also facilitate the evaluation of uMLIPs in property prediction tasks like the full adsorption isotherms, which are critical for evaluating uMLIPs in host-guest modeling.

\section{Discussion}

Our main goal in this study was to evaluate \acp{umlip} in modeling challenging nanoporous materials, with a focus on MOFs. The chemical diversity and structure complexity inherent to these materials, driven by their porosity and coordination chemistry, pose a challenge for any interatomic potential. Through a comprehensive benchmark, spanning critical tasks, including geometry optimization, molecular dynamics stability, property prediction, and host-guest interactions, we identified both promising advances in \acp{umlip} and key factors influencing their performance and robustness. 

Our results highlight substantial progress in the development of \acp{umlip}, demonstrating their growing viability for complex simulations of MOFs and related framework materials. The best-performing \acp{umlip} outperform the long-standing methods, represented by the Universal Force Field (UFF), suggesting the field is approaching a pivotal transition. These models enable accurate and efficient modeling of structural dynamics and framework flexibility, tasks that previously required expensive quantum mechanical methods, opening the door to a wide range of new research directions in materials discovery and design.

This transition to machine learning-based methods relates to the \enquote{the Bitter Lesson}\cite{sutton2019bitter} moment that was previously observed across many AI-disrupted disciplines such as text, image, video, and speech modeling. The availability of large, diverse datasets and scalable learning algorithms has enabled data-driven models to overtake rule-based, hand-crafted approaches. In the context of this work, the shift is from classical force fields, which are interpretable yet rigid, to uMLIPs, which trade some transparency for greater accuracy and transferability. With large data and compute, learning from first-principles reference calculations increasingly outperforms functional forms encoded by expert intuition. Yet, it is important to emphasize that this advancement primarily builds on decades of experts' insight from quantum chemistry to traditional force fields. 
The progress we report would not be possible without this foundational knowledge and community effort. Furthermore, our evaluations and previous studies have elucidated critical aspects that require attention when further scaling models and datasets, including the importance of out-of-equilibrium samples and conservative forces, highlighting that simply adding more data and compute does not necessarily result in the desired outcomes.

\begin{figure*}[t]
    \centering
    \includegraphics[width=\linewidth]{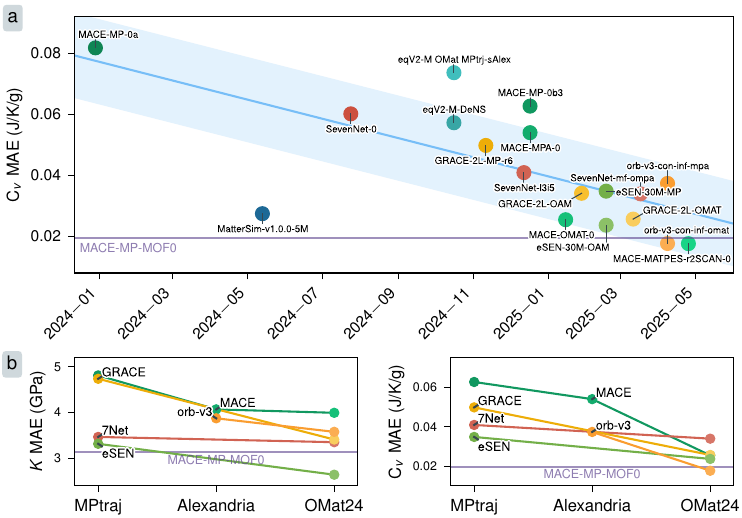}
    \caption{\textbf{Evolution of uMLIP performance.}
    a) Evolution of mean absolute error (MAE) for heat capacity predictions of 231 MOFs, COFs, and zeolites compared with DFT references. b) MAE for heat capacity ($C_V$) and bulk modulus ($K$) predictions, grouped by the largest training dataset used. 
    Models can be categorized based on the most extensive dataset included in training: MPtraj-only, MPtraj + Alexandria/sAlex, and OMat24-based (with or without additional datasets).
    }
    \label{fig:evolution}
\end{figure*}

The landscape of \acp{umlip} is evolving with impressive speed. 
A performance curve for this progress is illustrated in Figure~\hyperref[fig:evolution]{\ref*{fig:evolution}a}, which tracks the accuracy in predicting the heat capacity in successive uMLIP releases, showing a clear, rapid increase in performance over a short period. 
While the fine-tuned MACE-MP-MOF0 initially offered improved performance by compromising universality over the universal MACE potential for MOF-specific tasks, several recent general-purpose \acp{umlip} now match or surpass this specialized baseline. 

From a benchmarking perspective, we find that while the classification metric for material discovery reported in MatBench Discovery does not correlate well with performance in MOF-specific modeling tasks (see Figure \ref{fig:matbench_corr}; correlation coefficient close to zero), the accuracy of thermal conductivity predictions \cite{póta_thermalconductivitypredictionsfoundation_2024_ksrme} is a good performance indicator. 
This underscores the supplementary role of domain-specific evaluation frameworks in assessing the practical utility of uMLIPs in targeted applications and materials classes.

From a \ac{umlip} development point of view, the performance curve of \acp{umlip} highlights the central role of training data. The open release of the OMat24 dataset (late 2024) has been particularly impactful, leading to significant improvements in models trained after its release. This relationship between dataset quality and model performance is explored further in Figure~\hyperref[fig:evolution]{\ref*{fig:evolution}b}. Early uMLIPs were typically trained on MPtraj alone \cite{deng_chgnet_2023_mptraj}, while subsequent models incorporated broader datasets such as the Alexandria database \cite{schmidt_machinelearningassisted_2023_alexandria} and the comprehensive OMat24 dataset \cite{barroso-luque_open_2024_omat24}. Figure~\hyperref[fig:evolution]{\ref*{fig:evolution}b} shows that for both bulk modulus ($K$) and heat capacity ($C_V$), prediction errors consistently decrease across all conservative architectures as training dataset size and diversity increase. While there are differences between models trained on the same data, jumps in performance correlate strongly with larger, more diverse training datasets. This indicates that further gains are likely to come mostly from continued expansion and curation of comprehensive, out-of-equilibrium datasets.

From an architectural point of view, non-conservative forces deliver the least competitive performance. We observed that even small inconsistencies between predicted forces and the true energy gradient can lead to instabilities in structural optimizations, MD, and property predictions. This highlights the importance of computing forces as exact energy derivatives and supports the findings of a recent study that showed the negative effects of non-conservative forces on minimization and \ac{md} workflows \cite{bigi_2025_darkforcesassessingnonconservative}. For the tasks investigated, the accuracy of gradient-based forces is crucial and outweighs the computational overhead compared to models that directly predict forces.

Moreover, a critical architectural choice is the inclusion of inductive biases to yield meaningful, physically plausible internal representations of the structures, e.g., with spherical harmonics or equivariance. The orb architecture presents an exception, with an underlying GNS architecture \cite{sanchezgonzalez2020learningsimulatecomplexphysics_gns} approach and strictly learned equivariances through regularization and data augmentation. Our results show that this architecture is competitive with the best-performing, more constrained frameworks on all tasks. Furthermore, it exceeds the accuracy of the fine-tuned MACE-MP-MOF0 model in the heat capacity prediction. This presents a valuable case for architectures that rely on learned symmetries from the data, instead of pre-defined inductive biases that often come at the cost of worse computational scaling \cite{rhodes_orb-v3_2025}. In contrast, the eSEN architecture was developed through rigorous evaluation of architectural decisions to ensure smooth potential surfaces. This approach has also proven successful for modeling nanoporous structures across all investigated tasks, leading to excellent accuracy and outperforming MACE-MP-MOF0 on several tasks. Overall, throughout the study, no architecture paradigm among GNNs, GTs, GNS, or graph-basis functions could be established as superior, highlighting the validity of all approaches for accurate modeling of atomistic systems.

Our investigation of the fine-tuned MACE-MP-MOF0 model showed that fine-tuning over a well-designed, yet relatively small dataset of 127 structures and 4764 DFT data points can adjust the \ac{pes} to a subset of \acp{mof} with high accuracy. Although the fine-tuned model outperforms the original universal MACE variants, its advantage has narrowed as newer, more powerful universal \acp{umlip} achieve comparable or better accuracy. This raises the question of the effectiveness of fine-tuning, especially when the resulting model is no longer applicable to most elements. We conclude that fine-tuning can be an effective approach for domains with limited training data available for the training of universal models or when one is interested in studying specific systems or tasks rather than high-throughput investigations. Still, the reduced applicability of the new model must be carefully evaluated. Furthermore, training or fine-tuning on D3-corrected data can significantly reduce the runtime for inference, and MACE-MP-MOF0 successfully demonstrated its effectiveness.

Our modular benchmark is publicly available and designed for extensibility, allowing easy integration of new \acp{umlip} models and new DFT reference data. We expect this open resource will support continued development and systematic evaluation of universal machine learning interatomic potentials for MOFs and beyond, accelerating progress toward robust, general-purpose atomistic modeling tools.


\section{Methods}

All computations use the Atomic Simulation Environment (ASE) \cite{HjorthLarsen2017_ase}, which enables a reproducible pipeline for all \acp{umlip}. All model tasks are evaluated on Nvidia V100 32GB SXM, A100 80GB SXM, or H100 94GB SXM GPUs. Speed benchmarking is performed using a single H100 94GB SXM GPU.

\subsection{Structure Curation}

Our analysis focuses on \acp{mof} but additionally includes \ac{cof} and zeolite structures, which have also been investigated for \ch{CO2} capture applications. We curate structures from several sources. Our curated set is divided into five subsets:

\begin{enumerate}[label=\arabic*.]
    \item Prototypical \acp{mof}: Four MOF structures ubiquitous in the literature: IRMOF-1, IRMOF-10, UiO-66, and HKUST-1.
    \item Main structure set: In addition to the prototypical MOFs, we curated a set of 96 structures comprising 52 \acp{mof} from MOSAEC-DB \cite{gibaldi_mosaec-db_2024}, 27 \acp{mof} from QMOF \cite{rosen_machine_2021_qmof}, 10 zeolites, and 7 \acp{cof} from the CURATED-COF dataset \cite{ongari_building_2019}. Structures from MOSAEC and QMOF were selected using stratified sampling across the largest pore diameter, computed with Poreblazer \cite{sarkisov_materials_2020_poreblazer}, to feature diverse chemistries and metal variance. Zeolites and COFs were selected based on overlap with Ref. \cite{moosavi_data-science_2022}\,. 
    \item Copper \acp{mof}: We curated a set of 13 Cu \acp{mof} from QMOF \cite{rosen_machine_2021_qmof} with varying copper coordination numbers. 
    \item Heat capacity set: 231 MOF, COF, and zeolite structures from Ref. \cite{moosavi_data-science_2022} with available DFT heat capacity.
    \item GoldDAC \cite{Lim_2025_DAC-SIM}: 26 MOFs with inserted molecules at different distances, resulting in a test set of 312 structures with available DFT interaction energies. D3 dispersion corrections were applied to match the universal potentials.
\end{enumerate}

To ensure high quality of the structures, we performed error checks with MOFChecker 2.0 \cite{jin_mofchecker_2025} and only selected structures with no irregularities. Several structures were reduced to primitive unit cells to reduce the computational complexity of the reference DFT calculations.

\subsection{Machine Learning Interatomic Potentials}

We evaluate 9 \acp{umlip} and MACE-MP-MOF0 in the main text (shown in bold) with 11 more shown in the SI:

\begin{enumerate}[label=\arabic*.]
    \item \textbf{MACE-MOF-0}\cite{elena2024machinelearnedpotentialhighthroughput_mace_mof_0}: A fine-tuned MACE-MP-0b (medium) \cite{batatia_foundation_2024_mace} model on a dataset of 127 MOFs and 4764 DFT calculations. In contrast to the original MACE-MP-0b model, it supports fewer elements and cannot compute all structures from our selection.
    \item \textbf{MACE-MP-0a}\cite{batatia_foundation_2024_mace}: The first universal MACE \cite{batatia_macehigherorderequivariant_2023} model trained on MPtraj \cite{deng_chgnet_2023_mptraj}.
    \item MACE-MP-0b3 (medium)$^\ast$ \cite{batatia_foundation_2024_mace}: A MACE \cite{batatia_macehigherorderequivariant_2023} model with improved pair repulsion, correct isolated atoms, and better stability at high pressures. Trained only on MPtraj\cite{deng_chgnet_2023_mptraj}.
    \item MACE-MPA-0\cite{batatia_foundation_2024_mace}: A MACE \cite{batatia_macehigherorderequivariant_2023} model trained on the MPtraj \cite{deng_chgnet_2023_mptraj} and sAlex \cite{schmidt_machinelearningassisted_2023_alexandria, barroso-luque_open_2024_omat24} dataset.
    \item \textbf{MACE-OMAT-0}\cite{batatia_foundation_2024_mace}: A MACE \cite{batatia_macehigherorderequivariant_2023} model trained on the OMat24 \cite{barroso-luque_open_2024_omat24} dataset.
    \item MACE-MATPES-r2SCAN-0\cite{batatia_foundation_2024_mace}: A MACE \cite{batatia_macehigherorderequivariant_2023} model trained on the MATPES-r2SCAN \cite{kaplan_foundational_2025_matpes} dataset.
    \item \textbf{MatterSim-v1 (5M)} \cite{yang_mattersim_2024}: Based on the M3GNet architecture \cite{Chen2022_m3gnet} that models three-body interactions using a graph-neural network approach. Its training uses an uncertainty-aware active-learning pipeline using model ensembles for uncertainty estimation.
    \item \textbf{orb-d3-v2} \cite{neumann2024orbfastscalableneural}: A Graph Network-based Simulator (GNS) architecture \cite{sanchezgonzalez2020learningsimulatecomplexphysics_gns} trained on the MPtraj \cite{deng_chgnet_2023_mptraj} and Alexandria \cite{schmidt_machinelearningassisted_2023_alexandria} datasets. Notably, the training was performed on D3-corrected energies and forces, eliminating the need to compute dispersion corrections at inference time.
    \item orb-mptraj-only-v2$^\ast$: A GNS trained only on MPtraj \cite{deng_chgnet_2023_mptraj}. In contrast to orb-d3-v2, this model does not predict D3-corrected outputs.
    \item \textbf{orb-v3-con-inf-omat} \cite{rhodes_orb-v3_2025}: The third-generation orb model with uncapped neighbor limit and conservative forces, trained on the OMat24 \cite{barroso-luque_open_2024_omat24} dataset.
    \item orb-v3-con-inf-mpa \cite{rhodes_orb-v3_2025}: The third-generation orb model with uncapped neighbor limit and conservative forces, trained on the MPTraj \cite{deng_chgnet_2023_mptraj} and Alexandria \cite{schmidt_machinelearningassisted_2023_alexandria} datasets.
    \item \textbf{eqV2-M OMat MPtrj-sAlex} \cite{barroso-luque_open_2024_omat24}: A model based on the EquiformerV2 architecture \cite{liao2024equiformerv2improvedequivarianttransformer}, trained on the OMat24 dataset \cite{barroso-luque_open_2024_omat24}, and fine-tuned on the MPtraj \cite{deng_chgnet_2023_mptraj} and sAlex \cite{barroso-luque_open_2024_omat24, schmidt_machinelearningassisted_2023_alexandria} datasets.
    \item eqV2-M-DeNS$^\ast$\cite{barroso-luque_open_2024_omat24}: An EquiformerV2 \cite{liao2024equiformerv2improvedequivarianttransformer} model that only uses the MPtraj training data.
    \item \textbf{eSEN-30M-OAM} \cite{fu2025learningsmoothexpressiveinteratomic}: Based on the eSEN architecture which ensures smooth and expressive potential energy surfaces, trained on OMat24 \cite{barroso-luque_open_2024_omat24}, MPtraj \cite{deng_chgnet_2023_mptraj}, and sAlex \cite{barroso-luque_open_2024_omat24, schmidt_machinelearningassisted_2023_alexandria}.
    \item eSEN-30M-MP$^\ast$ \cite{fu2025learningsmoothexpressiveinteratomic}: An eSEN model only trained on MPtraj \cite{deng_chgnet_2023_mptraj}.
    \item GRACE-2L-MP (r6)$^\ast$\cite{Bochkarev_graph_2024_grace}: The GRACE model extends the \ac{ace} \cite{drautz_atomic_2019} to incorporate graph basis functions. This model was only trained on MPtraj\cite{deng_chgnet_2023_mptraj}.
    \item \textbf{GRACE-2L-OMAT} \cite{Bochkarev_graph_2024_grace}: This GRACE model was only trained on OMat24 \cite{barroso-luque_open_2024_omat24}.
    \item GRACE-2L-OAM (r6) \cite{Bochkarev_graph_2024_grace}: This GRACE model was pre-fitted on OMat24 \cite{barroso-luque_open_2024_omat24} and fine-tuned on the sAlex\cite{barroso-luque_open_2024_omat24, schmidt_machinelearningassisted_2023_alexandria} and MPTraj\cite{deng_chgnet_2023_mptraj} datasets.
    \item SevenNet-0$^\ast$ \cite{park_scalable_2024_sevennet}: A model based on the NequIP architecture \cite{Batzner2022_nequip} trained on MPtraj \cite{deng_chgnet_2023_mptraj}.
    \item SevenNet-l3i5$^\ast$ \cite{park_scalable_2024_sevennet}: A SevenNet model with increased complexity trained on MPtraj \cite{deng_chgnet_2023_mptraj}.
    \item \textbf{SevenNet-ompa} \cite{park_scalable_2024_sevennet}: A SevenNet trained on OMat24 \cite{barroso-luque_open_2024_omat24}, sAlex \cite{barroso-luque_open_2024_omat24, schmidt_machinelearningassisted_2023_alexandria}, and MPtraj \cite{deng_chgnet_2023_mptraj}.
\end{enumerate}

\vspace{10pt}

Matbench Discovery-compliant models are marked with an asterisk ($^\ast$). While their performance is expected to lag behind models with broader training data, their inclusion enables direct architectural comparisons. Extended results using all model checkpoints are reported in the SI.

To the best of our knowledge, the training data sets that were used in the construction of these \acp{mlip} do not contain \ac{mof} structures, challenging them to make zero-shot predictions for this class of materials. The MACE-MP-MOF0 model is the only model that explicitly includes a training procedure on \acp{mof}.

\textbf{Dispersion Correction}:
Dispersion corrections are critical for accurate modeling of \acp{mof} \cite{formalik_benchmarking_2018}. Therefore, all models either predict D3-corrected outputs (orb-d3-v2 and MACE-MP-MOF0) or the D3 correction is computed at inference time using the torch-dftd \cite{takamoto2021pfp_dftd} package with \texttt{dispersion\_xc=pbe}, \texttt{dispersion\_cutoff=40}\,Bohr, \texttt{damping=bj}.

\subsection{Structural minimization}
Our energy minimization optimizes atom positions and cell parameters simultaneously and is performed using the \texttt{FrechetCellFilter} and the \texttt{LBFGS} optimizer. All structures are relaxed until a force convergence criterion of $10^{-3}$\,eV/\AA\ or a maximum of 5,000 optimizer steps is reached. UFF/UFF4MOF simulations were performed using the  LAMMPS package\cite{thompson_lammps_2022}. Force field parameters were applied by running \texttt{lammps-interface}\cite{boyd_force-field_2017} without \texttt{-{}-fix-metal} and with \texttt{-{}-fix-metal}. The minimizer style in LAMMPS was set to \texttt{cg}, all other parameters and convergence settings were kept the same as those in the ASE \cite{HjorthLarsen2017_ase} calculations.

\subsection{Molecular Dynamics Simulations}
We perform three types of \ac{md} simulations: generic NpT at ambient pressure and temperature, NpT at slightly increased temperature, and NpT with gradually increasing temperatures up to 1000\,K. All simulations use a time step of 1\,fs and are performed with the ase IsotropicMKTNPT driver \cite{HjorthLarsen2017_ase}. We tested simulation performance with the ASE NPT driver, ASE NPTBerendsen, and the LAMMPS  NPT driver for MACE and found no qualitative differences in the results using the MACE-MP-0b3 model.

All simulations start with atom position minimization using the \texttt{LBFGS} optimizer until a force convergence criterion of $10^{-3}$\,eV/\AA\ or 1,000 optimizer steps is reached. In LAMMPS, the minimizer style was set to \texttt{cg} in atomic position minimization. Velocities are then initialized from a Maxwell-Boltzmann distribution at 300\,K, and adjusted for zero center-of-mass momentum and zero total angular momentum. The structures are equilibrated in an NVT ensemble with Langevin dynamics at 300\,K using a friction coefficient of $0.01$\,fs$^{-1}$ for $1$\,ps. Subsequent NpT simulations use $\text{tdamp} = 100$\,fs, $\text{pdamp} = 1000 \text{\,fs}$, and an external stress of 1\,bar. 

In our first set of stability simulations, the NVT-equilibrated structures are simulated in an NpT ensemble at 300\,K for 50\,ps.

In simulations of copper \acp{mof}, the NVT-equilibrated structures are first modeled in an NpT ensemble at 300\,K for 10\,ps, then at 400\,K for 10\,ps, and again at 300\,K for 10\,ps. At the end of the simulation, atom positions are relaxed until a force convergence criterion of $10^{-3}$\,eV/\AA\ or 1,000 optimizer steps is reached to ensure accurate representation of the coordination environments. All the NVT and NpT settings in LAMMPS for stability and Cu coordination number simulations were consistent with those implemented in the ASE evaluations. Coordination numbers were computed using pymatgen \cite{Ong2013_pymatgen} CrystalNN with parameters \texttt{weighted\_cn=True}, \texttt{x\_diff\_weight=1.5}, \texttt{search\_cutoff=4.5} after the initial structural optimization and after the final structural optimization.

For simulations at increasing temperatures, the NVT-equilibrated structures are modeled in an NpT ensemble with temperatures from 300\,K to 1000\,K in 100\,K steps for 20\,ps each, resulting in a total NpT length of 160\,ps.

\subsection{Bulk Modulus}

All structures were optimized with atom position and cell minimization using the \texttt{LBFGS} optimizer until a force convergence criterion of $10^{-3}$\,eV/\AA\ or 1,000 optimizer steps was reached. We apply a volumetric strain of $\pm 4\,\%$ in 11 evenly spaced steps. The resulting structures were optimized using the \texttt{FIRE} optimizer until a force convergence criterion of $10^{-3}\,$eV/\AA\ or 1,000 optimizer steps was reached. The bulk modulus was computed from a fitted Birch-Murnaghan equation of state. Structures for which the EOS volume minimum deviates more than 1\% from the volume minimum of the initial optimization procedure are considered failed and excluded. The EOS fits of these structures were unstable in our experiments, leading to wrongly predicted bulk moduli, often as extreme outliers. Such filtering does not rely on ground-truth data and can be applied to reduce uncertainty in the predictions.

The DFT bulk moduli were calculated using the CP2K ver.9.1 package~\cite{kuhne2020cp2k}.
Structures were fully optimized with respect to atomic positions and cells. Given the high computational cost of the DFT calculations, the volumetric strain of $\pm 4\,\%$ was applied in 5 evenly spaced steps. Structures with strains were optimized with respect to atomic positions. All calculations were performed through the Automated Interactive Infrastructure and Database for Computational Science, AiiDA~\cite{pizzi2016aiida}, employing the \textit{Cp2KMultistageWorkChain} workflow from the \textit{aiida-lsmo} plugin. 

The Quickstep code~\cite{vandevondele2005quickstep} was used in the CP2K calculation. The Perdew-Burke-Enzerhof (PBE) exchange-correlation functional~\cite{perdew_generalized_1996} was employed along with DFT-D3(BJ) dispersion corrections~\cite{Grimme2010_dftd3}. The GTH pseudopotentials\cite{goedecker1996separable}, DZVP-MOLOPT-SR basis sets, and Gaussian plane wave were used. The cutoff energy of plane waves was set to 800\,Ry. The energy and force convergences in the self-consistent filed were $1\mathrm{E}{-8}$\,Ry and $0.00015\,\mathrm{bohr}^{-1} \times \mathrm{hartree}$, respectively. All other settings were set to the default defined in the \textit{Cp2KMultistageWorkChain}. These settings have been demonstrated to be robust enough to find energetically stable configurations by comparing with QMOF relaxed structures in Supplementary Figure \ref{fig:opt_boydwoo_str_m3_o23_o28_pcu_sym_68_primitive} and \ref{fig:opt_core_MUNNUF_freeONLY}.

\subsection{Heat Capacity}

All structures were first optimized with atom position and cell minimization using the \texttt{LBFGS} optimizer until a force convergence criterion of $10^{-3}$\,eV/\AA\ or 1,000 optimizer steps was reached. The heat capacity of the obtained structure was computed using Phonopy \cite{phonopy-phono3py-JPCM, phonopy-phono3py-JPSJ}. 
Force constants are computed using no supercells and the finite difference method with a distance of $0.01\,$\AA. 
The mesh sampling phonon calculation is performed with $\texttt{mesh = 100}$. The heat capacity is extracted for $T=300$\,K.

\subsection{GoldDAC}

Interaction energies are computed from the test set of the GoldDAC \cite{Lim_2025_DAC-SIM} dataset which features 312 structures from 26 MOFs with inserted gas molecules, either \ch{H2O} or \ch{CO2}, at different positions, including repulsive (R), equilibrium (E), and
weak-attraction near pore center (W) regions of the potential energy surface. Single-point calculations are performed to compute the potential energy of the framework with inserted molecule, the framework only, and the molecule only. Interaction energies are computed as $E_\text{interaction}~=~E_\text{total}~-~E_\text{MOF}~-~E_\text{gas}$, where $E_\text{total}$ is the potential energy of the MOF with an inserted gas molecule, $E_\text{MOF}$ is the potential energy of the MOF framework only, and $E_\text{gas}$ is the potential energy of the gas molecule.
Forces are computed for the frameworks with inserted molecules. 
Errors for interaction energies and forces are computed using the GoldDAC DFT references. Dispersion corrections were computed and added to the DFT references to match the uMLIPs.

\begin{acknowledgement}

This research was enabled by support provided by the SciNet HPC Consortium and the Digital Research Alliance of Canada, as well as support from the state of Baden-Württemberg through bwHPC. The project received financial support from the University of Toronto's Acceleration Consortium through the Canada First Research Excellence Fund under Grant number CFREF-2022-00042. S.M.M.\ research program receives financial support from Natural Sciences and Engineering Research Council of Canada (NSERC) through the discovery program. J.H.\ acknowledges support from the Eric and Wendy Schmidt AI in Science Postdoctoral Fellowship, a program of Schmidt Sciences.

\end{acknowledgement}

\section*{Author contributions}
Conceptualization: H.K., J.H., \& S.M.M.
; Data curation: H.K. \& J.H.
; Formal analysis: H.K., J.H., \& S.M.M.
; Funding acquisition: S.M.M.
; Investigation: H.K., J.H., \& S.M.M.
; Methodology: H.K., J.H., \& S.M.M.
; Software: H.K. \& J.H.
; Supervision: S.M.M.
; Validation: H.K. \& J.H.
; Visualization: H.K. \& J.H.
; Writing – original draft: H.K., J.H., \& S.M.M.
; Writing – review \& editing: H.K., J.H., \& S.M.M.

\section*{Competing interest}
The authors declare no competing interests.

\section{Data availability}
Supplementary Information, containing extended results for 21 MLIPs for optimization, molecular dynamics, heat capacity, and bulk modulus predictions, is available for this paper. In addition, the DFT and UFF results are available on GitHub at \url{https://github.com/AI4ChemS/mofsim-bench}

\section{Code availability}
The benchmarking code is available at \url{https://github.com/AI4ChemS/mofsim-bench}. 

\bibliography{Bibliography}

\appendix
\setcounter{figure}{0}
\renewcommand\thefigure{S.\arabic{figure}} 
\renewcommand\thetable{S.\arabic{table}} 

\newpage
\section{Supplemental Information} \label{sec:si}

\subsection{Speed Benchmarking}

\begin{figure}[H]
    \centering
    \includegraphics[width=\linewidth]{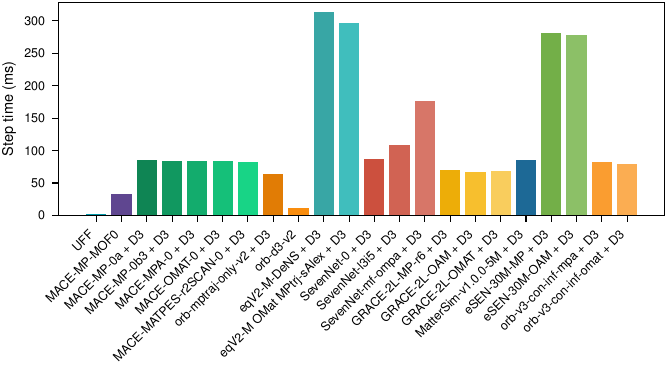}
    \caption{\textbf{Speed comparison of 1,000 optimization steps.}
    All models computed 1,000 optimization steps using FrechetCellFilter and the FIRE optimizer on MOF-5 (424 atoms). Results show the average runtime per optimization step. UFF was computed on 4 CPU cores, uMLIPs were computed using a NVIDIA H100 94GB SXM GPU.}
    \label{fig:speed}
\end{figure}

\subsection{QMOF Potential Energy}

\begin{figure}[H]
    \centering
    \includegraphics[width=\linewidth]{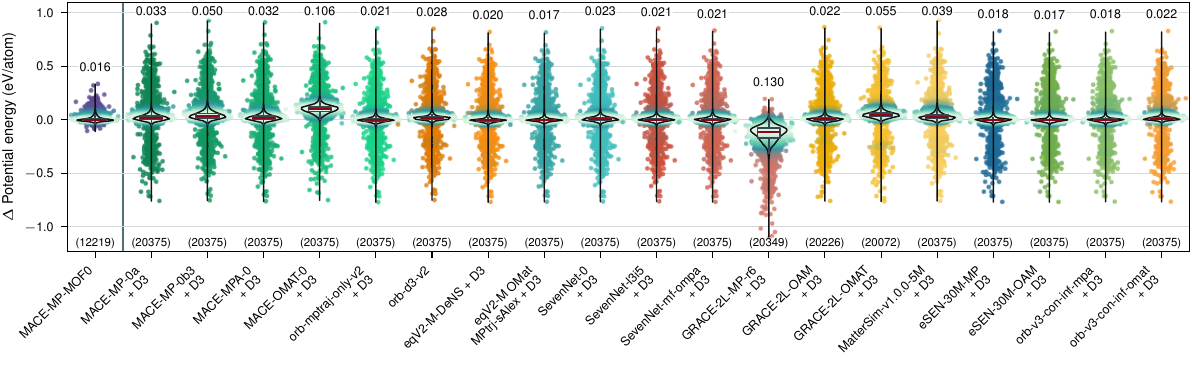}
    \caption{\textbf{Potential energy predictions vs. QMOF DFT references.}
    Numbers above the violin plots indicate the mean absolute error in eV/atom, numbers below show the number of computed structures. Highlighted regions indicate dense areas in the error distribution.}
    \label{fig:qmof_pot_eng_all}
\end{figure}

\begin{figure}[H]
    \centering
    \includegraphics[width=\linewidth]{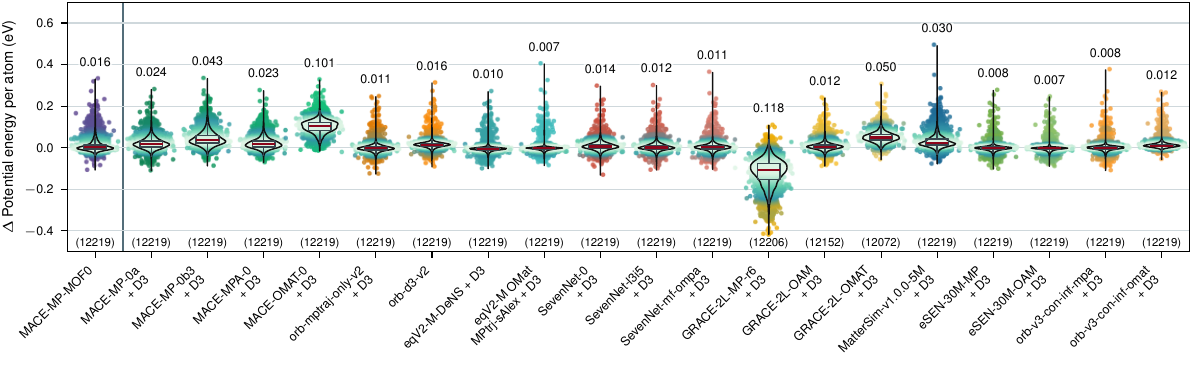}
    \caption{\textbf{Filtered potential energy predictions vs. QMOF DFT references.}
    Filtered to include only structures that MACE-MP-MOF0 can compute.
    Numbers above the violin plots indicate the mean absolute error in eV/atom, numbers below show the number of computed structures. Highlighted regions indicate dense areas in the error distribution.}
    \label{fig:qmof_pot_eng_all_intersection}
\end{figure}

\begin{figure}[H]
    \centering
    \includegraphics[width=\textwidth]{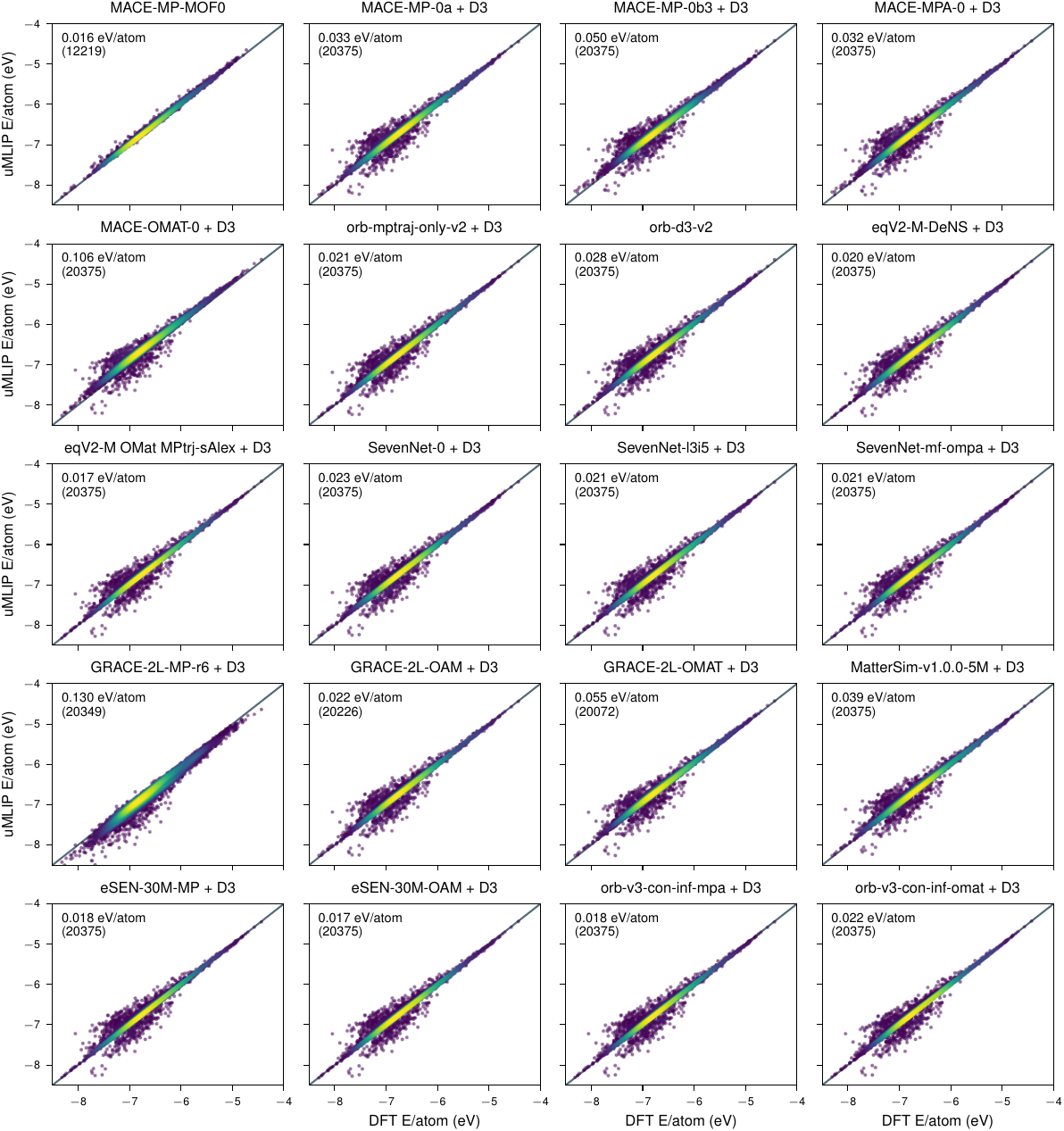}
    \caption{\textbf{Parity plots of uMLIP-predicted energy/atom of 20375 QMOF \cite{rosen_machine_2021_qmof} structures vs. DFT reference.} Brighter areas indicate a higher density of points. Each point represents one structure.}
    \label{fig:qmof_pot_eng_parity}
\end{figure}

\begin{figure}[H]
    \centering
    \begin{subfigure}[b]{0.3\textwidth}
        \centering
        \includegraphics[width=\textwidth]{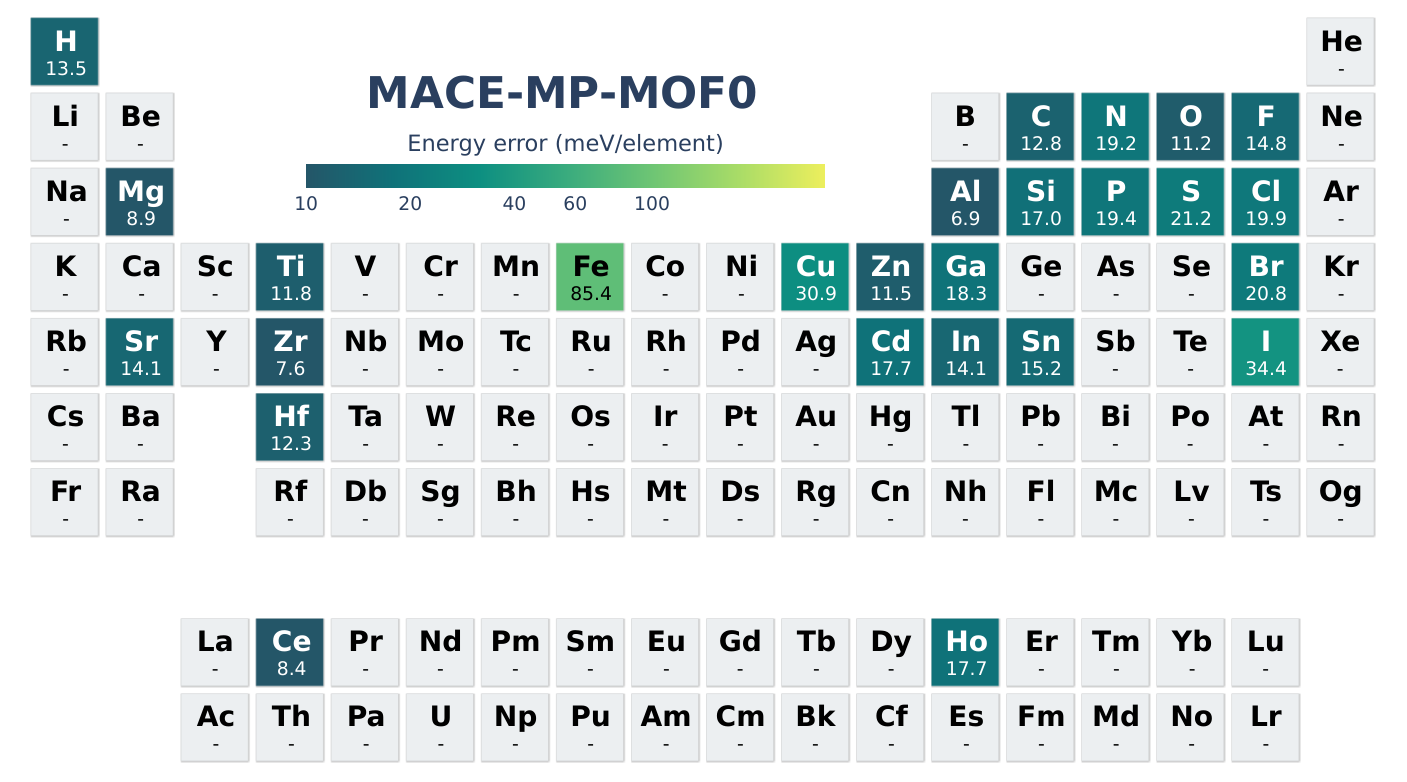}
    \end{subfigure}
    \hfill
    \begin{subfigure}[b]{0.3\textwidth}
        \centering
        \includegraphics[width=\textwidth]{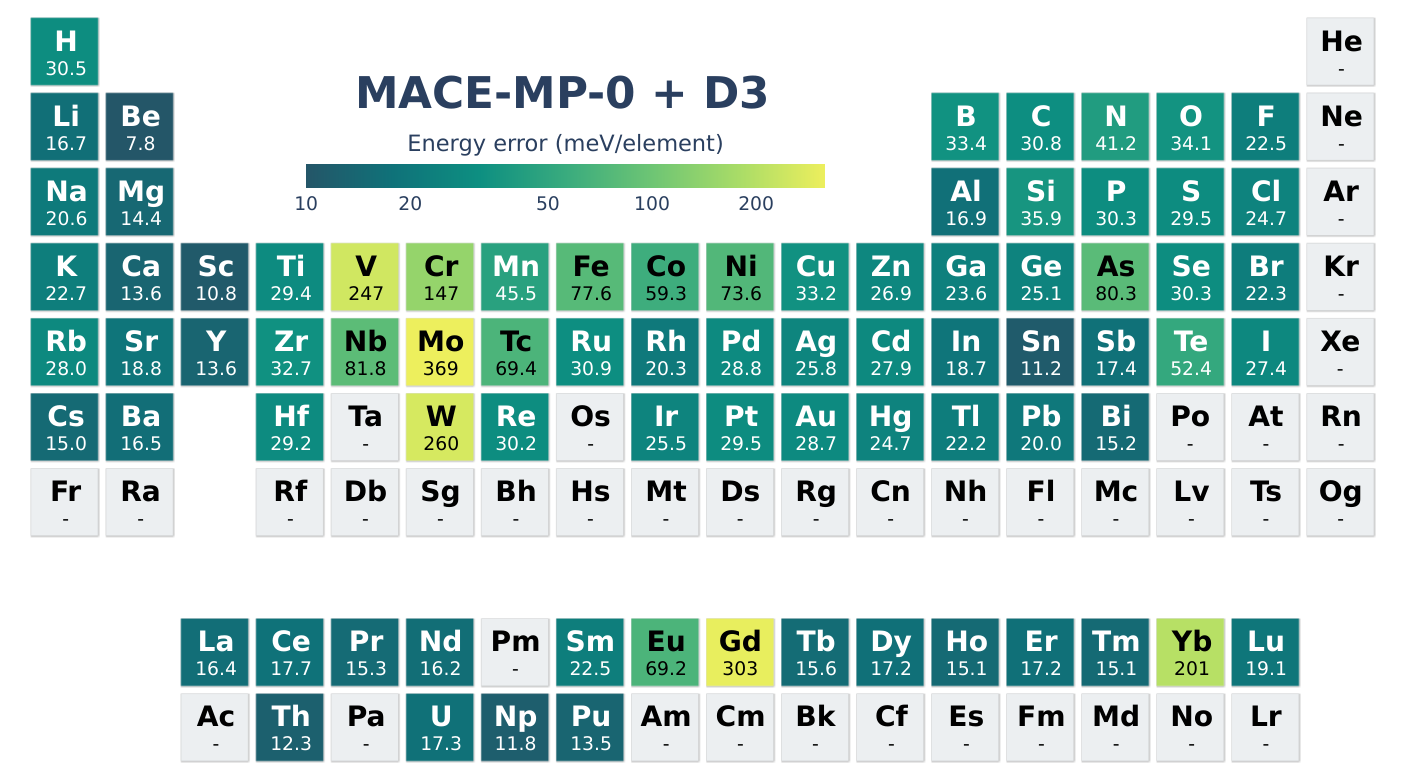}
    \end{subfigure}
    \hfill
    \begin{subfigure}[b]{0.3\textwidth}
        \centering
        \includegraphics[width=\textwidth]{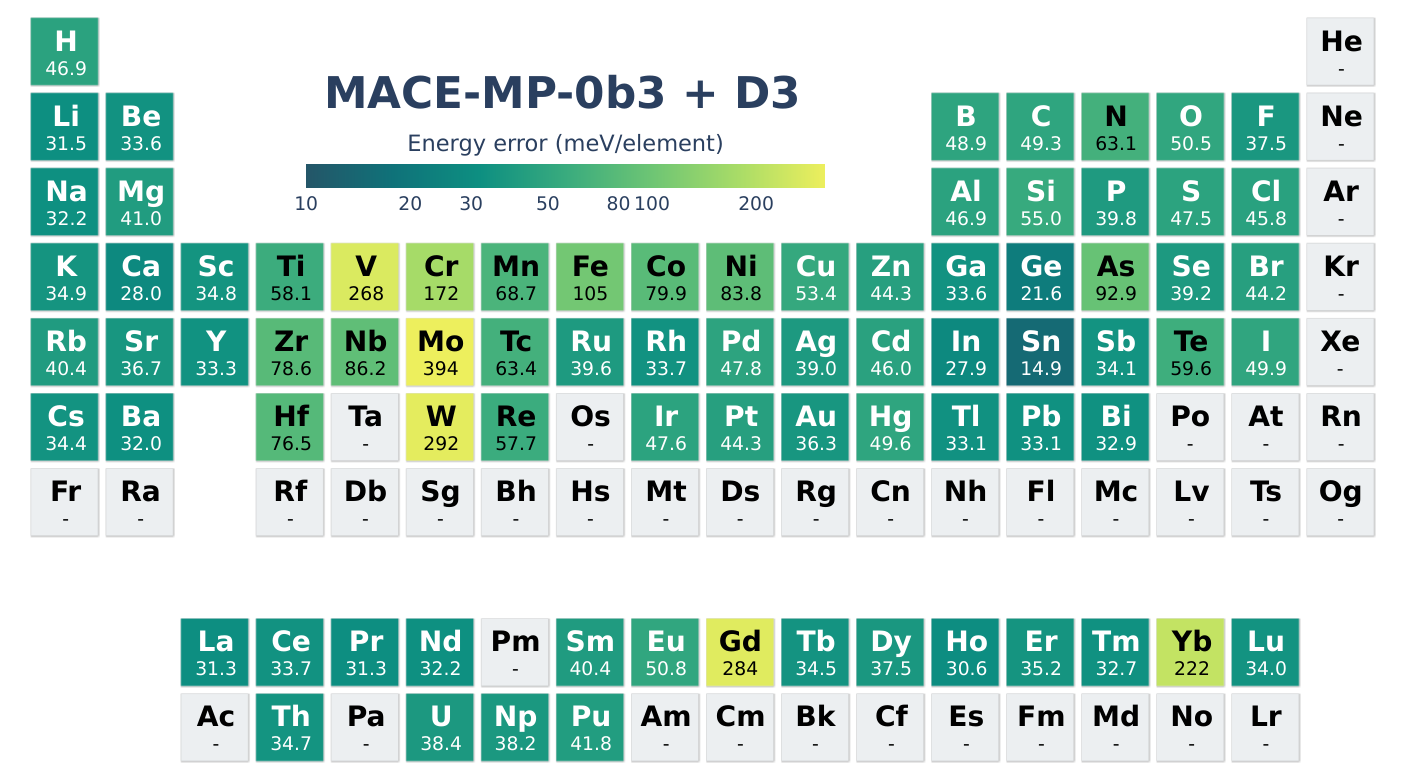}
    \end{subfigure}
    \hfill
    \begin{subfigure}[b]{0.3\textwidth}
        \centering
        \includegraphics[width=\textwidth]{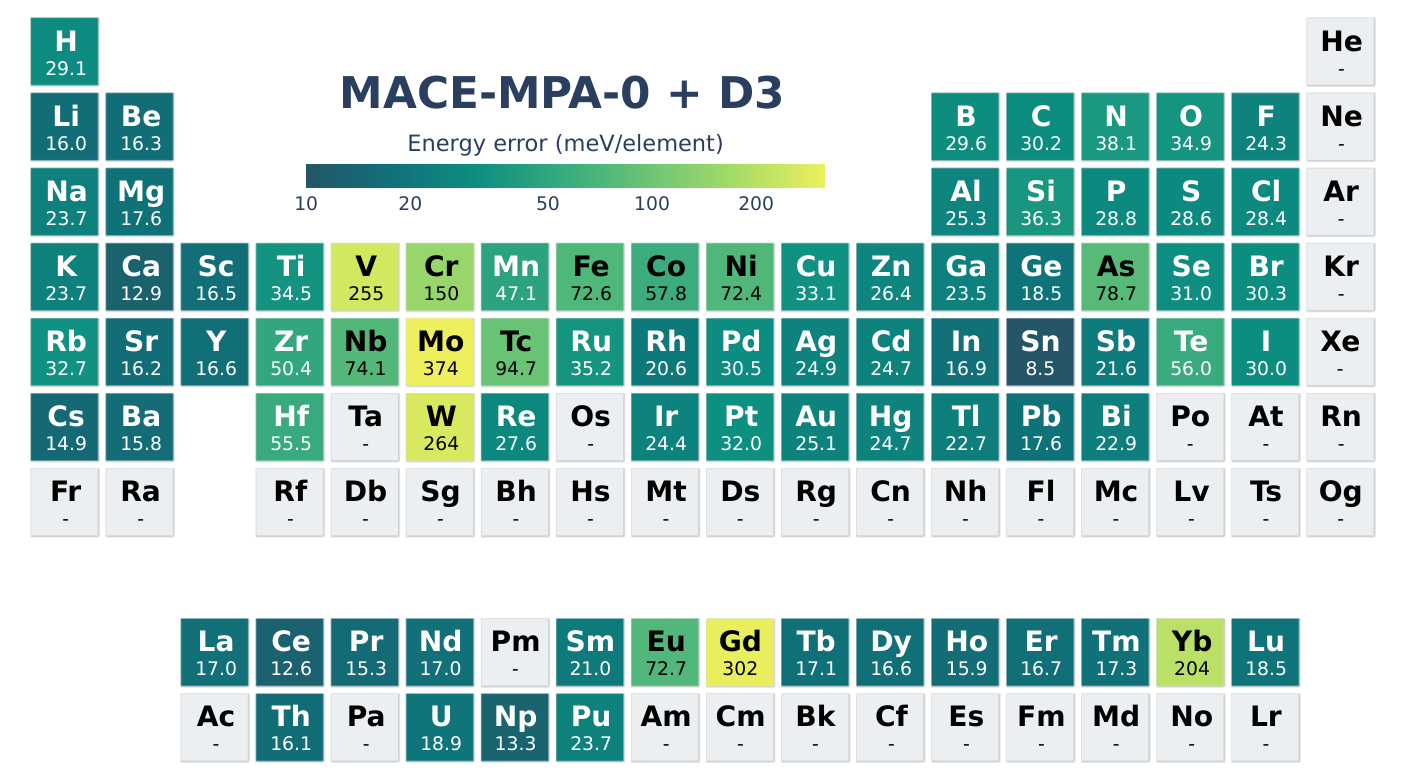}
    \end{subfigure}
    \hfill
    \begin{subfigure}[b]{0.3\textwidth}
        \centering
        \includegraphics[width=\textwidth]{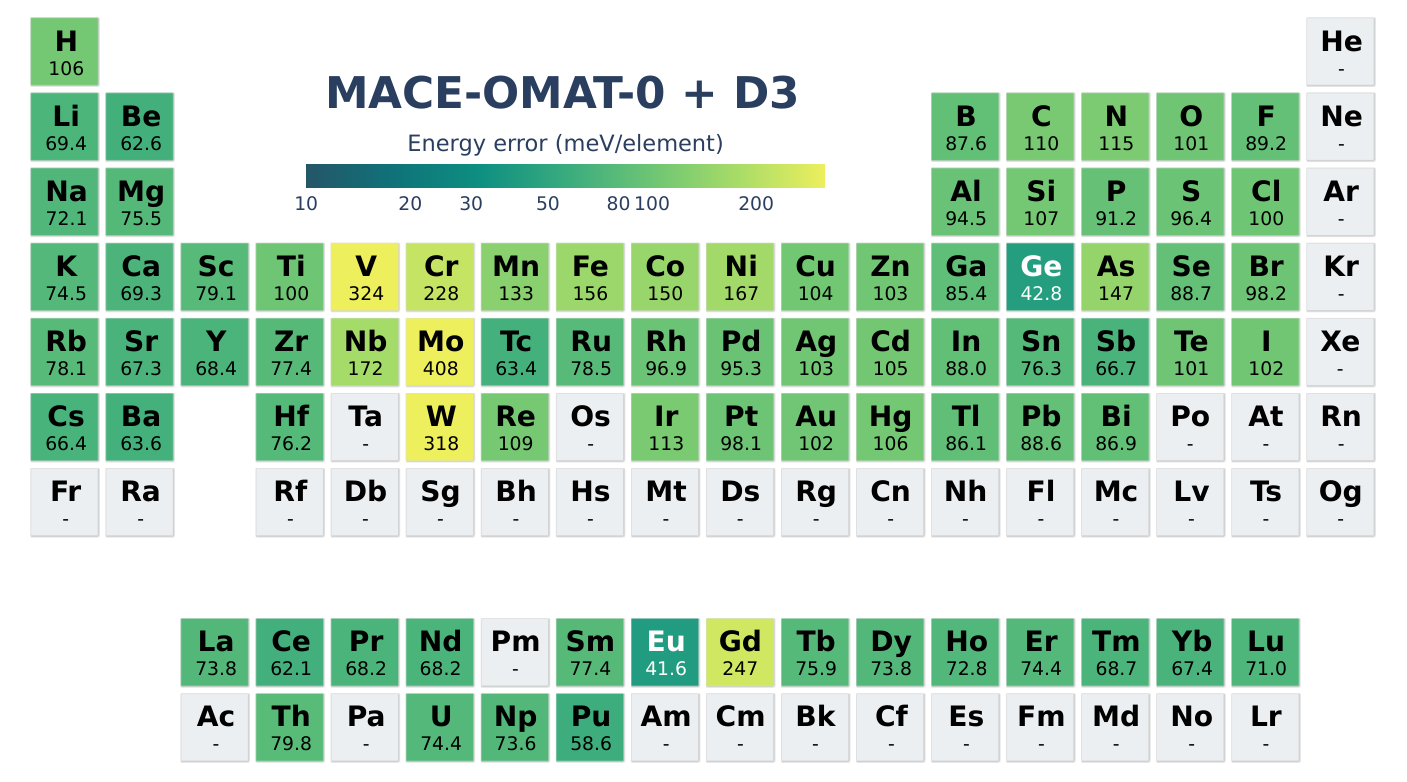}
    \end{subfigure}
    \hfill
    \begin{subfigure}[b]{0.3\textwidth}
        \centering
        \includegraphics[width=\textwidth]{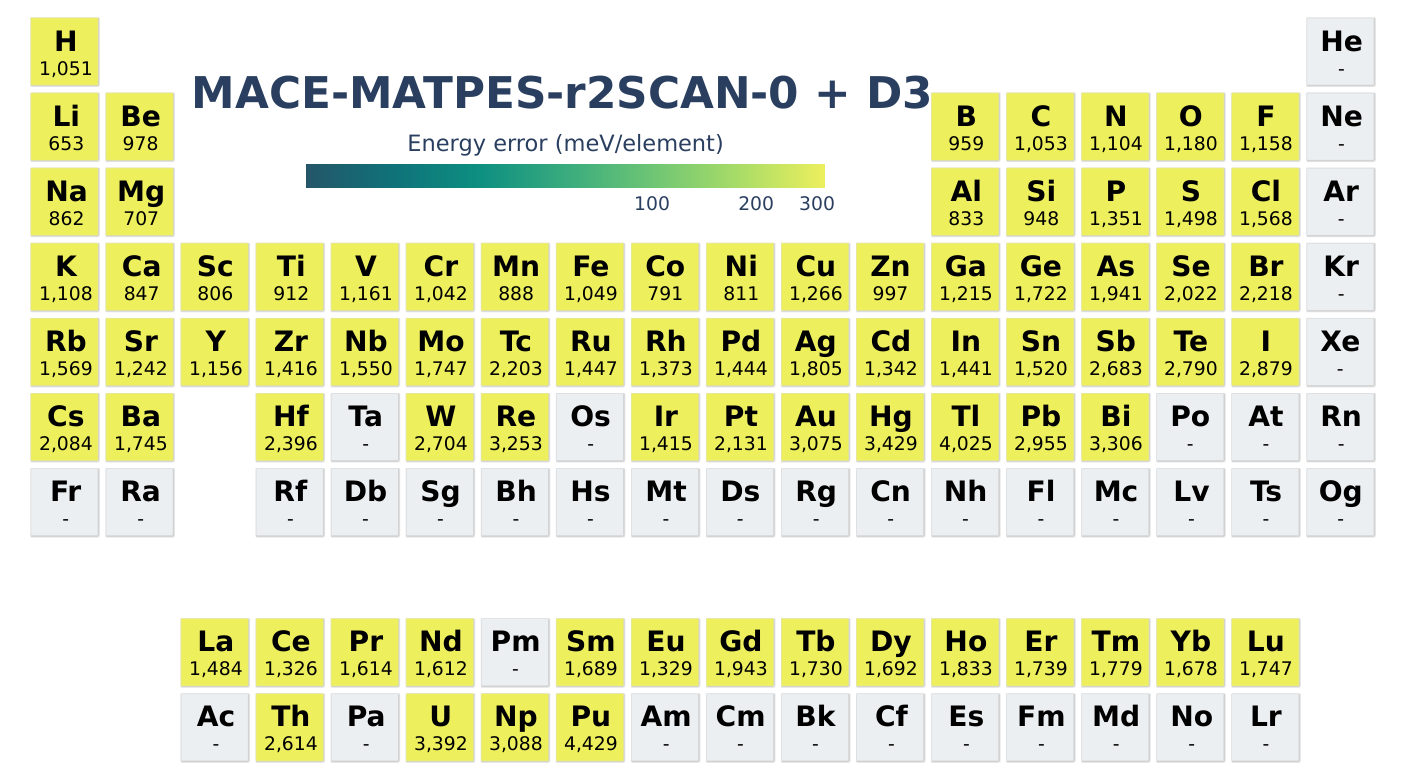}
    \end{subfigure}
    \hfill
    \begin{subfigure}[b]{0.3\textwidth}
        \centering
        \includegraphics[width=\textwidth]{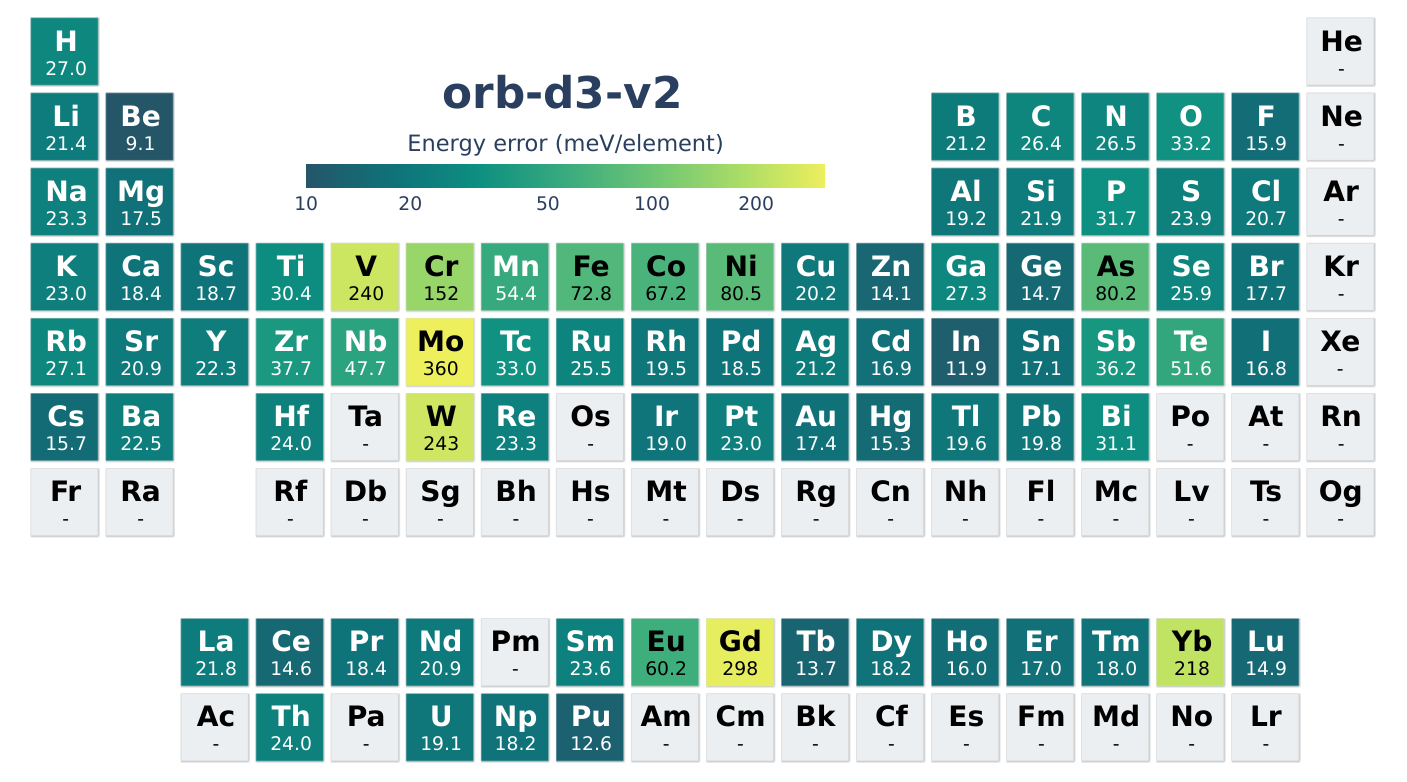}
    \end{subfigure}
    \hfill
    \begin{subfigure}[b]{0.3\textwidth}
        \centering
        \includegraphics[width=\textwidth]{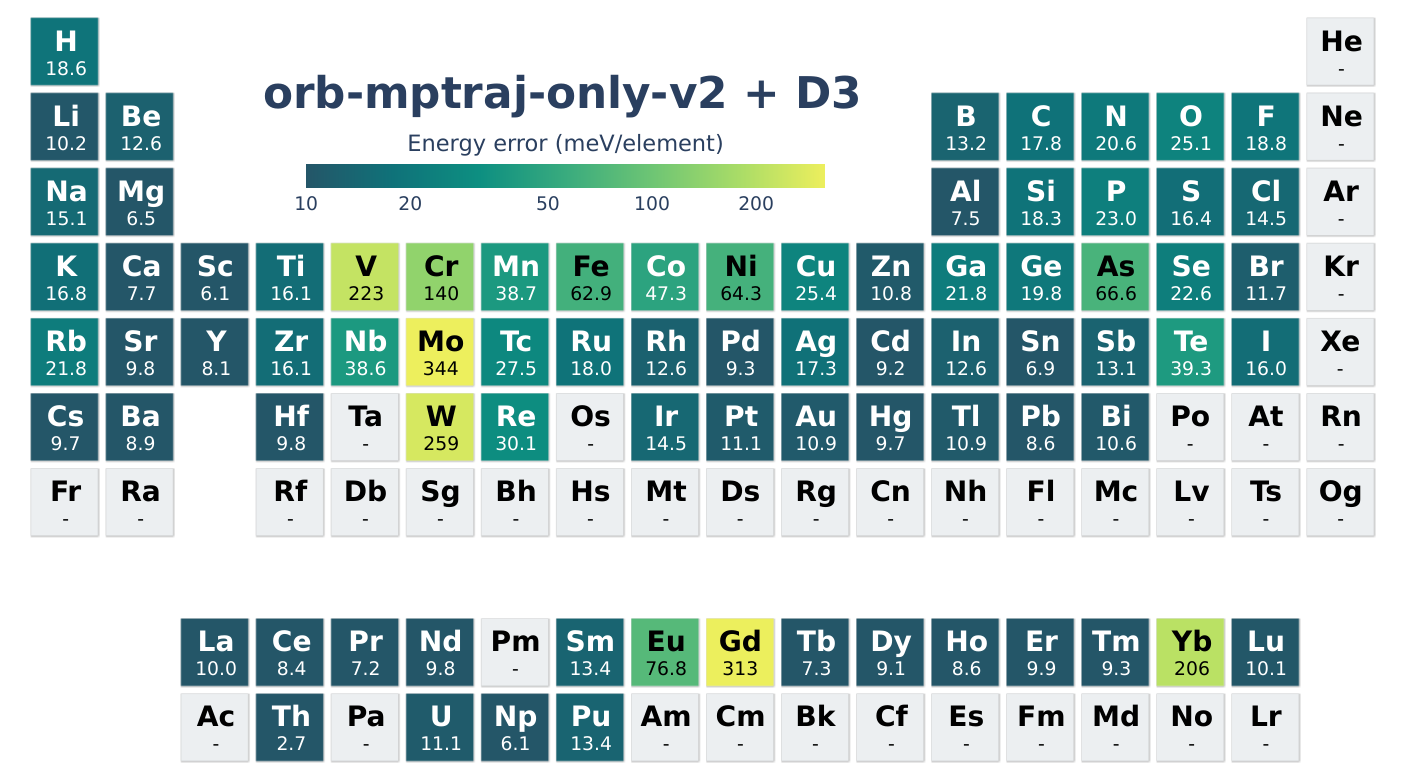}
    \end{subfigure}
    \hfill
    \begin{subfigure}[b]{0.3\textwidth}
        \centering
        \includegraphics[width=\textwidth]{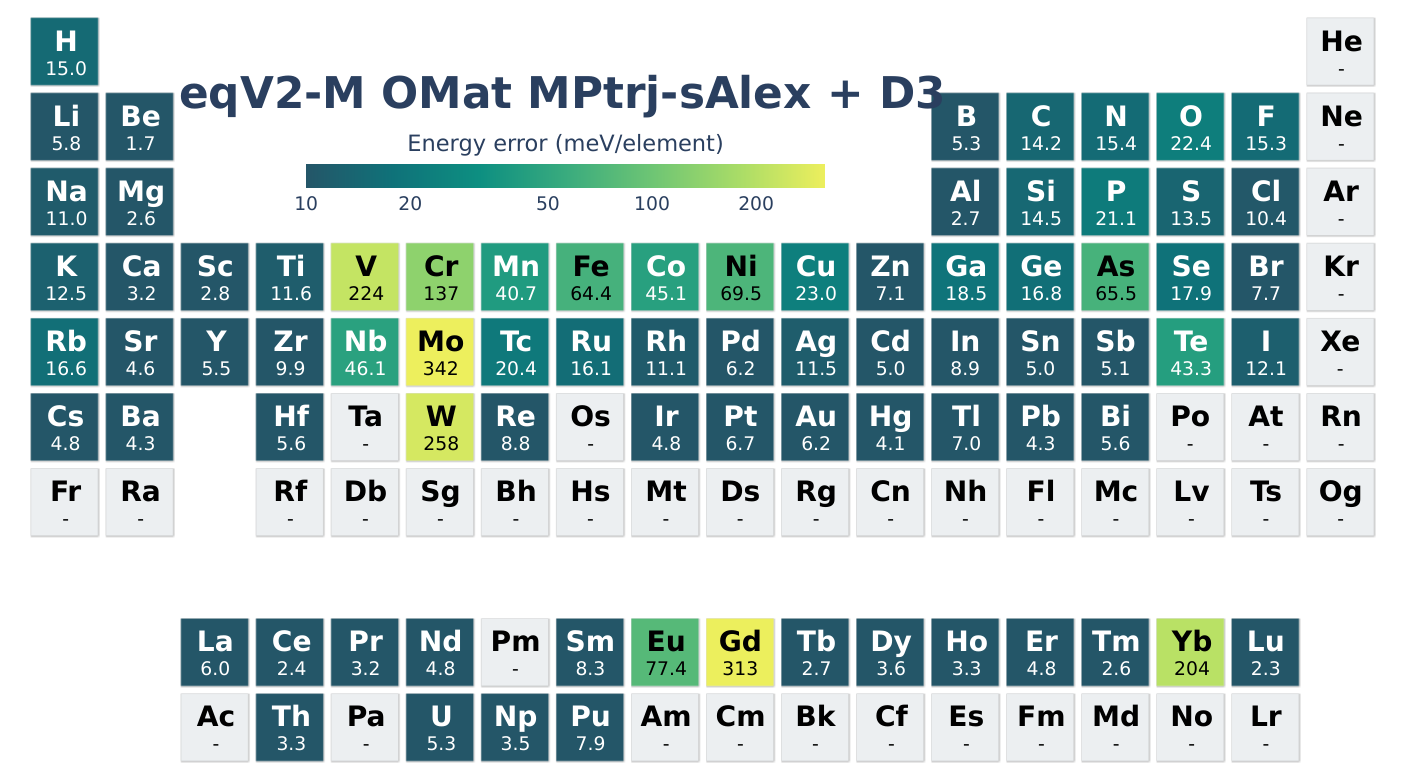}
    \end{subfigure}
    \hfill
    \begin{subfigure}[b]{0.3\textwidth}
        \centering
        \includegraphics[width=\textwidth]{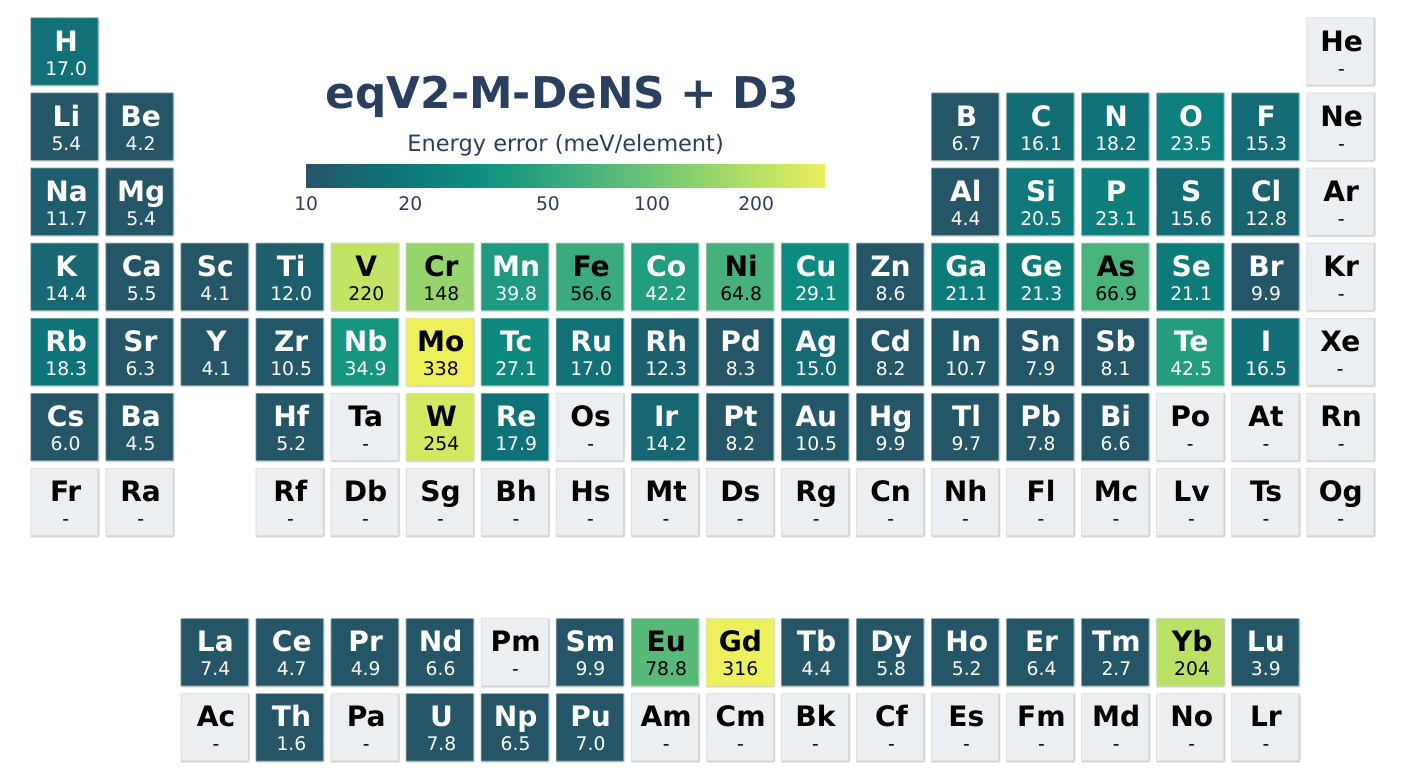}
    \end{subfigure}
    \hfill
    \begin{subfigure}[b]{0.3\textwidth}
        \centering
        \includegraphics[width=\textwidth]{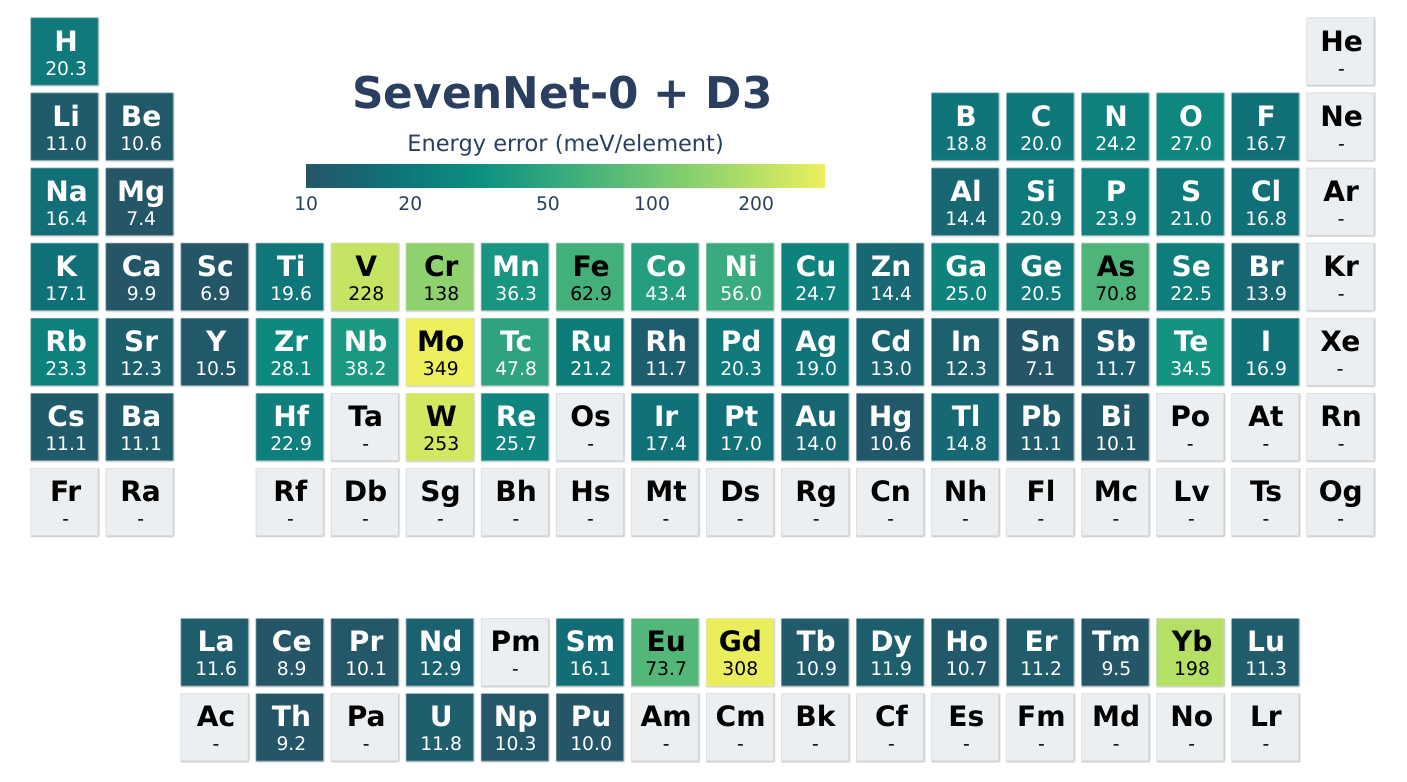}
    \end{subfigure}
    \hfill
    \begin{subfigure}[b]{0.3\textwidth}
        \centering
        \includegraphics[width=\textwidth]{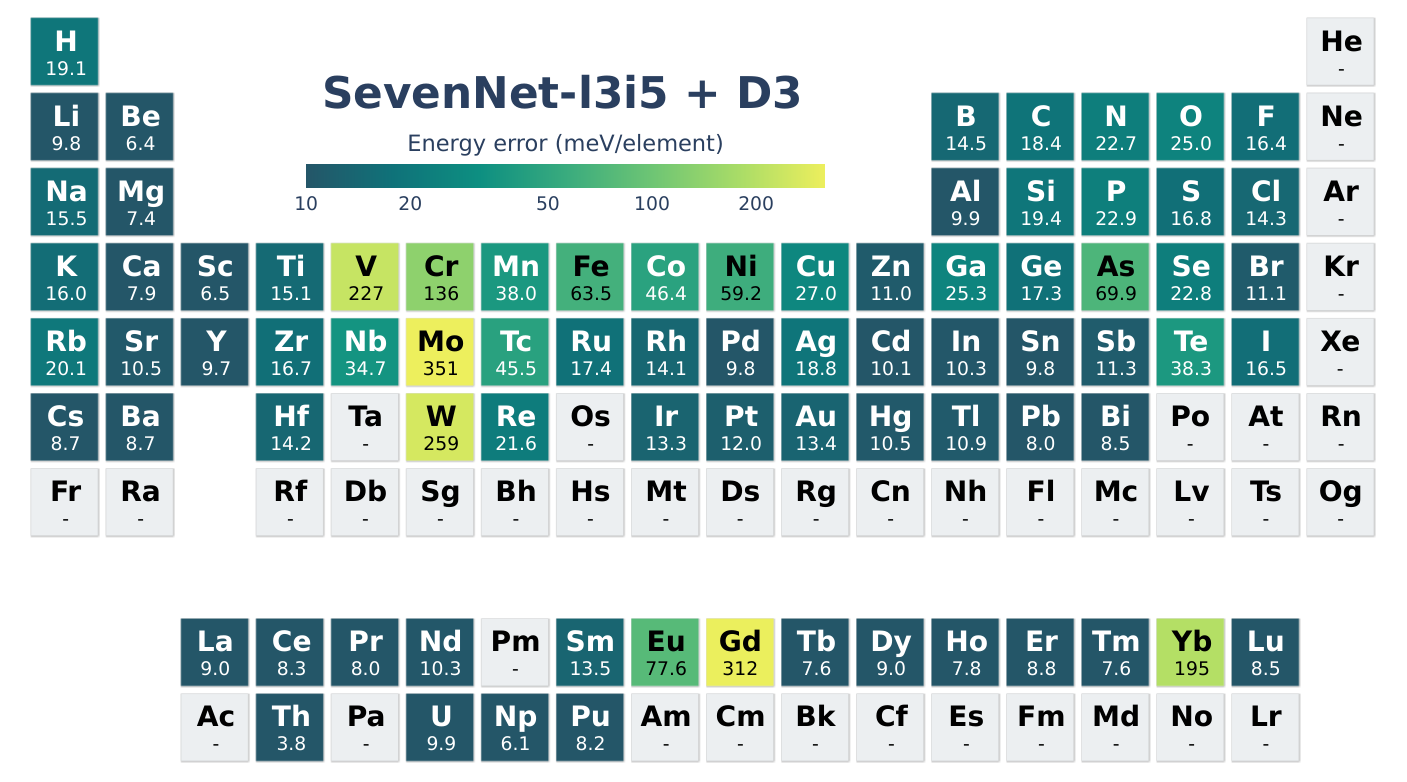}
    \end{subfigure}
    \hfill
    \begin{subfigure}[b]{0.3\textwidth}
        \centering
        \includegraphics[width=\textwidth]{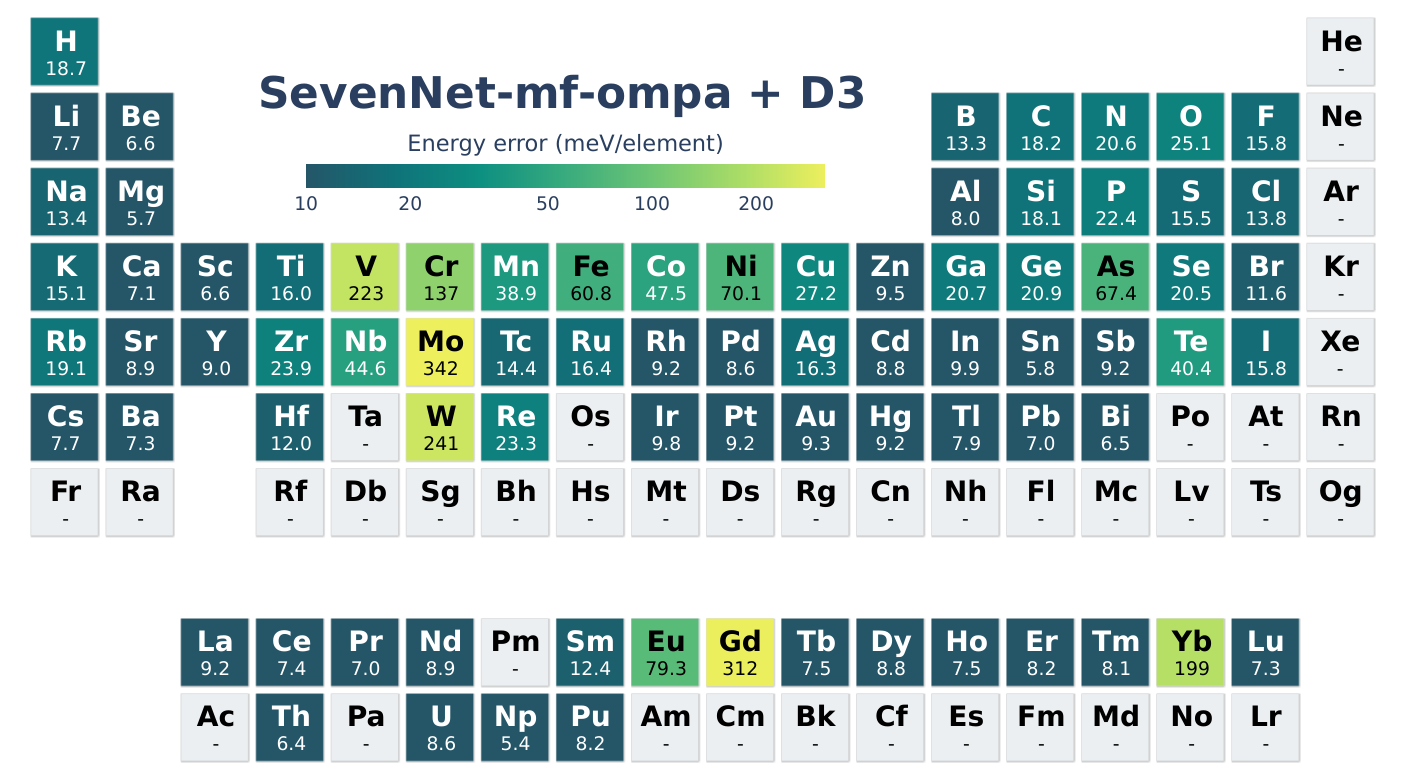}
    \end{subfigure}
    \hfill
    \begin{subfigure}[b]{0.3\textwidth}
        \centering
        \includegraphics[width=\textwidth]{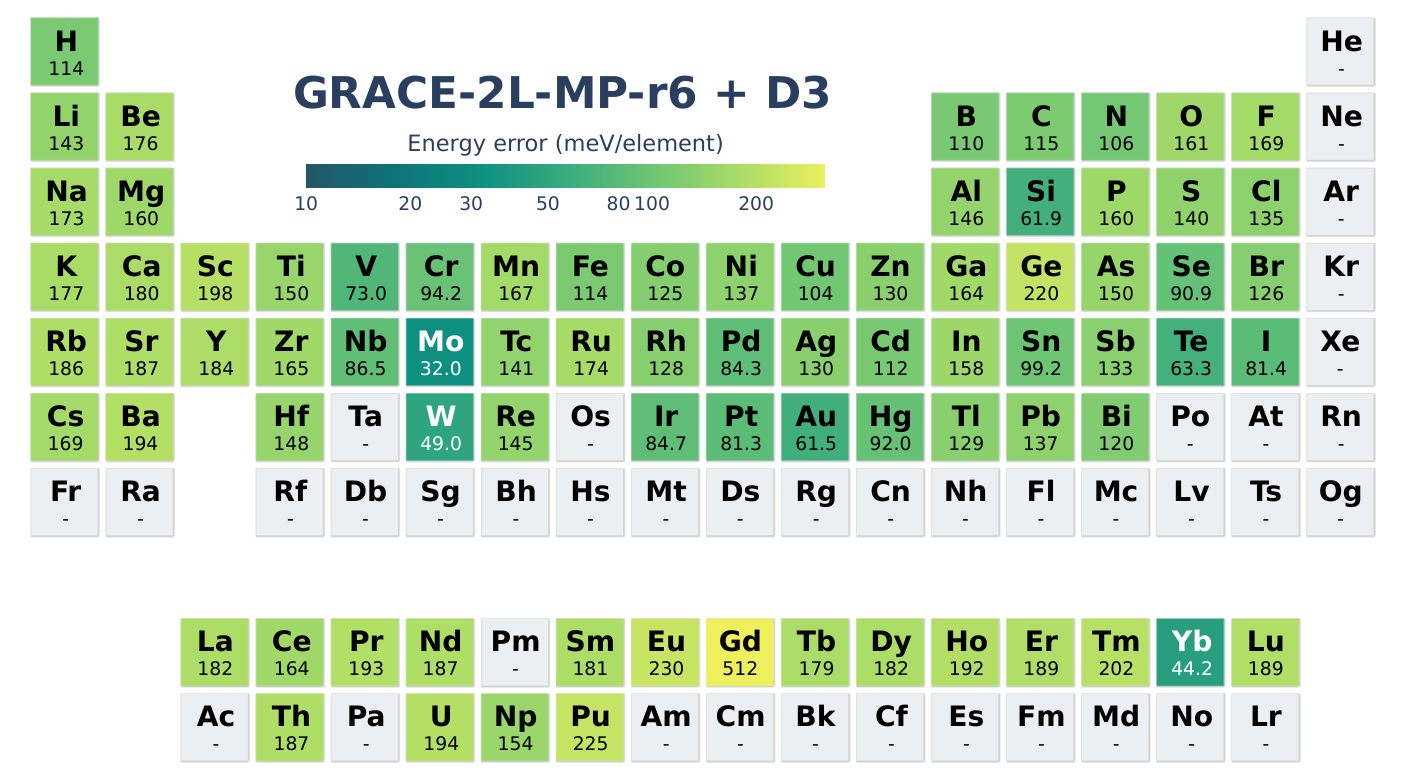}
    \end{subfigure}
    \hfill
    \begin{subfigure}[b]{0.3\textwidth}
        \centering
        \includegraphics[width=\textwidth]{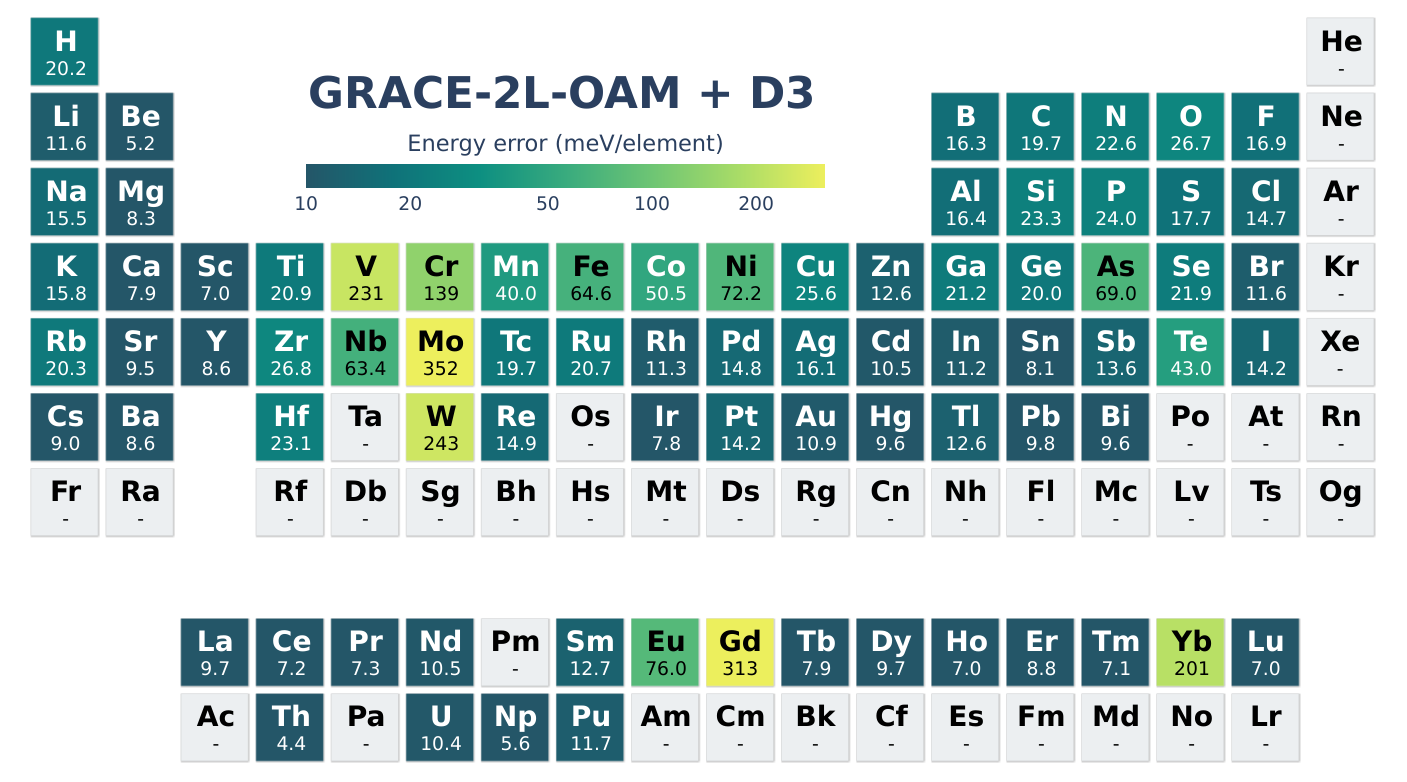}
    \end{subfigure}
    \hfill
    \begin{subfigure}[b]{0.3\textwidth}
        \centering
        \includegraphics[width=\textwidth]{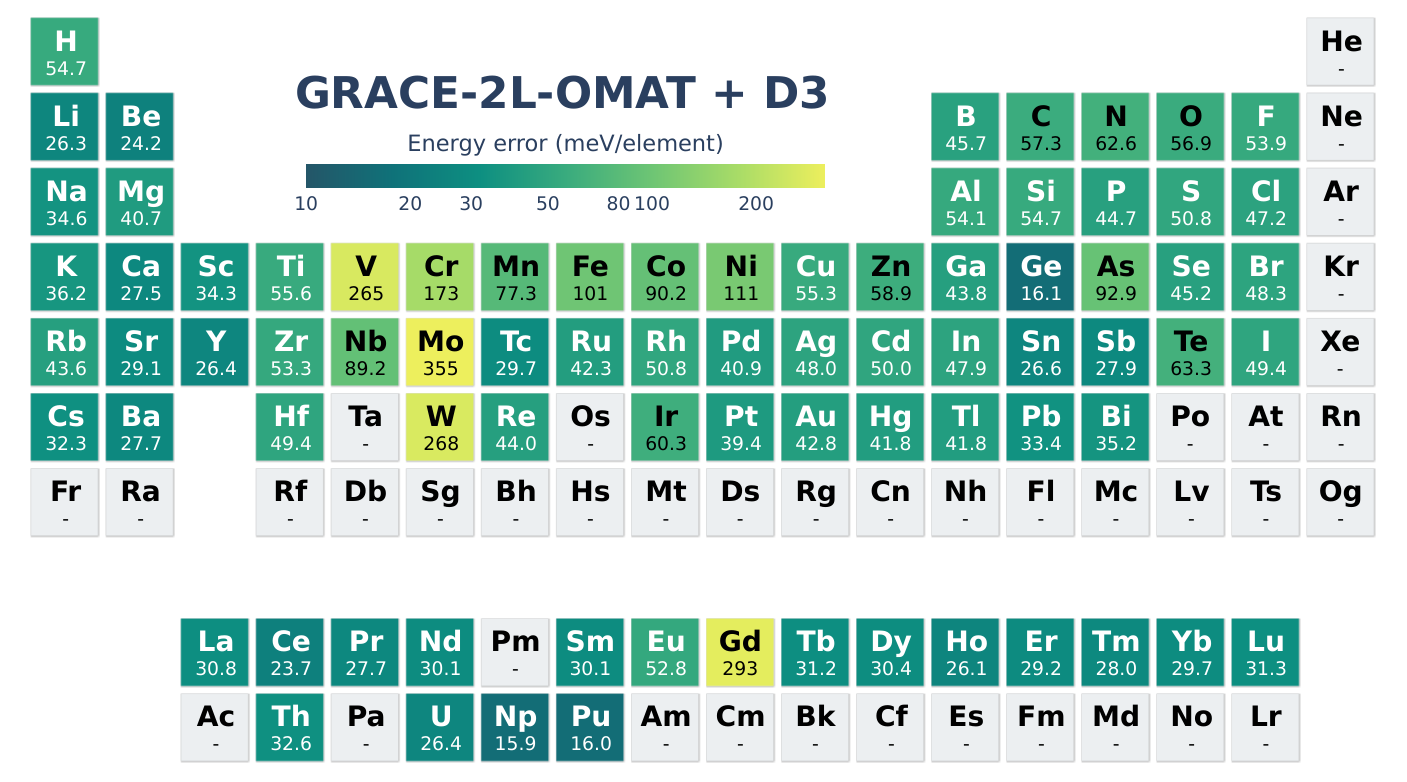}
    \end{subfigure}
    \hfill
    \begin{subfigure}[b]{0.3\textwidth}
        \centering
        \includegraphics[width=\textwidth]{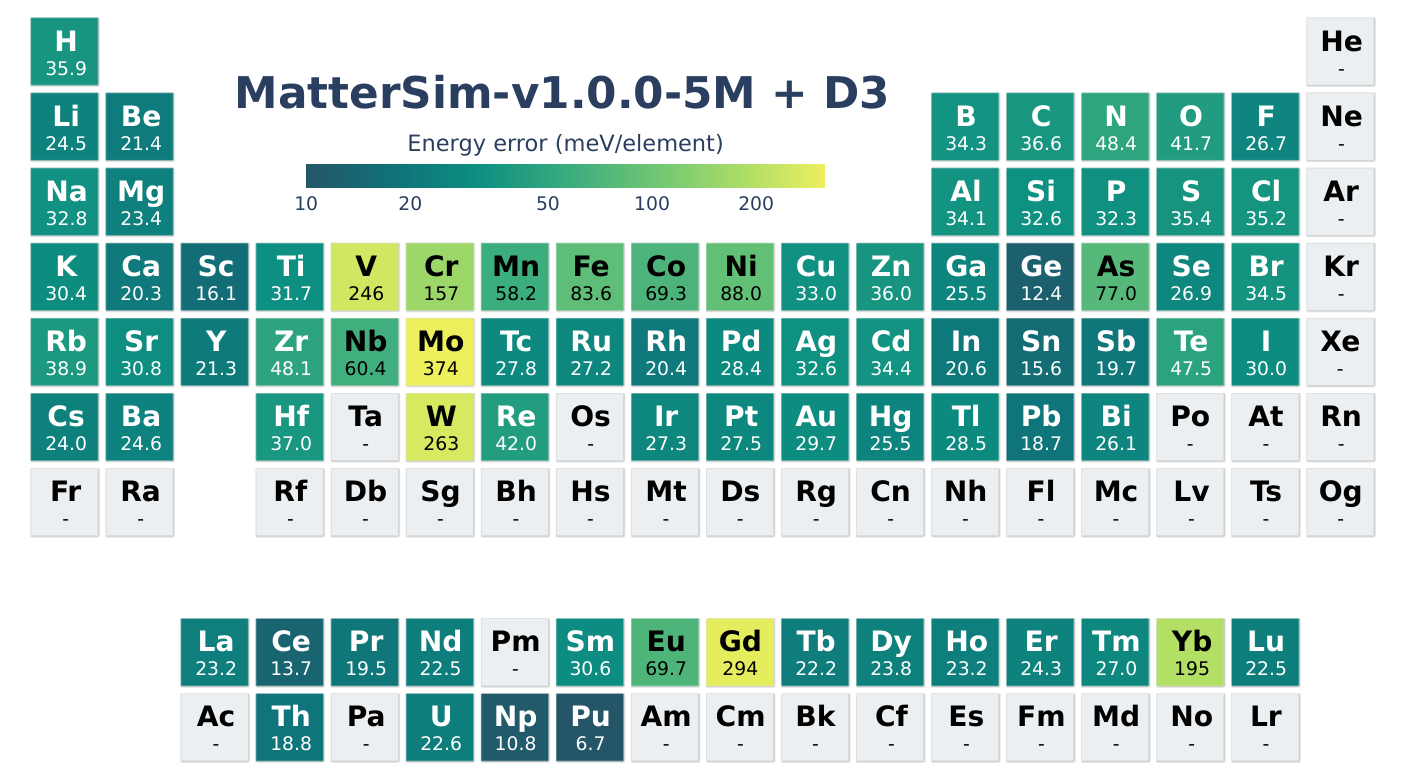}
    \end{subfigure}
    \hfill
    \begin{subfigure}[b]{0.3\textwidth}
        \centering
        \includegraphics[width=\textwidth]{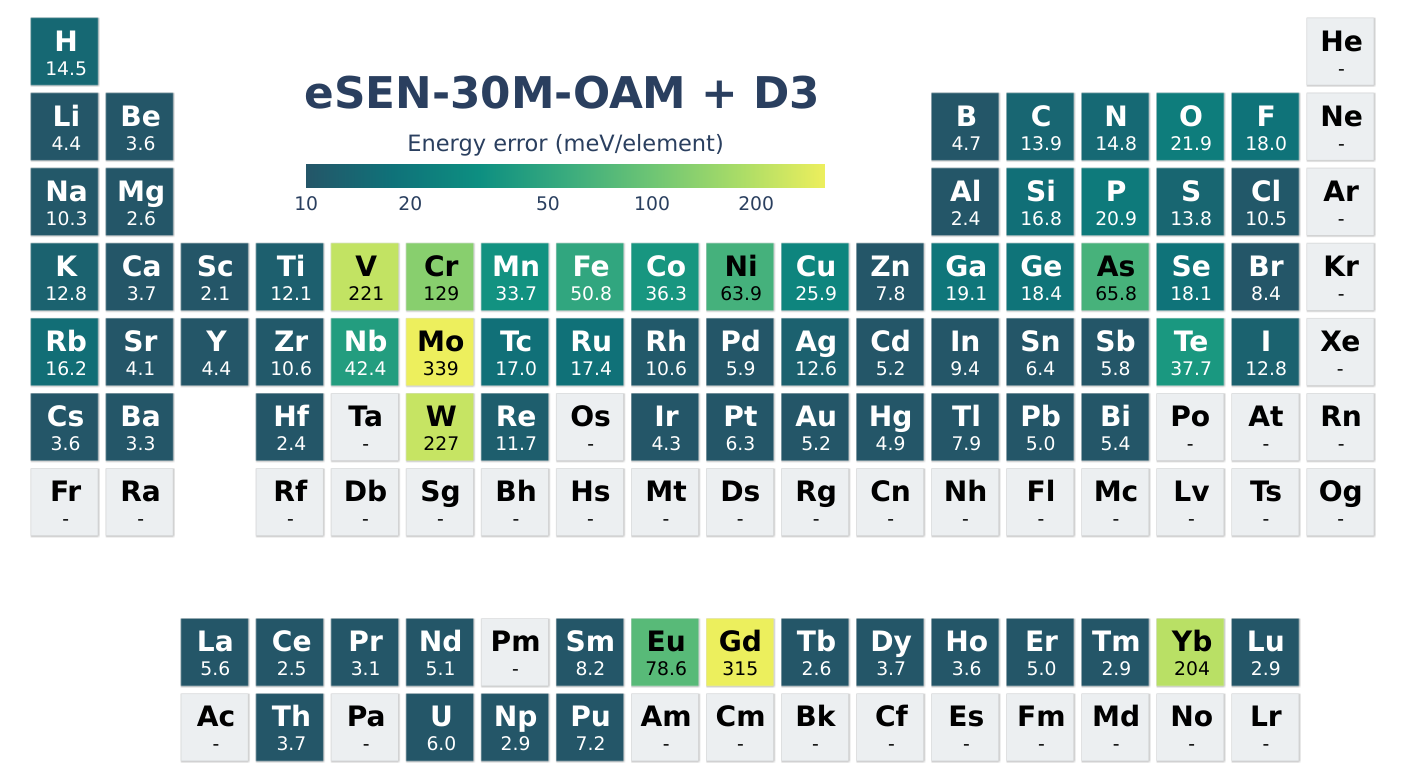}
    \end{subfigure}
    \hfill
    \begin{subfigure}[b]{0.3\textwidth}
        \centering
        \includegraphics[width=\textwidth]{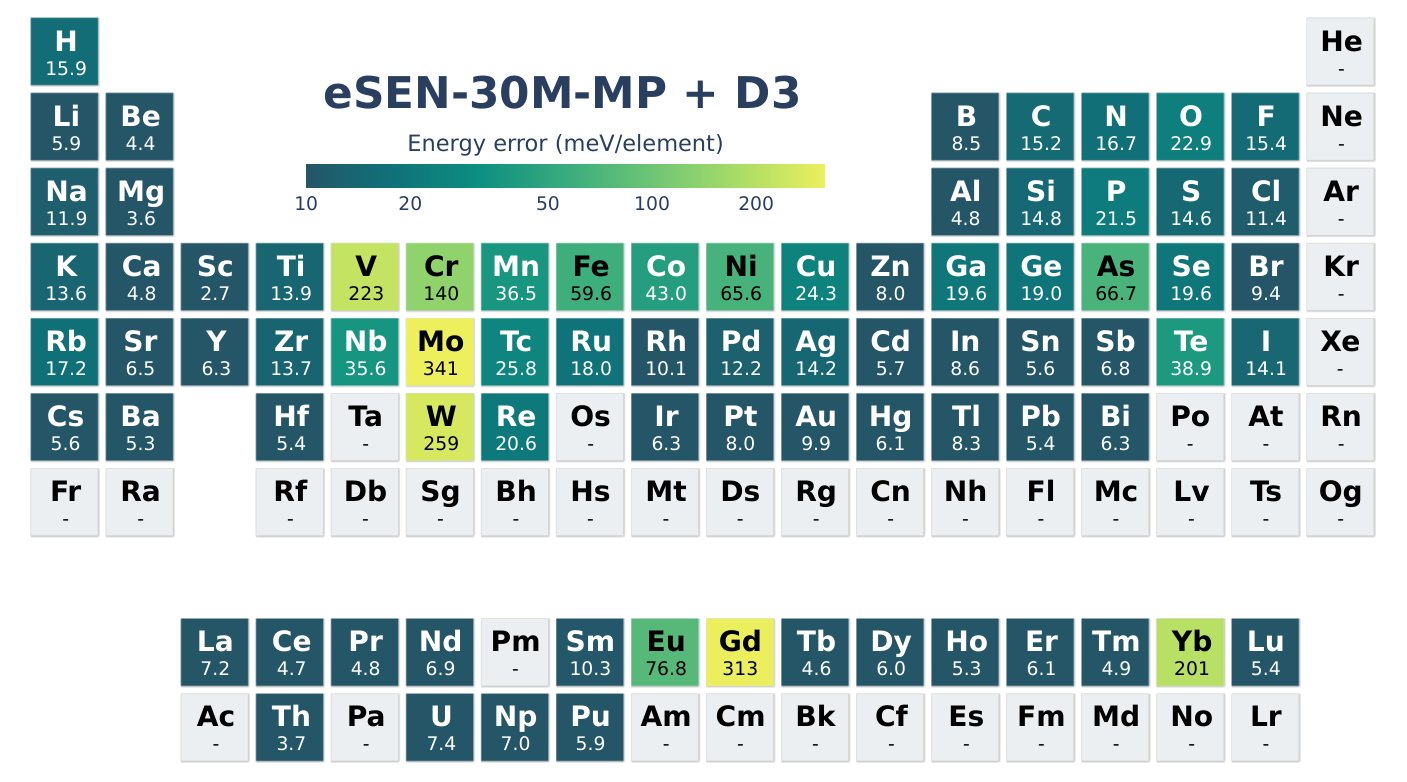}
    \end{subfigure}
    \hfill
    \begin{subfigure}[b]{0.3\textwidth}
        \centering
        \includegraphics[width=\textwidth]{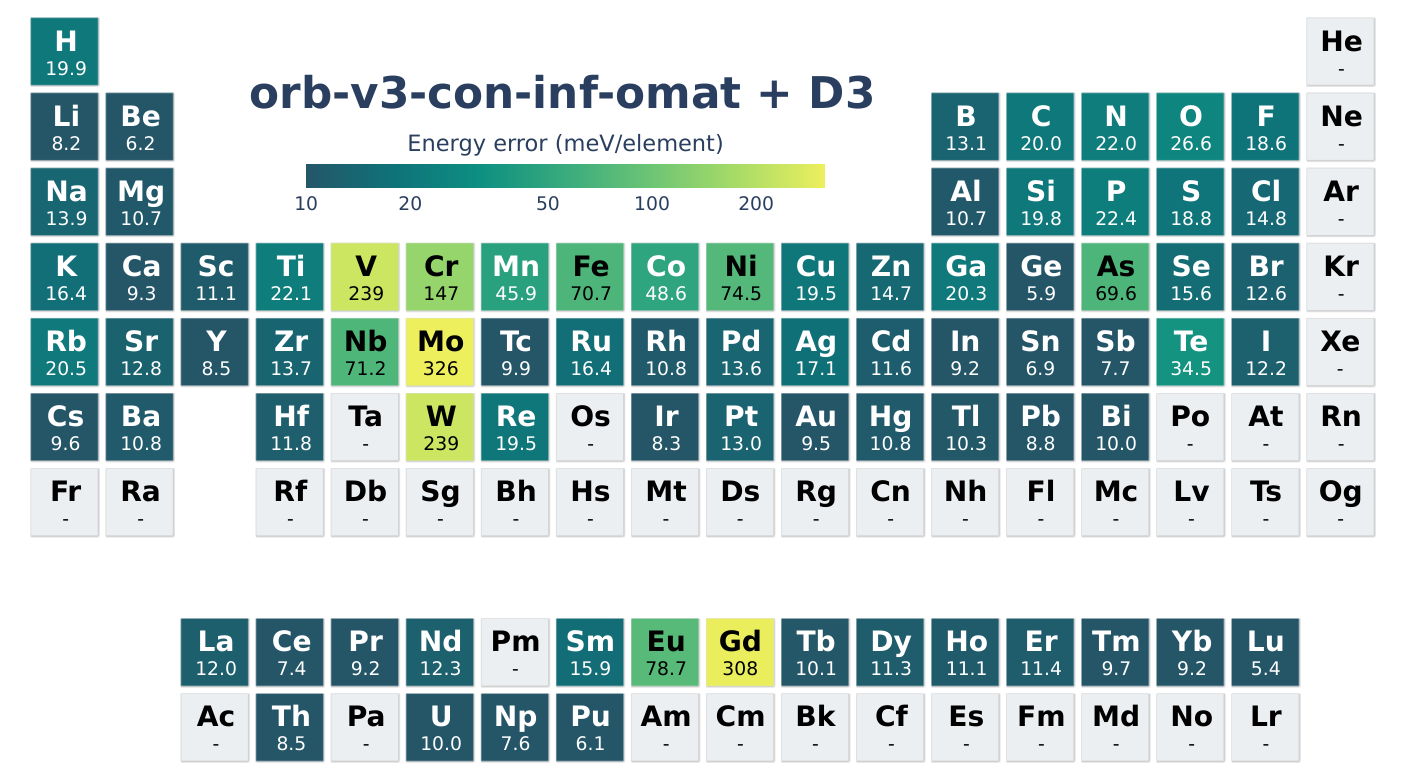}
    \end{subfigure}
    \hfill
    \begin{subfigure}[b]{0.3\textwidth}
        \centering
        \includegraphics[width=\textwidth]{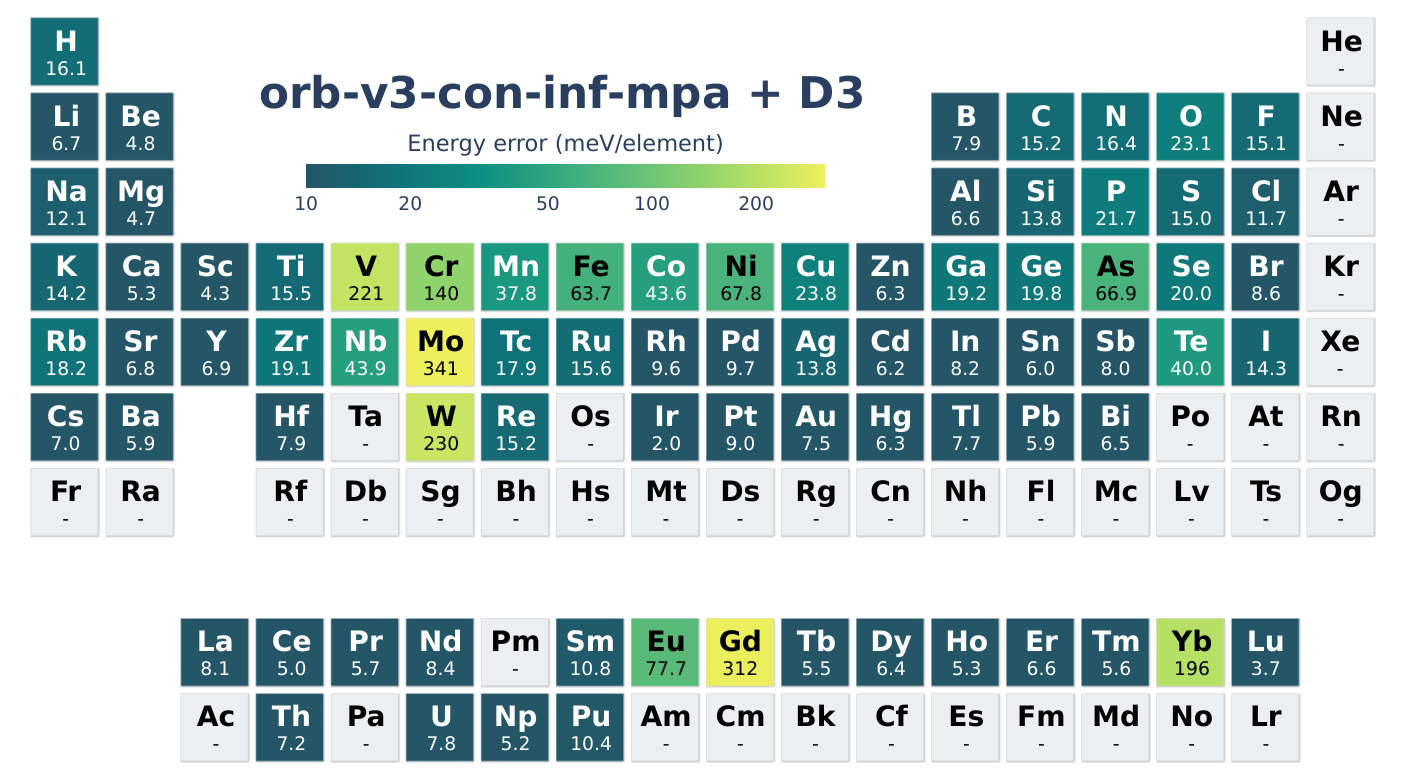}
    \end{subfigure}
    \hfill
    \caption{\textbf{Element-wise mean absolute energy errors of uMLIP-predicted energies compared to PBE energies from QMOF in meV.} The absolute energy error of each structure is distributed over all constituent elements. Gray backgrounds indicate elements not present in the computed set of structures of that model.}
    \label{fig:periodic_tables}
\end{figure}

\subsection{Structural Optimization}

\begin{figure}[H]
    \centering
    \includegraphics[width=\linewidth]{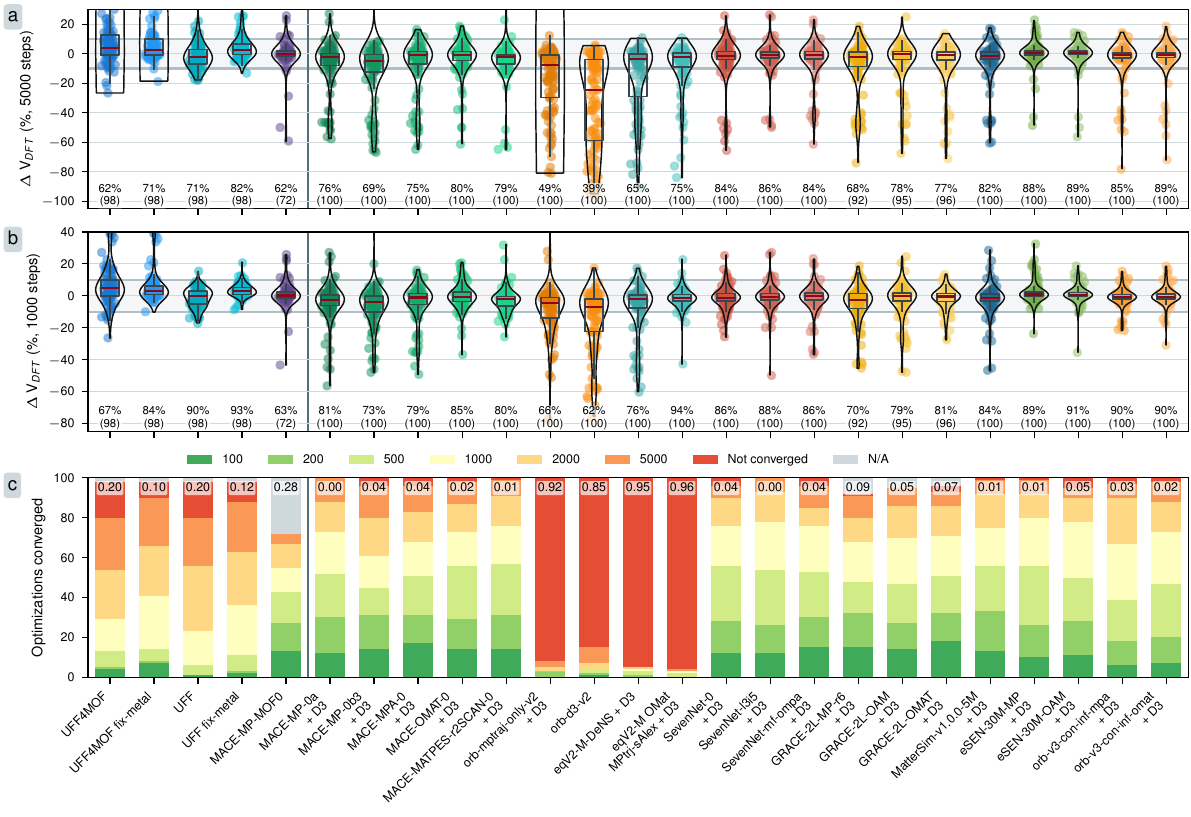}
    \caption{\textbf{Structural optimization behavior of uMLIPs.}
    a) Difference between optimized volumes after up to 5,000 optimization steps and DFT-optimized volumes. b) Difference between optimized volumes after up to 1,000 optimization steps and DFT-optimized volumes. c) Converged structures after up to 5,000 optimization steps, grouped by number of steps until convergence. N/A structures indicate out-of-memory errors or unsupported elements by the model (MACE-MP-MOF0).}
    \label{fig:opt_all}
\end{figure}

\begin{figure}[H]
    \centering
    \includegraphics[width=\linewidth]{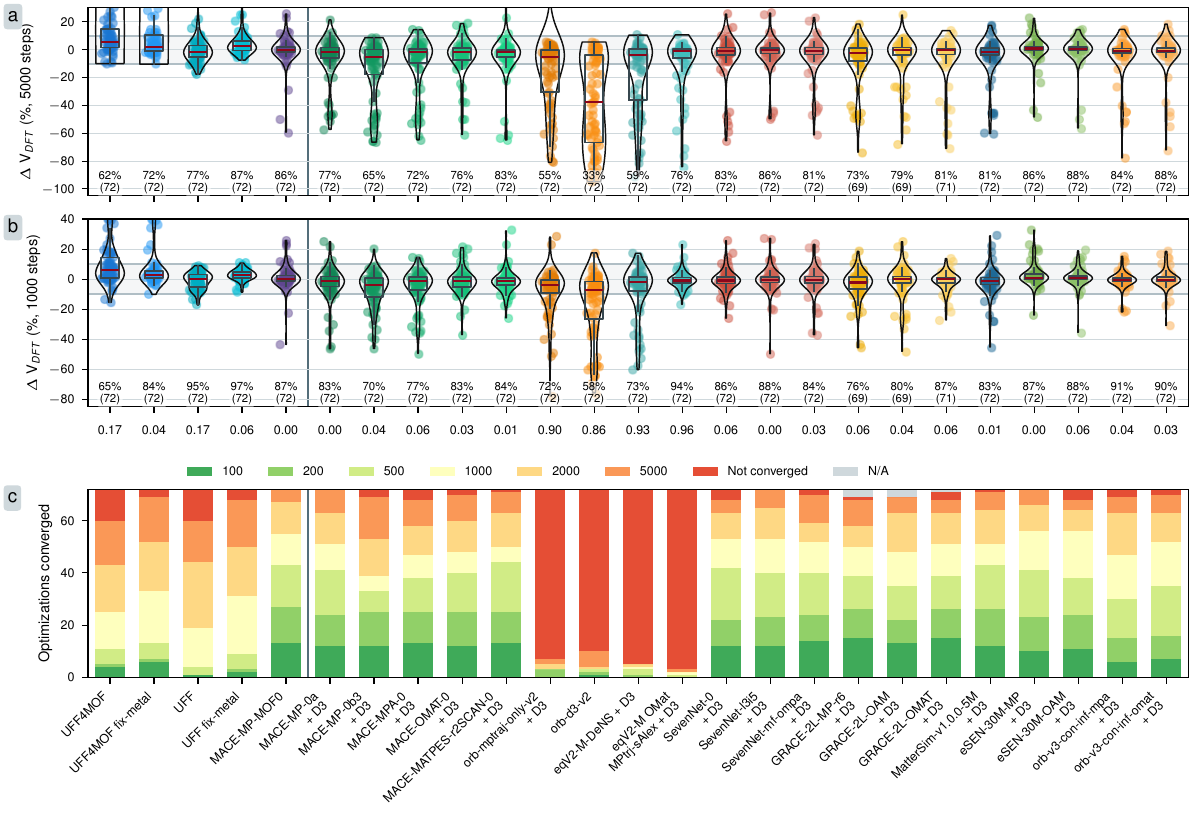}
    \caption{\textbf{Filtered structural optimization behavior of uMLIPs.} Filtered to only include structures that MACE-MP-MOF0 could compute.
    a) Difference between optimized volumes after up to 5,000 optimization steps and DFT-optimized volumes. b) Difference between optimized volumes after up to 1,000 optimization steps and DFT-optimized volumes. c) Converged structures after up to 5,000 optimization steps, grouped by number of steps until convergence. N/A structures indicate out-of-memory errors or unsupported elements by the model (MACE-MP-MOF0).}
    \label{fig:opt_all_intersection}
\end{figure}

\begin{figure}[H]
    \centering
    \includegraphics[width=\linewidth]{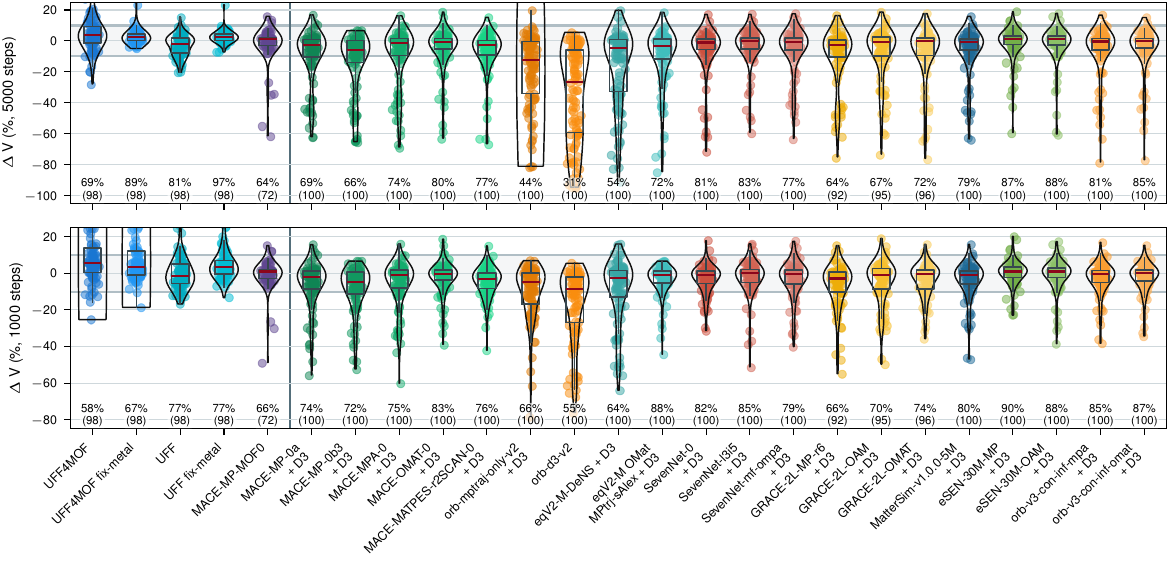}
    \caption{\textbf{Optimization volume delta between initial and final structures.} a) Difference between optimized volumes after up to 5,000 optimization steps and initial volumes. b) Difference between optimized volumes after up to 1,000 optimization steps and initial volumes.}
    \label{fig:opt_volume_change_all}
\end{figure}

\begin{figure}[H]
    \centering
    \includegraphics[width=\linewidth]{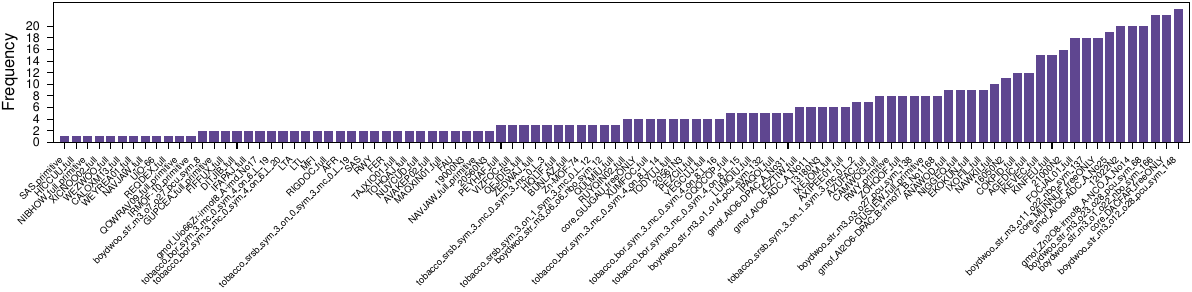}
    \caption{\textbf{Histogram of structures with significant volume changes.} Number of models for which a given structure shows significantly changed volume (>10\,\%) compared to DFT after up to 5,000 structural optimization steps.}
    \label{fig:failed_opt_structure_hist}
\end{figure}

\begin{figure}[H]
    \centering
    \includegraphics[width=\linewidth]{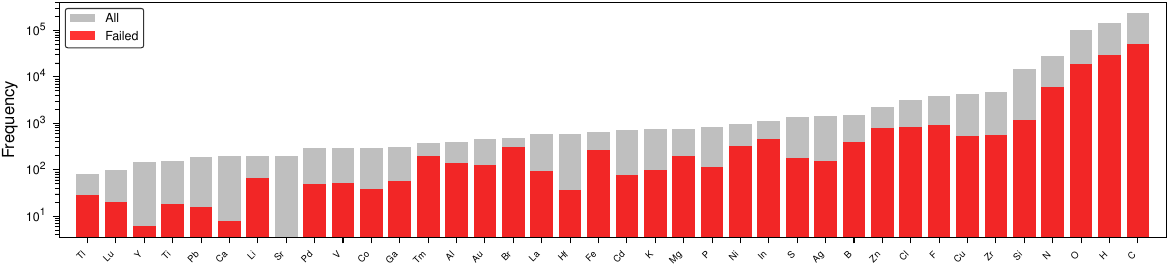}
    \caption{\textbf{Histogram of elements present in structures.} Number of elements present in structures with significant volume changes, shown with the total number of elements present in all structures.}
    \label{fig:failed_symbol_distribution_hist}
\end{figure}

\begin{figure}[H]
    \centering
    \includegraphics[width=0.8\linewidth]{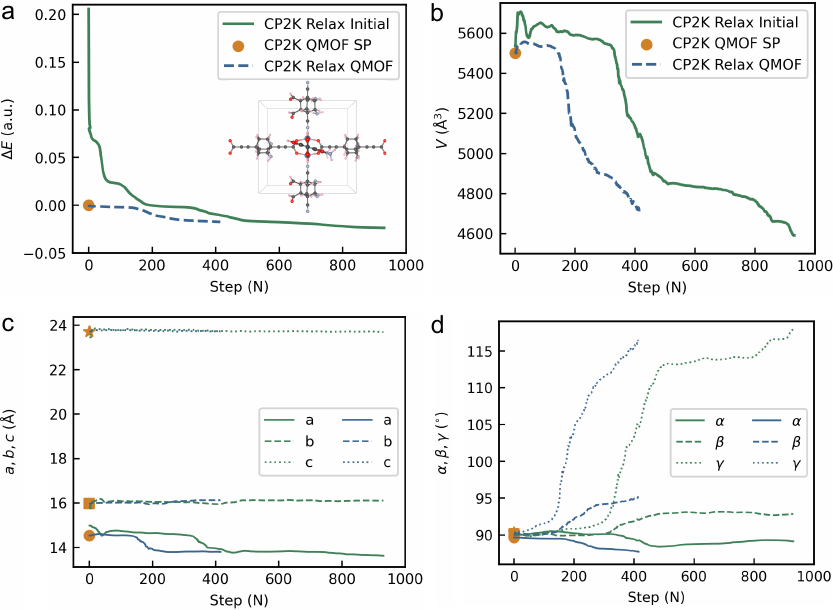}
    \caption{\textbf{CP2K structural optimizations of MOF in this work which identify more stable configurations than that in QMOF DFT optimization.} Structural relaxations were performed using CP2K starting from both the initial structure and from the relaxed structure of \texttt{boydwoo\_str\_m3\_o23\_o28\_pcu\_sym\_68} from QMOF dataset. (\textbf{a}) Energy ($\Delta E$) of the two CP2K optimization trajectories, with energies normalized to the single-point (SP) energy of the QMOF-relaxed structure. (\textbf{b}-\textbf{d}) Volume (V), lattice parameters (a, b, c), and lattice angles ($\alpha$, $\beta$, $\gamma$) during the optimizations. Atom color scheme: \ce{C} (dark grey), \ce{H} (light pink), \ce{O} (red), \ce{N} (light blue), \ce{Zn} (light purple). This example shows how the CP2K settings in this work will find a more energetically favorable configuration compared to that from QMOF DFT optimization.}
    \label{fig:opt_boydwoo_str_m3_o23_o28_pcu_sym_68_primitive}
\end{figure}

\begin{figure}[H]
    \centering
    \includegraphics[width=0.8\linewidth]{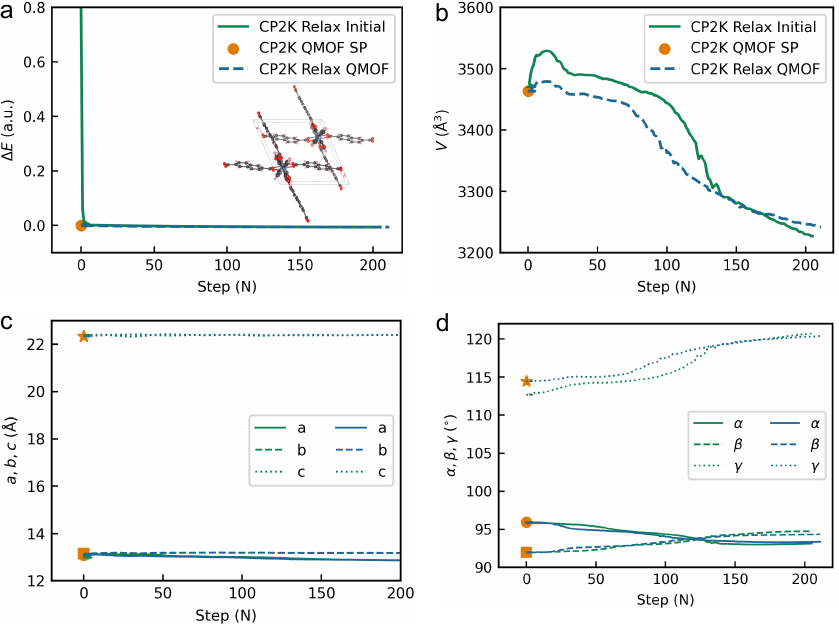}
    \caption{\textbf{CP2K structural optimizations of MOFs in this work with an energetically equivalent but more lattice tilting configuration compared to that from QMOF DFT optimization.} Structural relaxations were performed using CP2K starting from both the initial structure and from the relaxed structure of \texttt{core\_MUNNUF\_freeONLY} from QMOF dataset. (\textbf{a}) Energy ($\Delta E$) of the two CP2K optimization trajectories, with energies normalized to the single-point (SP) energy of the QMOF-relaxed structure. (\textbf{b}-\textbf{d}) Volume (V), lattice parameters (a, b, c), and lattice angles ($\alpha$, $\beta$, $\gamma$) during the optimizations. Atom color scheme: \ce{C} (dark grey), \ce{H} (light pink), \ce{O} (red), \ce{N} (light blue), \ce{Zn} (light purple). This example shows how the CP2K settings in this work will find a more energetically equivalent configuration compared to that from QMOF DFT optimization, but yields a more compact framework from the tilting in the lattice angles.}
    \label{fig:opt_core_MUNNUF_freeONLY}
\end{figure}

\subsection{MD Stability}

\begin{figure}[H]
    \centering
    \includegraphics[width=\linewidth]{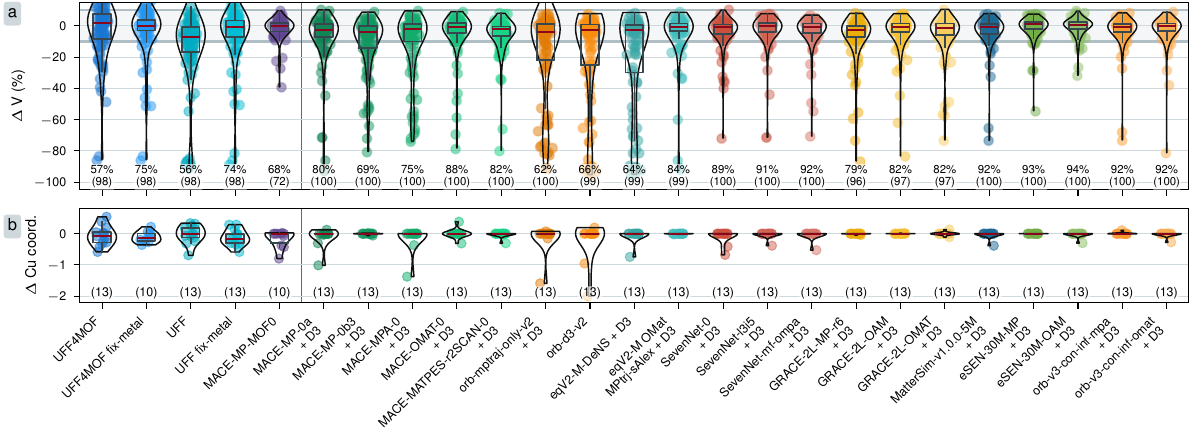}
    \caption{\textbf{NpT simulation stability of uMLIPs.}
    a) Difference between volume measured at the end of 50\,ps long NpT simulations and the initial volume, relative to the initial volume. Grey shaded area indicates a $\pm10$\,\% threshold for outliers. Numbers indicate the percentage of successful and accurate simulations and the number of successfully computed structures. b) Change in copper coordination number from the first snapshot of the simulation to the last snapshot. Each data point shows the difference in average coordination for a single simulated material. Numbers below the violin plots indicate the number of computed structures. }
    \label{fig:stability_volume_copper_coordination_all}
\end{figure}

\begin{figure}[H]
    \centering
    \includegraphics[width=\linewidth]{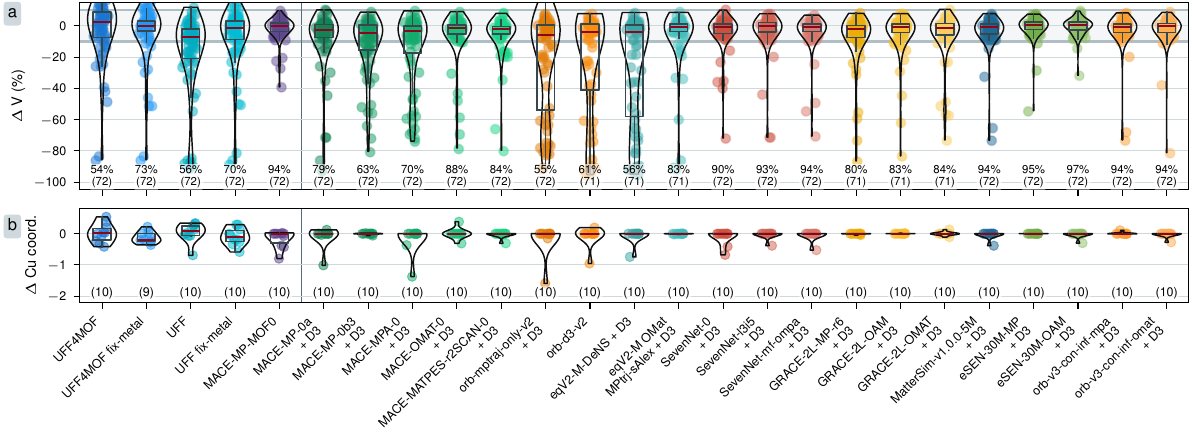}
    \caption{\textbf{Filtered NpT simulation stability of uMLIPs.}
    Filtered only to include structures that MACE-MP-MOF0 could compute. a) Difference between volume measured at the end of 50\,ps long NpT simulations and the initial volume, relative to the initial volume. Grey shaded area indicates a $\pm10$\,\% threshold for outliers. Numbers indicate the percentage of successful and accurate simulations and the number of successfully computed structures. b) Change in copper coordination number from the first snapshot of the simulation to the last snapshot. Each data point shows the difference in average coordination for a single simulated material. Numbers below the violin plots indicate the number of computed structures. }
    \label{fig:stability_volume_copper_coordination_all_intersection}
\end{figure}

\begin{figure}[H]
    \centering
    \includegraphics[height=\dimexpr\textheight-39pt\relax]{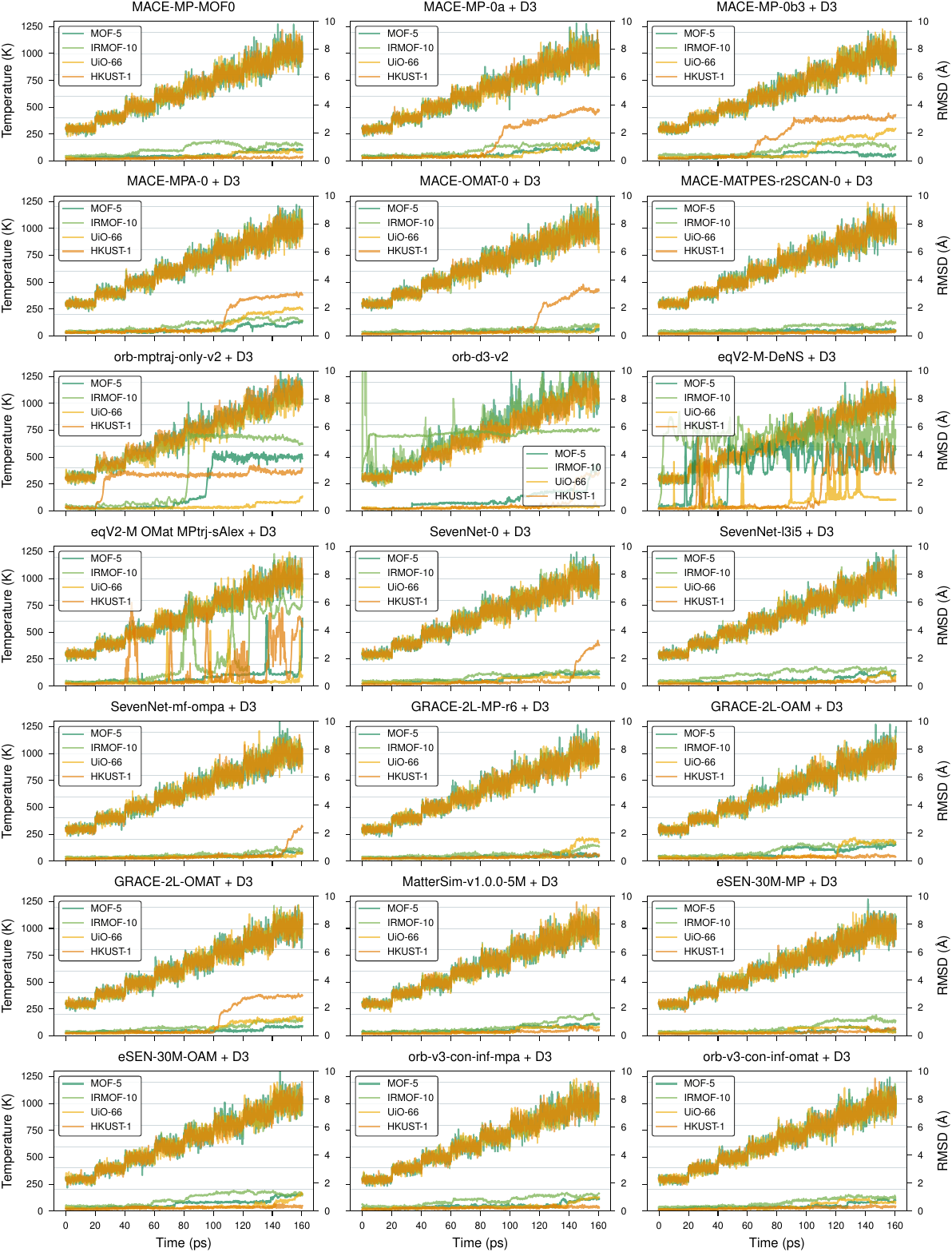}
    \caption{\textbf{Temperature and Root Mean Square Distance of four prototypical MOFs at increasing temperatures.}}
    \label{fig:stability_temp}
\end{figure}

\subsection{Bulk Properties}

\begin{table}[H]
    \centering
    \begin{tabular}{l|c|c|c|c}
        Potential & MOF-5 & IRMOF-10 & UiO-66 & HKUST-1 \\ \hline\hline
        MACE-MP-MOF0 & 15.88 & 9.27 & 36.70 & 22.50 \\
        MACE-MP-0b3 + D3 & 14.61 & 8.28 & 16.77 & 18.38 \\
        MACE-MPA-0 + D3 & 14.62 & 8.77 & 32.51 & 17.18 \\
        MACE-OMAT-0 + D3 & 15.64 & 9.55 & 37.79 & 20.79 \\
        MACE-MATPES-r2SCAN-0 + D3 & 17.31 & 10.21 & 38.63 & 24.73 \\
        orb-d3-v2 & 29.85 & 0.07 & 47.49 & 11.28 \\
        orb-mptraj-only-v2 + D3 & 26.31 & 10.88 & 28.84 & 21.33 \\
        eqV2-M OMat MPtrj-sAlex + D3 & 20.88 & 12.08 & 54.56 & 37.61 \\
        eqV2-M-DeNS + D3 & 21.45 & 11.50 & 40.87 & 24.46 \\
        SevenNet-0 + D3 & 10.54 & 5.34 & 32.98 & 14.35 \\
        SevenNet-l3i5 + D3 & 13.11 & 6.30 & 28.58 & 17.98 \\
        SevenNet-mf-ompa + D3 & 15.18 & 8.76 & 34.66 & 21.07 \\
        GRACE-2L-MP-r6 + D3 & 15.35 & 8.85 & 29.51 & 18.59 \\
        GRACE-2L-OAM + D3 & 14.88 & 8.74 & 33.25 & 20.00 \\
        GRACE-2L-OMAT + D3 & 14.69 & 8.86 & 33.00 & 18.06 \\
        MatterSim-v1.0.0-5M + D3 & 16.02 & 9.20 & 32.06 & 20.69 \\
        eSEN-30M-OAM + D3 & 14.98 & 7.39 & 36.84 & 21.35 \\
        eSEN-30M-MP + D3 & 15.39 & 7.35 & 35.69 & 21.35 \\
        orb-v3-con-inf-omat + D3 & 13.04 & 7.52 & 36.69 & 21.62 \\
        orb-v3-con-inf-mpa + D3 & 12.97 & 6.91 & 35.42 & 19.99 \\
        \hline\hline
        DFT & 16.06 & 9.4 & 37.5 & 23.58 \\
        UFF & 14.5 & 7.6 & 28.7 & 42.4 \\
    \end{tabular}
    \caption{\textbf{Bulk modulus computed with different uMLIPs for MOF-5, IRMOF-10, HKUST-1, and UiO-66.}}
    \label{tab:bulk_modulus_esmof}
\end{table}

\begin{figure}[H]
    \centering
    \includegraphics[width=\linewidth]{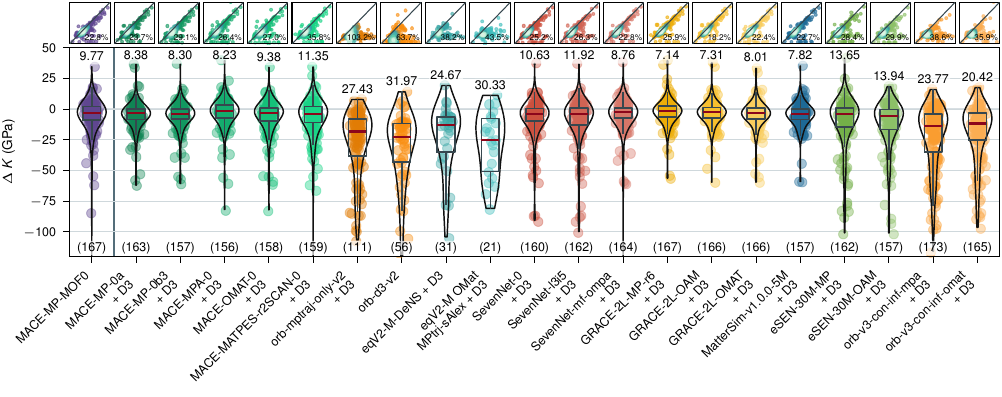}
    \caption{\textbf{Bulk modulus prediction vs. DFT on IZA zeolites.} uMLIP bulk modulus predictions on 177 IZA zeolites, unstable EOS fits as described in the methods section are excluded. DFT references are obtained from Ref.\cite{sours_predicting_2023}\,. Numbers in the parity plots show the MAPE, numbers above the violin plots show the MAE, and numbers at the bottom indicate the number of successfully computed structures.}
    \label{fig:bm_combined_zeolite_all}
\end{figure}

\begin{figure}[H]
    \centering
    \includegraphics[width=\linewidth]{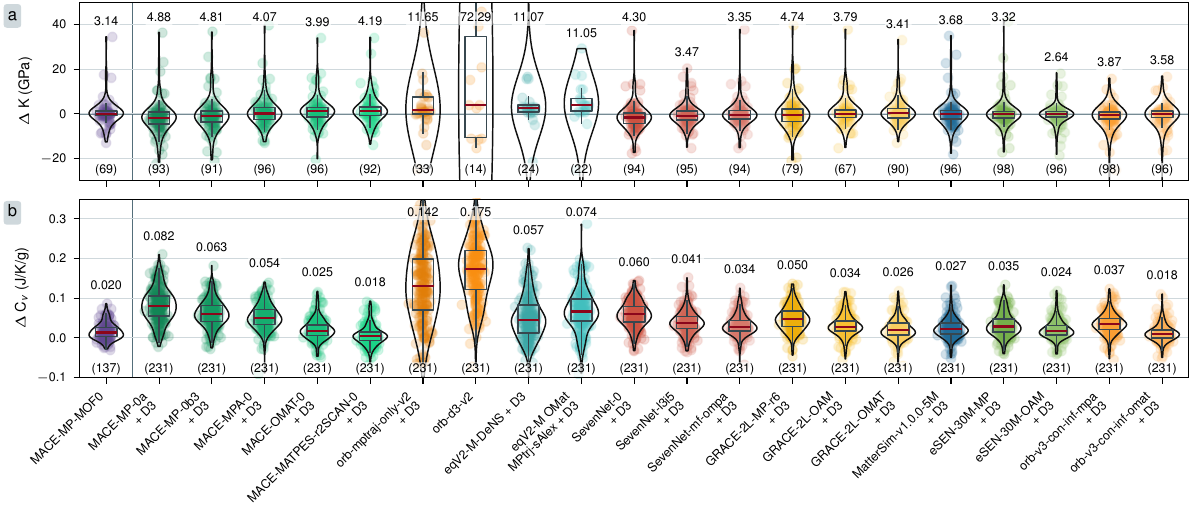}
    \caption{\textbf{Comparison of bulk modulus and heat capacity predictions of uMLIPs vs DFT.} Unstable EOS fits as described in the methods section are excluded.}
    \label{fig:heat_capacity_comparison_all}
\end{figure}

\begin{figure}[H]
    \centering
    \includegraphics[width=\linewidth]{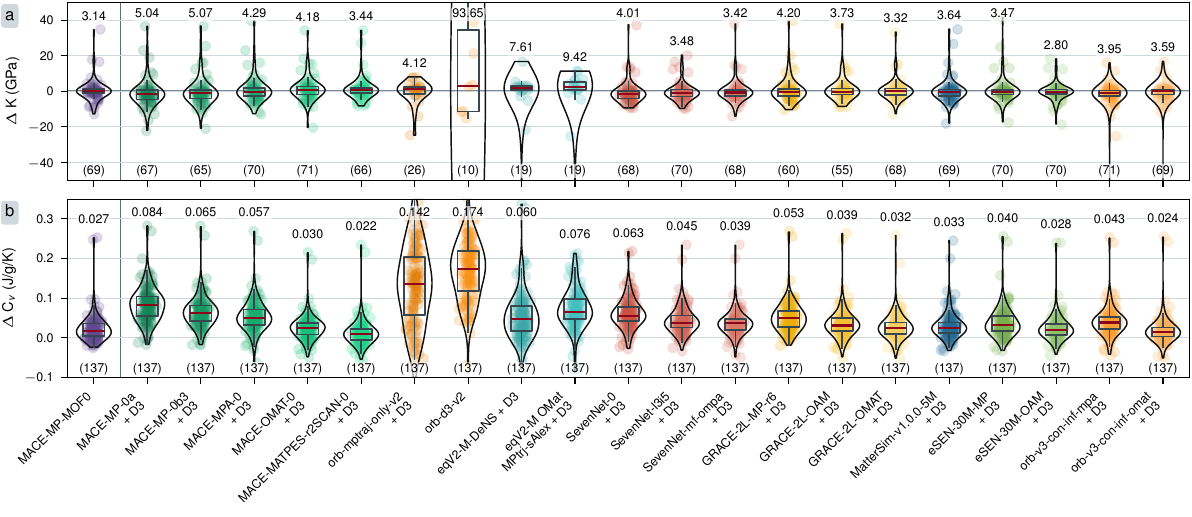}
    \caption{\textbf{Filtered Comparison of bulk modulus and heat capacity predictions of uMLIPs vs DFT, filtered to include only structures MACE-MP-MOF0 can compute.} Structures with unstable EOS fits as described in the methods are excluded.}
    \label{fig:bm_hc_all_intersection}
\end{figure}

\begin{figure}[H]
    \centering
    \includegraphics[width=\linewidth]{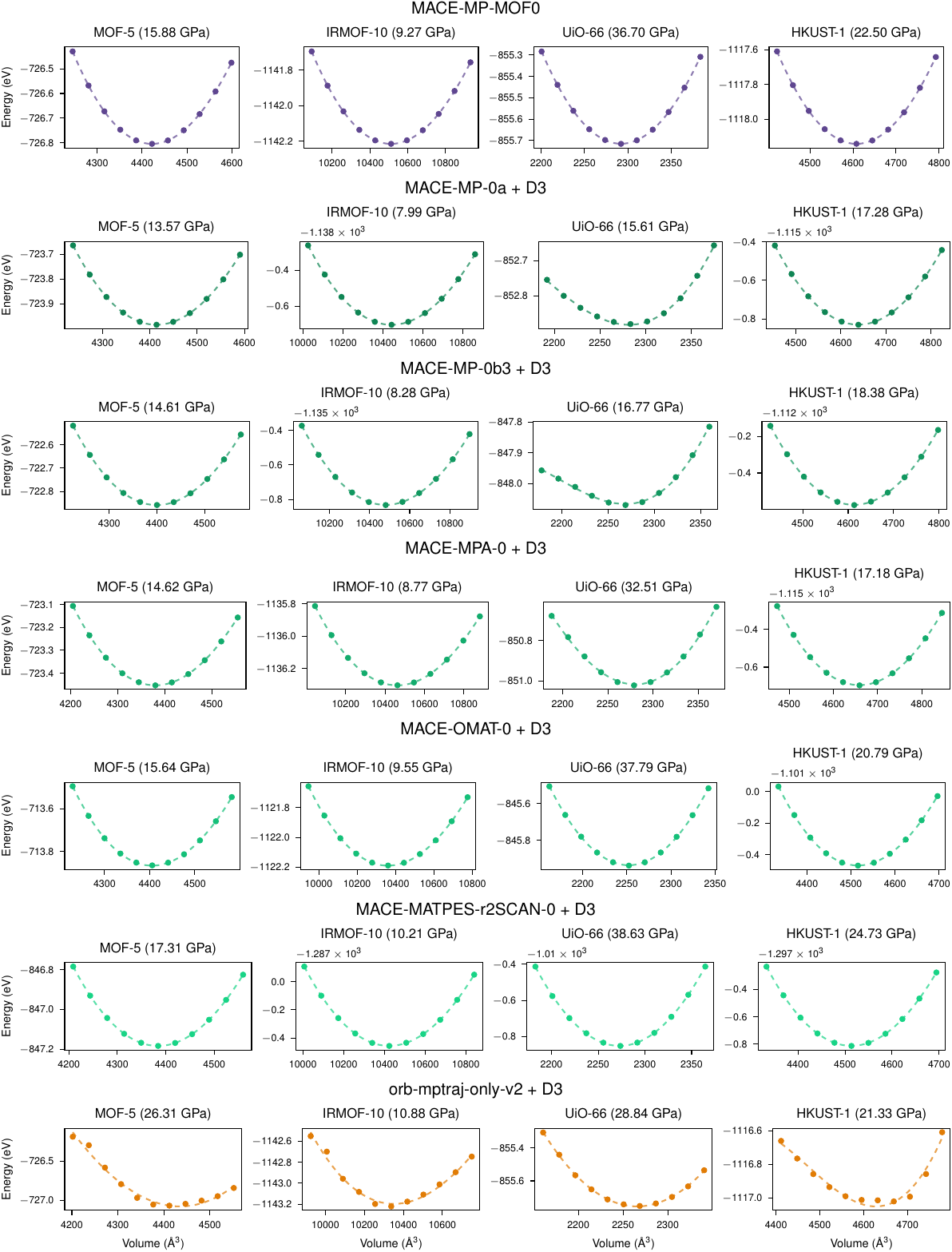}
    \caption{\textbf{Equation of state of four prototypical MOFs.}}
    \label{fig:bulk_modulus_eos_all_0}
\end{figure}

\begin{figure}[H]
    \centering
    \includegraphics[width=\linewidth]{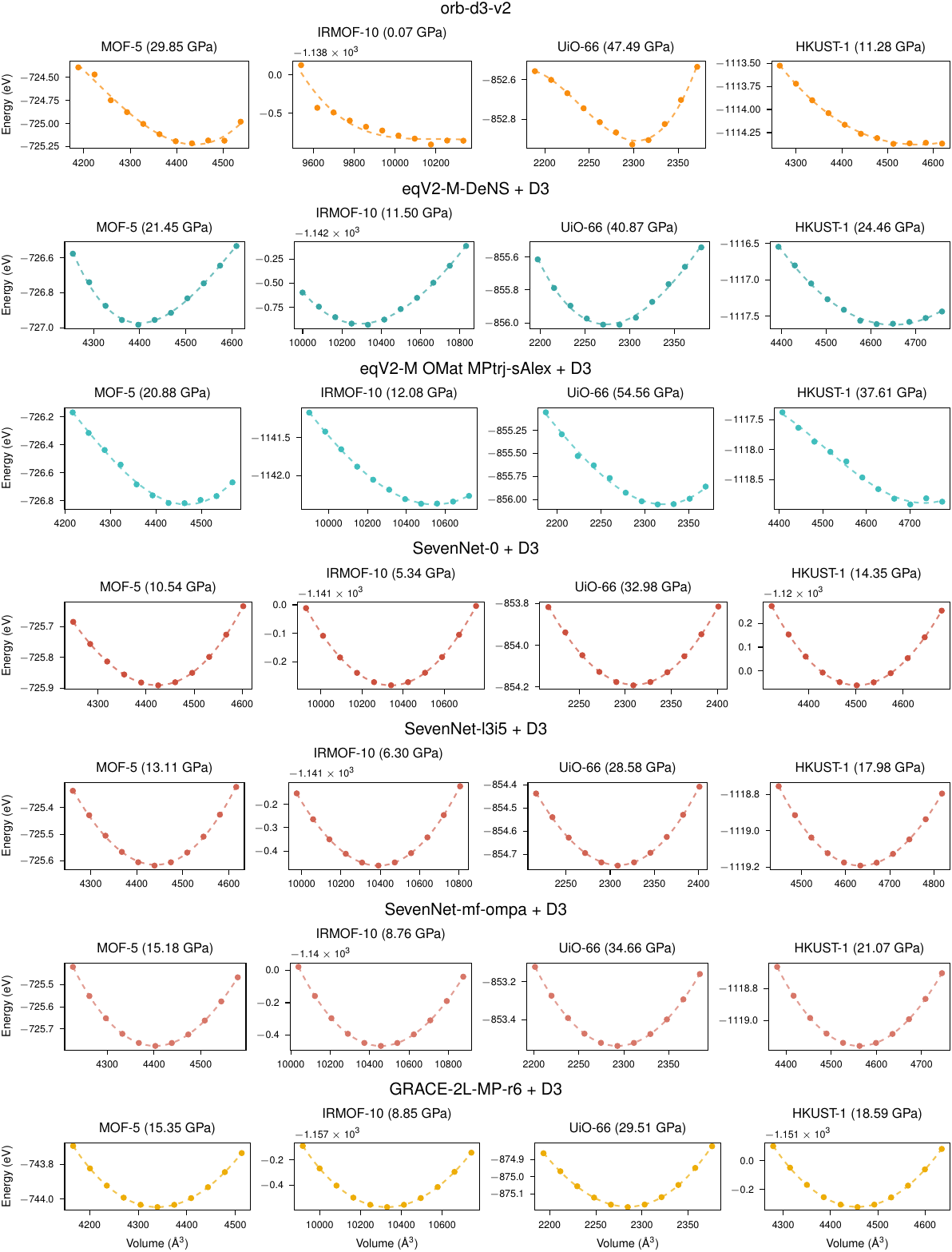}
    \caption{\textbf{Equation of state of four prototypical MOFs.}}
    \label{fig:bulk_modulus_eos_all_1}
\end{figure}

\begin{figure}[H]
    \centering
    \includegraphics[width=\linewidth]{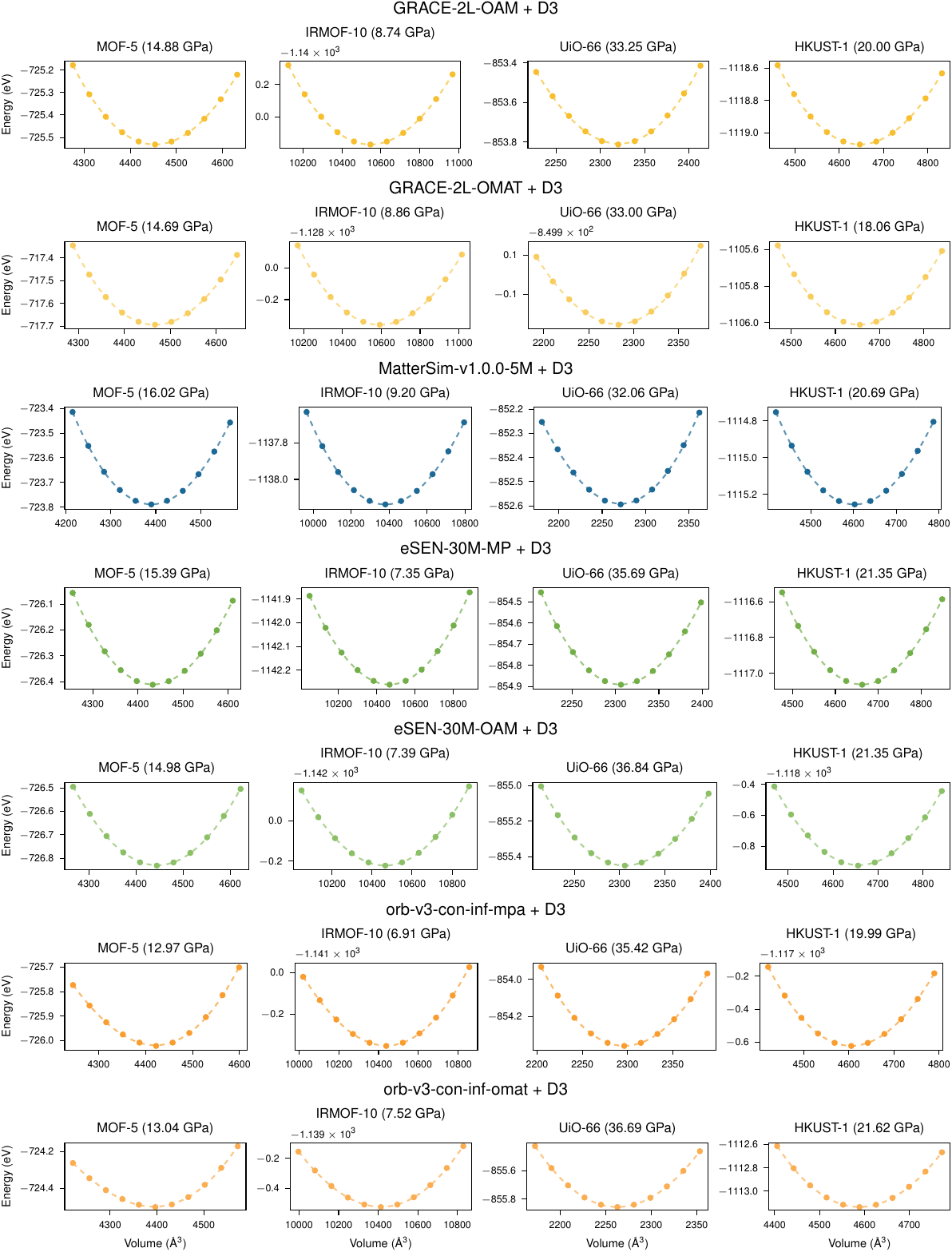}
    \caption{\textbf{Equation of state of four prototypical MOFs.}}
    \label{fig:bulk_modulus_eos_all_2}
\end{figure}

\begin{figure}[H]
    \centering
    \includegraphics[height=\dimexpr\textheight-66pt\relax]{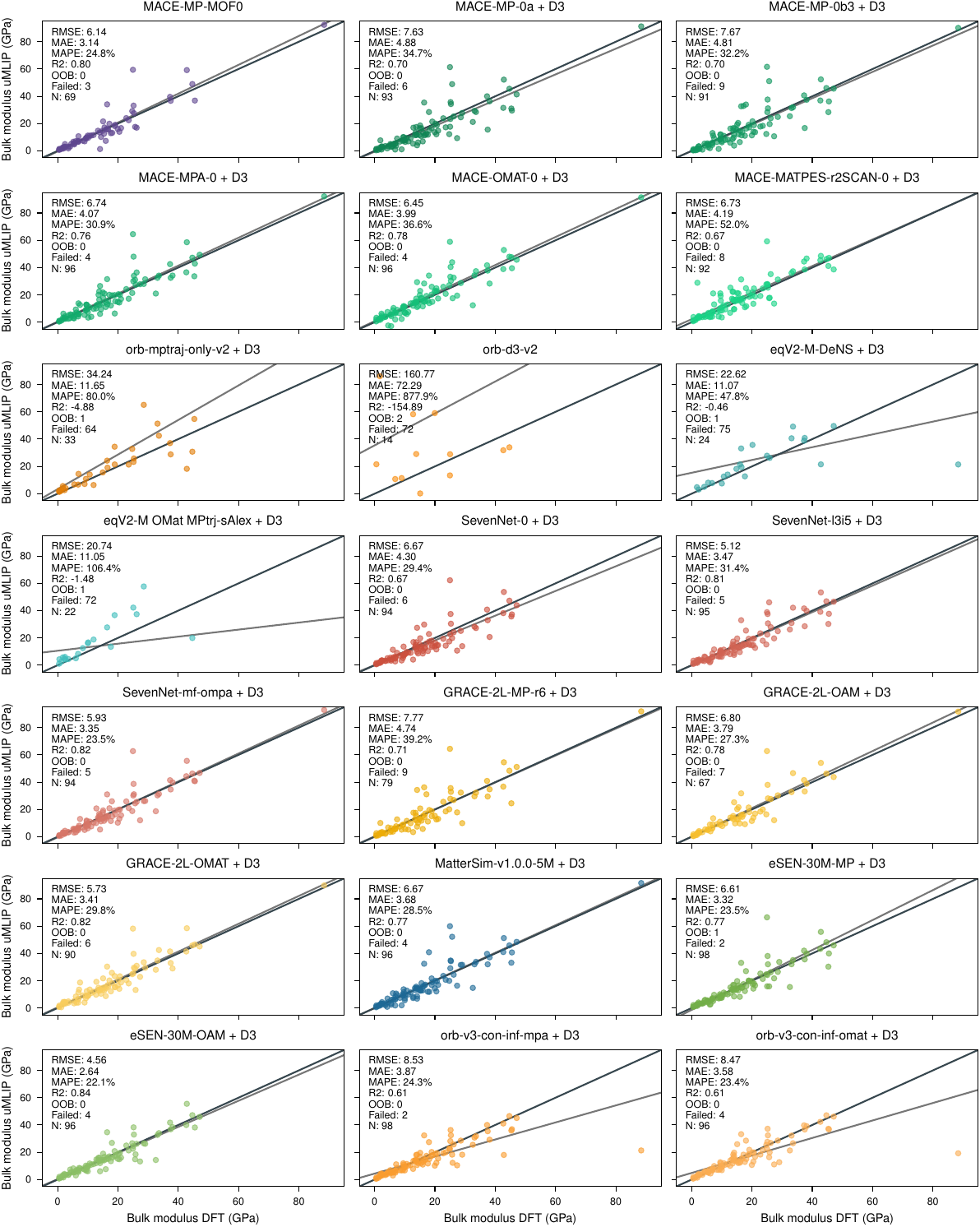}
    \caption{\textbf{Parity plots of uMLIP bulk modulus vs. DFT.}
    OOB (out of bounds) indicates the number of data points outside the shown range. N indicates the number of successfully computed bulk moduli. Errors can include OOM or failed EOS fitting. Failed indicates structures where the Birch-Murnaghan-fitted optimal volume differs more than 1\,\% from the MLIP-optimized volume, which are not shown in the plot. A linear best fit is shown in gray.}
    \label{fig:bulk_modulus_parity_all}
\end{figure}

\begin{figure}[H]
    \centering
    \includegraphics[width=\textwidth]{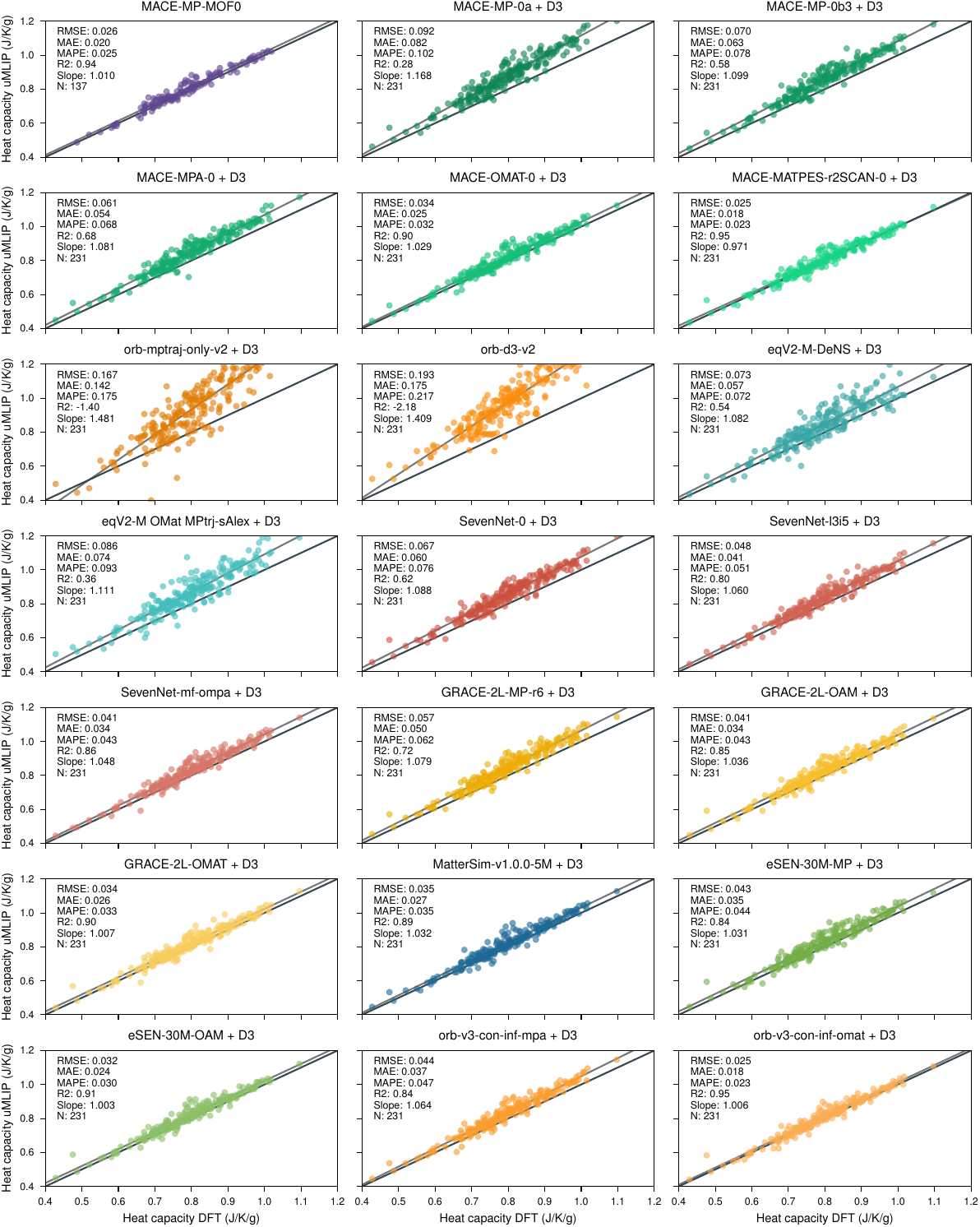}
    \caption{\textbf{Parity plots of uMLIP heat capacity vs. DFT.}
    N indicates the number of successfully computed bulk moduli. A linear best fit is shown in gray.}
    \label{fig:heat_capacity_parity_all}
\end{figure}

\subsection{Interaction Energy}

\begin{figure}[H]
    \centering
    \includegraphics[width=\textwidth]{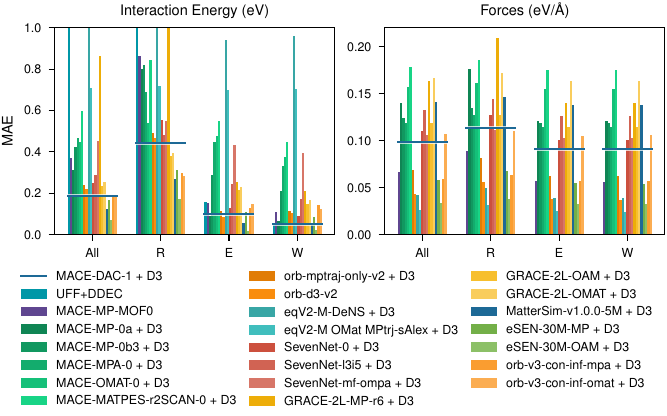}
    \caption{\textbf{Comparison of interaction energy errors and force errors.} Results were computed on the test set of GoldDAC \cite{Lim_2025_DAC-SIM}. Interaction energies are computed as $E_\text{interaction} = E_\text{total} - E_\text{MOF} - E_\text{gas}$, where $E_\text{total}$ is the potential energy of the MOF with an inserted gas molecule, $E_\text{MOF}$ is the potential of the MOF framework only, and $E_\text{gas}$ is the energy of the gas molecule, either \ch{CO2} or \ch{H2O}, generated with ase. The framework structures were obtained from the GoldDAC test structures, and the inserted molecules were removed.}
    \label{fig:dac_all}
\end{figure}

\begin{figure}[H]
    \centering
    \includegraphics[width=\linewidth]{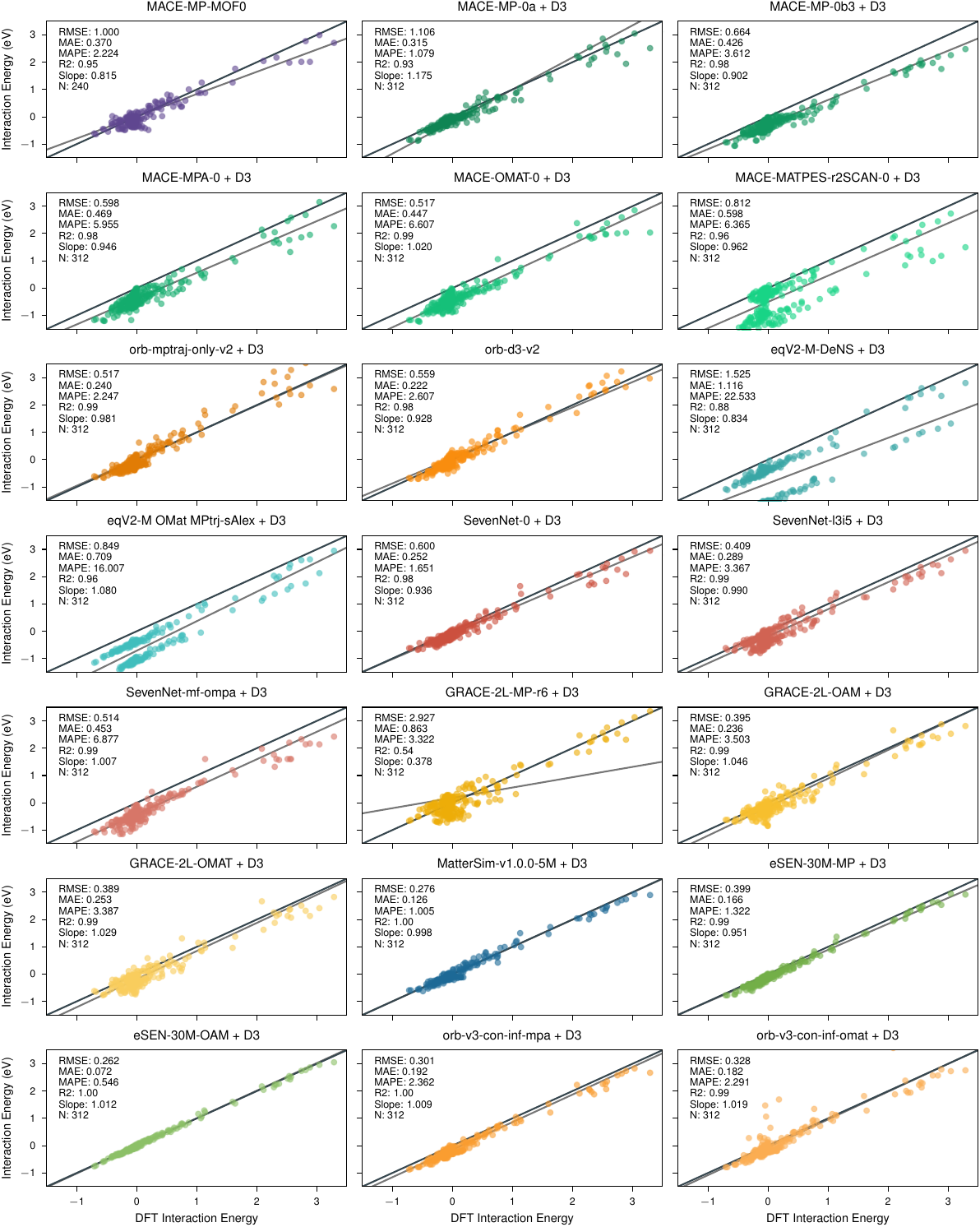}
    \caption{\textbf{Interaction Energy Parity Plot.} Comparison of DFT interaction energies with uMLIP interaction energies. A gray line indicates a linear best fit.}
    \label{fig:dac_parity_all}
\end{figure}

\begin{figure}[H]
    \centering
    \includegraphics[width=\linewidth]{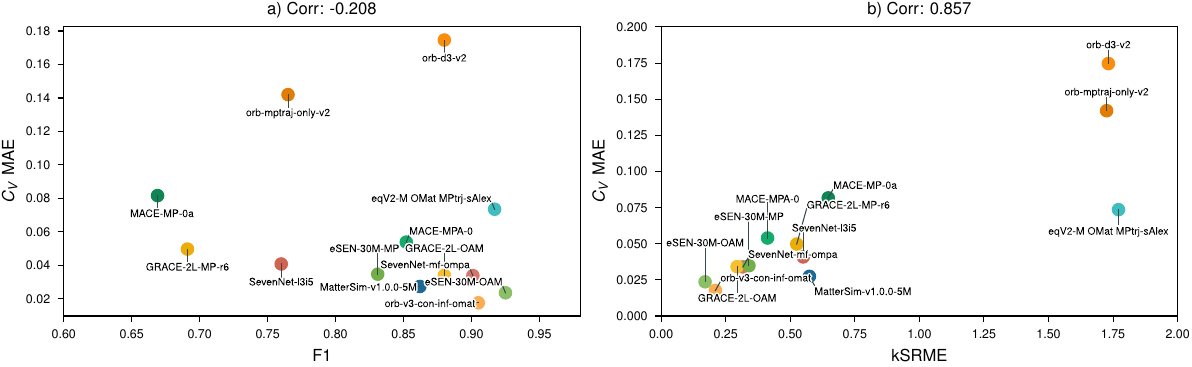}
    \caption{\textbf{Comparison of heat capacity $C_V$ MAE against Matbench Discovery \cite{riebesell_matbench_2023} F1 and $\kappa$SRME scores \cite{póta_thermalconductivitypredictionsfoundation_2024_ksrme}.}
    Better models exhibit lower $C_V$ MAE and $\kappa$SRME, and higher F1 scores. A perfect correlation would correspond to a value of $-1$ between $C_V$ MAE and F1, and $+1$ between $C_V$ MAE and $\kappa$SRME.
    }
    \label{fig:matbench_corr}
\end{figure}

\end{document}